\documentclass[aps,prb, onecolumn]{revtex4-2}

\usepackage[utf8]{inputenc}
\usepackage{amsthm}
\usepackage{amsmath}
\usepackage{color}
\usepackage{hyperref}
\usepackage{graphicx,xcolor}
\usepackage{amssymb}
\usepackage{bbold}
\usepackage{gensymb}
\usepackage[export]{adjustbox}
\usepackage{soul}
\newcommand{\ii}{\mathrm{i}} %imaginary unit
\newcommand{\eul}{\mathrm{e}} %exp(1)
\newcommand{\diff}{\mathrm{d}} %d
\newcommand{\id}{\mathbb{1}} %d
\newcommand{\ket}[1]{|#1\rangle} %ket
\newcommand{\bra}[1]{\langle#1|} %bra
\newcommand{\braket}[1]{\langle#1\rangle} %braket
\newcommand{\ketbra}[2]{|#1\rangle\!\langle #2|} %ketbra
\newcommand{\tr}{\mathrm{tr}} %trace
\newcommand{\emphquote}[1]{‘#1’} %quote

\begin{document}

\title{Dynamical Phase Transitions, Caustics, and Quantum Dark Bands}
\author{Valentin Link}
\affiliation{Institut f{\"u}r Theoretische Physik, Technische Universit{\"a}t Dresden, 
D-01062, Dresden, Germany}
\date{\today}

\author{Walter T. Strunz}
\affiliation{Institut f{\"u}r Theoretische Physik, Technische Universit{\"a}t Dresden, 
D-01062, Dresden, Germany}

\author{D. H. J. O'Dell}
\affiliation{Department of Physics and Astronomy, McMaster University, 1280 Main St. W., Hamilton, Ontario, Canada L8S 4M1
}
\email{dodell@mcmaster.ca}

\begin{abstract}
We provide a new perspective on quantum dynamical phase transitions (DPTs) by explaining their origin in terms of caustics that form in the Fock space representation of the many-body state over time, using the fully connected transverse field Ising model as an example. In this way we establish a connection between DPTs in a quantum spin system and an everyday natural phenomenon: The dark band between the primary and seconday bows (caustics) in rainbows known as Alexander's dark band. The DPT occurs when the Loschmidt echo crosses the switching line between the evanescent tails of two back-to-back Airy functions that dress neighbouring fold caustics in Fock space and is the time-dependent analogue of what is seen as a function of angle in the sky. The structural stability and universal properties of caustics, as described mathematically by catastrophe theory, explains the generic occurrence of DPTs in the model and suggests that our analysis has wide applicability. Based on our thorough analytical understanding we propose a protocol which can be used to verify the existence of a DPT in a finite system experiment.
\end{abstract}
\maketitle

\section{Introduction}

Phase transitions are characterised by singular changes in the properties of a macroscopic system as a control parameter is varied smoothly \cite{yanglee1952i,yanglee1952ii}. This non-analyticity is the source of the self-similar scaling laws and universality that occurs across very different physical systems close to continuous phase transitions. Whereas the traditional paradigm of phase transitions concerns changes to the equilibrium state, which in the case of quantum phase transitions usually implies the ground state \cite{Sondhi97}, the study of more general non-equilibrium phase transitions that occur in driven-dissipative quantum systems also has a long history \cite{Walls1978,Torre2013Feb}.
More recent developments are connected to the search for universality in general quantum dynamics \cite{Polkovnikov11,Eisert15,Prufer2018Nov,Eigen18,Erne18}, and include excited state quantum phase transitions (ESQPTs) \cite{Ribeiro07,Caprio2008,Brandes13,Santos16,Cejnar2021,Marino_2022} that arise from the diverging density of states that occurs in the semiclassical regime at energies corresponding to a separatrix in classical phase space \cite{Smerzi97,Cataliotti01,Albiez2005Jun,Levy2007Oct,Zibold10,Leblanc2011,Gerving12,Trenkwalder16}. The control parameter in ESQPTs is the energy relative to the ground state and can be set by a well controlled initial quench in which a parameter in the Hamiltonian is suddenly changed.

In this paper we focus on another type of non-equilibrium phase transition where the control parameter is time itself. In this case, which has been categorized by Corps \textit{et al}  \cite{Corps2023Mar} as a type II dynamical phase transition, an observable changes non-analytically at a finite critical time following the initial quench. We refer to this scenario simply as a \textit{dynamical phase transition} (DPT). A DPT is different to an ESQPT where, once launched at a particular energy, the system remains in the same phase over time (although there can be situations where there is an initial transient \cite{Marino_2022,Corps2023Mar}).

\begin{figure}
    \centering
    \includegraphics{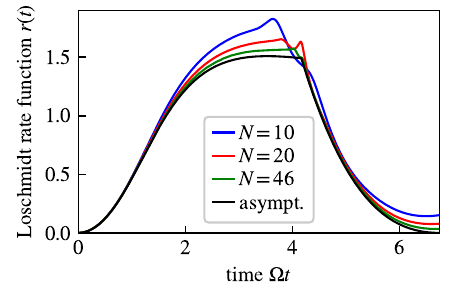}    \caption{Paradigmatic example of a dynamical phase transition. The Loschmidt rate function acquires a nonanalytic kink in the limit of large system size $N\rightarrow\infty$ at a finite time \cite{Lang2018Sep}. The model used here and throughout this paper is the fully-connected transverse field Ising model given in Eq.\ \eqref{eq:FCTFI_Hamiltonian}. In this example we have choosen the ground state at zero transverse field ($\Omega=0$)  as the initial state, which we denote as $\ket{\psi(0)}=\ket{0}^{\otimes N}$, and at $t=0$ we quench the transverse field through the quantum phase transition point from $\Omega=0$ to $\Omega=G$.  }
    \label{fig:fctfi_1}
\end{figure}

DPTs have been primarily studied  in the context of quantum many-body systems, both theoretically \cite{Heyl2013Mar,Diehl2010Jul,Heyl2019Feb,Hunyadi04,Eisler18,Trapin2018May,Lacki2019Mar,Khasseh2020Jul,Halimeh2021Sep,Pallister22,VanDamme2023Aug,Cheraghi2023Oct} and experimentally \cite{Jurcevic2017Aug,Zhang17,Flaschner2018Mar}, although they also appear in classical physics \cite{Janas2016Sep,Smith2018Apr,Baek2019Oct}.  In contrast to equilibrium transitions, which give rise to a global reorganization of the system that can be detected by macroscopic observables such as the magnetization of a bulk metal, the DPT phenomenon is rather subtle and difficult to detect directly in experiments because it involves measuring an exponentially suppressed quantity such as an overlap of many-body quantum states or the exponential tail of a classical probability distribution.
In keeping with most works on DPTs, we focus in this paper on the probability to return to the initial state via the dynamics after time $t$. In the quantum case the return probability is known as the Loschmidt echo. Specifically, let $\ket{\psi(t)}$ be the time evolved state of a many-body system consisting of $N$ particles. Then the Loschmidt echo is given by
\begin{equation}
    L(t)=|\braket{\psi(0)|\psi(t)}|^2\equiv\eul^{-Nr(t)}.
    \label{eq:loschmidt_defn}
\end{equation}
The overlap of two distinct many-body states is generally expected to be exponentially small in the system size, a property known as the \emphquote{orthogonality catastrophe}---almost all states in Hilbert space become orthogonal in the large-$N$ limit. This makes experimentally measuring the Loschmidt echo very difficult in a many-body context. Mathematically, we can describe the Loschmidt echo of a many-body system with the rate function $r(t)$ (see Eq.~\eqref{eq:loschmidt_defn}) which typically has a finite limit as $N\rightarrow\infty$. DPTs are characterized by nonanalytic behaviour of this function.  A typical example of the dynamics of the rate function is displayed in Fig.~\ref{fig:fctfi_1}. We see that the  return function develops a kink as the system size is increased, and the limiting function for $N\rightarrow\infty$ is clearly nonanalytic. This characteristic `sharkfin' shape is common to many DPTs, yet, the shape of the curve and the position of the kink (critical time) depend on the specific model in a highly non-trivial way \cite{Heyl2019Feb}.

The main innovation of the present paper is to point out a surprising connection between DPTs in a quantum system and rainbows. Specifically, we show that in a paradigmatic spin model, the DPTs occur in a special window of times 
that is closely analogous to Alexander's dark band (DB), which is the name given to the dark region in the sky between the primary and secondary bows of an ordinary optical rainbow, and is named after Alexander of Aphrodisias who described it around A.D. 200 \cite{Nussenzveig1977}. 
In the context of the quantum spin system we call the analogoue of this phenomenon the \textit{quantum dark band}. 
At the heart of our treatment is the observation that both Alexander's DB and the quantum DB are bounded by caustics, which are singularities of the classical theory where the probability density diverges (in the rainbow case the two caustics are precisely the primary and secondary bows). Caustics have the special property of structural stability meaning that they are stable against perturbations and hence have universal qualitative properties that are independent of the quantitative details of the system under consideration \cite{Nye_natural_focusing,berry81}.

Our key finding is that the nonanalytic behaviour of the Loschmidt echo can be linked to a switching in exponential dominance of the evanescent tails from the two bounding caustics.  This interpretation not only directly implies the structural stability of DPTs (via the structural stability of caustics) but can also be utilized for analytic calculations using methods such as Poisson resummation, and in this way we are able to obtain exact limiting expressions for the Loschmidt echo and to propose a measurement scheme for the DPTs with finite resources.

The universal properties of caustics can be explained by catastrophe theory \cite{Thom1975,Arnold1975} (not to be confused with the orthogonality catastrophe, which is unrelated). This shows that structurally stable singularities are partitioned into distinct equivalence classes; in the two dimensions relevant to this work the two classes are folds and cusps, or $A_{2}$ and $A_{3}$ in the ADE classification scheme. Each equivalence class comes with its own set of scaling exponents under which the catastrophes display self-similar scaling. This is reminiscent of the universality classes and critical exponents of equilibrium phase transitions, except that in our treatment the non-analyticity occurs in the action rather than the free energy and is therefore intrinsically dynamical. Thus, the catastrophe theory approach naturally emphasizes the generic structure and universal aspects of the dynamics associated with DPTs.  Furthermore, an appealing feature of catastrophe theory is that each class of catastrophe corresponds to a particular morphology (shape) of singularity and this allows for a particularly visual and hence intuitive interpretation of DPTs.  

The plan for the rest of this paper is as follows. In Section \ref{sec:model} we specify the model we use, which is the fully connected transverse field Ising model (FCTFIM), and Section \ref{sec:caustics} gives background information about the Fock state caustics that develop in the dynamics. Section \ref{sec:mainresults} gives an overview of our main results, including an explanation of how, in the vicinity of a DPT, the Loschmidt echo is analogous to the light intensity within Alexander's DB, but with the role of time replacing the angle to the sky. Section \ref{sec:rainbows} gives a more indepth treatment of rainbows, and Section \ref{sec:WKB} presents the main theoretical calculation in the paper which gives the rate function in the semiclassical regime. Section \ref{sec:measurement} discusses a possible measurement scheme, which is a tricky issue because the Loschmidt echo is exponentially small at the DPT. Our solution is to perform two different measurements on two halves of the system.  We give a general discussion of switching lines which are the essence of DPTs in Section \ref{sec:switching_lines}, and our conclusions are set out in Section \ref{sec:conclusion}.

\section{The model}
\label{sec:model}

In this article we focus on the FCTFIM \cite{Lang2018Sep,Corps2023Mar} which is the fully-connected version of the well-known transverse-field Ising model (TFIM).  This has been used as the minimal model for dynamics in a wide range of physical systems including bosonic Josephson junctions in ultracold atomic gases \cite{Albiez2005Jun,Levy2007Oct,Zibold10,Leblanc2011,Gerving12,Trenkwalder16}, weak links in superfluid $^3$He \cite{Backhaus98}, ferroelectrics such as KH$_{2}$PO$_{4}$ and ferromagnets such as Dy(C$_2$H$_5$SO$_4$)$_3$9H$_2$O \cite{Das2006}, exciton polaritons in semiconductors \cite{Lagoudakis10,Abbarchi13}, and light beams carrying angular momentum \cite{Gutierrez23}. The model is also relevant to TFIM systems that are not fully connected but have long range interactions such as those realized using trapped atomic ions \cite{Britton2012,Jurcevic14,Bohnet2016,Jurcevic2017Aug,Zhang17}, and the Dicke model describing cold atoms in optical cavities \cite{baumann2010,Klinder15,Defenu18,Muniz2020Apr}.  
Given that DPTs have been studied in the FCTFIM before \cite{Lang2018May,Lang2018Sep,Sehrawat2021Aug,Corps2022Jul,Corps2023Mar}, the goal of this paper is to establish a more intuitive explanation of the phenomenon in terms of caustics which are a generic feature of wave dynamics and which  occur in this model in the real-time dynamics of the state in the Fock space representation.

The FCTFIM \cite{Lang2018Sep,Corps2023Mar} consists of $N$ two-level systems (Pauli operators $\vec{\sigma}^i$) with Hamiltonian
\begin{equation}
    H=\frac{G}{2N}\sum_{i,j=1}^N\frac{\sigma_z^i\sigma_z^j}{4}+\Omega\sum_{i=1}^N\frac{\sigma_x^i}{2}  \ .
    \label{FCTFIM_hamiltonian0}
\end{equation}
$G$ determines the strength of the all-to-all interactions between the spins and $\Omega$ is the transverse field strength. 
The effective dimension of the problem can be drastically reduced by exploiting permutation symmetry. One can define a total spin operator
\begin{equation}
    \vec{J}=\frac{1}{2}\sum_{i=1}^N\vec{\sigma}^i
\end{equation}
such that the Hamiltonian can be expressed as
\begin{equation}\label{eq:FCTFI_Hamiltonian}
H=\frac{G}{2j}J_z^2+\Omega J_x
\end{equation}
where we have set $j=N/2$. In fact, the Hamiltonian preserves $J^2=j(j+1)$ and we can restrict the dynamics to this symmetry sector (Dicke subspace). For $\Omega=0$ the eigenstates of the Hamiltonian within the subspace are then simply the eigenstates of $J_z$
\begin{equation}
    J_z\ket{j,m}=m\ket{j,m},\quad m=-j,-j+1,...,j.
\end{equation}
These are commonly known as the Dicke states. However, we will refer to them as Fock states because they are the symmetrized single particle eigenstates of the number difference operator (giving the difference between the number of up and down spins, i.e.\ the magnetization).  For the sake of working with a specific example, in what follows we set $\Omega=G$, and choose the initial state to be a single Fock state $\ket{\psi_0}=\ket{j,m_0}$, with $m_0$ chosen as the closest integer (or half integer) to $0.6 j$.  Although a single Fock state might not appear to be a `typical' initial condition,  it is in fact not special because small variations in the initial state or the parameters in the Hamiltonian do not lead to qualitatively different outcomes in the wave function dynamics, as is to be expected from the structural stability of catastrophes. For example, choosing a narrow gaussian wavepacket of Fock states still leads to similar fold and cusp patterns \cite{Mumford2019May,Kirkby2019Nov}. Likewise, going to the opposite situation of an initial phase state (which is very broad in Fock space) also gives rise to caustics \cite{Kirkby2022Feb}, although of course they are in different positions. The key ingredient for caustics to appear is that the initial state must be energetically far above the ground state and such states can easily be generated in sudden quenches. Similar to the caustics, DPTs also occur generically in the FCTFIM for quenches from localized states (such as Fock states) to the regime with a strong transverse field (as for instance in Fig.~\ref{fig:fctfi_1}).

\begin{figure}\centering
\includegraphics{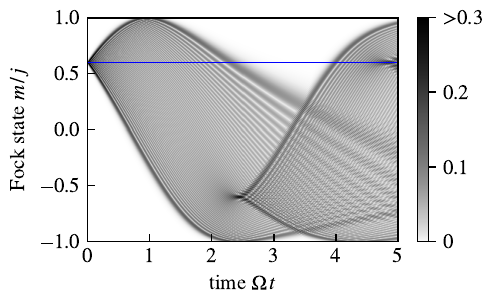}
\includegraphics{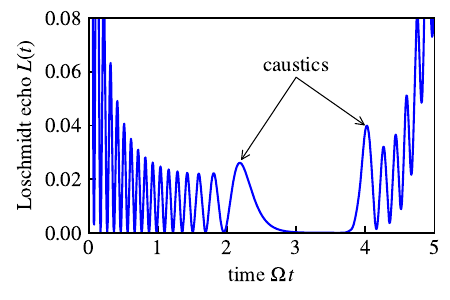}
\caption{\textbf{Left:} The modulus of the exact quantum wave function plotted in Fock space as a function of time.  The shading gives the magnitude of the overlap $|\braket{j,m|\psi(t)}|$ with each Fock state. The parameters are $\Omega=G$, $j=80$, and the initial state is the single Fock state $\vert j=80,m=48 \rangle$. The envelopes separating the bright and dark regions are fold caustics, and these meet at cusp caustics (in order to faciliate comparison with optical rainbows we use the terminology `bright' and `dark' although in the image colourmap higher amplitudes correspond to blacker shades and lower amplitudes to whiter shades). In the full quantum theory shown here, folds are dressed by Airy functions whose characteristic fringe pattern can be seen near the edges of the shaded (`bright') regions. Cusps are dressed by Pearcey functions as occurs near $m/j=-0.6$ and time $\Omega t \approx 2.4$. \textbf{Right:} The Loschmidt echo $L(t)$, or return probability, is given by the cut through the general wave function corresponding to the initial state. The cut is depicted by the horizontal blue line in the left panel. The Loschmidt echo shown here features a one dimensional slice through two fold caustics that bound a dark forbidden region where the wave function is exponentially small.}\label{fig:caustics}
\end{figure}

\section{Caustics in Fock space}
\label{sec:caustics}

Caustics are usually associated with waves or rays in real space and are the most intense regions of a wave field. In geometric optics they are the envelopes of families of rays and form quite generically through natural focusing, e.g.\ optical rainbows focused by water droplets in the sky and gravitational lensing by mass distributions in the cosmos \cite{Nye_natural_focusing}. Similarly, freak waves, tidal bores and tsunamis are caustics resulting from the natural focusing of water waves by the topography of the sea bed or estuary shape \cite{Hohmann10,metzger2014,Berry2018,Berry07,Degueldre16}.  
Caustics are singularities of the classical theory: they are regions where the wave amplitude diverges in the short wavelength (ray) limit $\lambda \rightarrow 0$, and their natural mathematical description is in terms of catastrophe theory \cite{Nye_natural_focusing,berry81}. Caustics in Fock space are the extension of these ideas to many-body quantum mechanics. A single ray in Fock space represents a single classical field configuration, while quantum configurations can be represented by the superposition of many rays. Whereas ordinary caustics are resolved by going to the wave theory where interference removes the singularity, caustics in Fock space are resolved by second-quantization where the discreteness of field excitations is the new feature needed to remove the singularity. The theory of Fock-space caustics has been developed over the past decade in the context of sudden quenches in cold atom systems \cite{ODell2012Oct,Mumford2017,Mumford2019May,Goldberg2019Dec,Kirkby2022Feb} and is also relevant to dissipative systems \cite{Link2020Sep}.

Catastrophe theory makes the remarkable prediction that there are only a finite number of distinct shapes that structurally stable singularities can take, depending on the dimension of the space in which they live \cite{Thom1975,Arnold1975}.  In the two dimensions which are relevant to the dynamics of the FCTFIM, namely the one dimension of Fock space that gives the total magnetization (difference between the number of up and down spins) plus time, the only structurally stable catastrophes are fold lines and cusp points. 
Furthermore, in the wave theory each type of catastrophe is dressed by its own characteristic local wave function known as a `diffraction integral' or `wave catastrophe' \cite{Berry&Upstill1980,Olver2010}. For the fold this is the Airy function and for the cusp it is the Pearcey function \cite{Pearcey46}. Both of these wave functions can be seen occurring in the left hand panel of Fig.~\ref{fig:caustics} that shows the development of a many-body wave function in Fock space as a function of time. Higher dimensional models display higher wave catastrophes \cite{Kirkby2022Feb} that form a hierarchy with the higher ones containing the lower ones. Each type of catastrophe forms an equivalence class with its own  scaling properties governed by exponents known as Arnold and Berry indices \cite{berry81}.

In fully-connected spin models the dynamics in Fock space are specified by the set of quantum amplitudes  
\begin{equation}
\{\braket{j,-j|\psi(t)}, \braket{j,-j+1|\psi(t)},\ldots , \braket{j,j|\psi(t)} \}
\end{equation}
given by the overlaps between the Fock states and the time-dependent wave function.  The results for our example system (see Sec.~\ref{sec:model}) are displayed in Fig.~\ref{fig:caustics}. The most striking feature is that there are quite sharply defined boundaries between regions where the amplitudes are finite and where they appear to be zero (more precisely, they are exponentially suppressed as these are the classically forbidden regions, see Sec.~\ref{sec:WKB}). The boundaries are the fold caustics and are particularly `bright', i.e.\ regions of high probability. The fold curves meet in pairs at cusps which are the brightest regions of all. %Caustics are associated with a breakdown of the corresponding classical ray theory, which will be introduced in detail later., but in the quantum theory they are smooth and dressed by characteristic oscillatory interference patterns which are given by diffraction integrals \cite{Berry&Upstill1980}. 
Caustics can also be seen in the Loschmidt echo. For the given initial condition the Loschmidt echo, displayed in the right panel of Fig.~\ref{fig:caustics}, is also a Fock-state overlap, and hence a cut through the more general Fock-space wave function shown in the left panel of Fig.~\ref{fig:caustics}. In a finite system the Loschmidt echo is an analytical function of time. We can identify oscillatory and exponentially decaying regions which are separated by caustics. The exponentially supressed \emphquote{dark} regions are of particular interest here because it is precisely where we find a nonanalytic limit as $N\rightarrow\infty$, and hence a dynamical phase transition, similar to Fig.\ \ref{fig:fctfi_1}.

\section{Main results: Rainbows and Dynamical Phase Transitions}
\label{sec:mainresults}

For the convenience of the reader, in this section we summarize the key insights and main results of this work, including the main physical intuition for the phenomenon of DPTs in the model. We provide more detailed technical steps and further connections in subsequent sections.

\begin{figure}
    \centering
    \includegraphics[height=1.5in,valign=t]{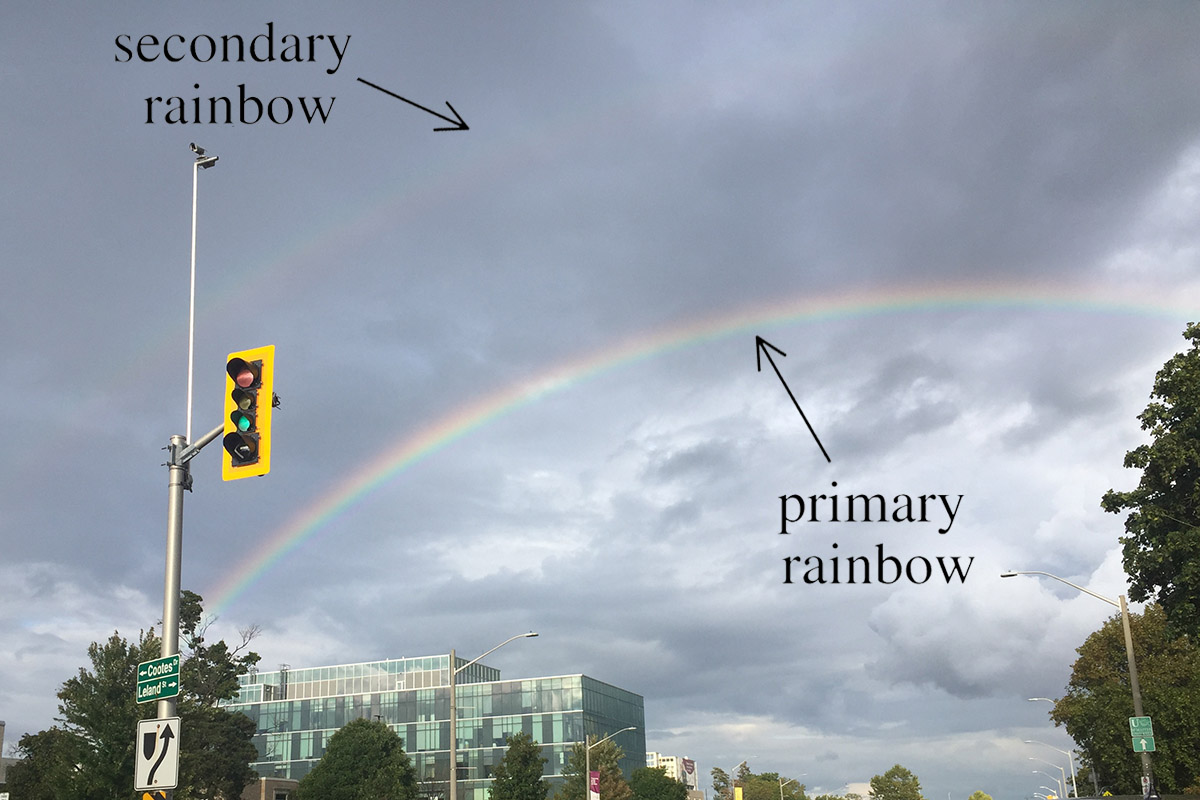}\qquad\qquad
    \includegraphics[height=1.5in,valign=t]{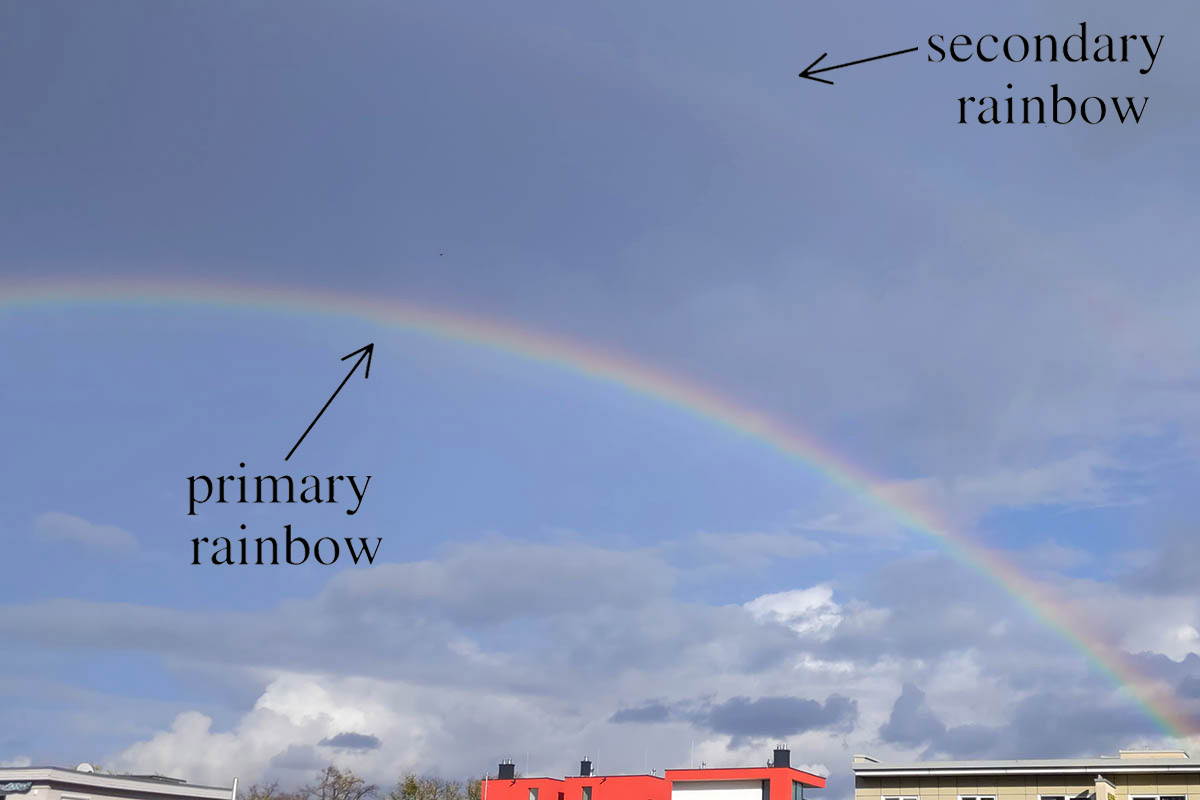}\qquad\qquad
    \includegraphics[valign=t]{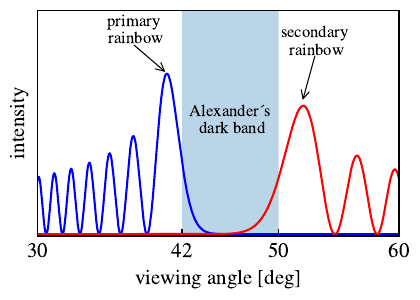}
    \caption{\textbf{Upper:} Examples of rainbows in the skies over Hamilton (left) and Dresden (right) showing primary and secondary bows as well as Alexander's DB. \textbf{Lower:} Airy theory of a rainbow for monochromatic light (schematic). One can distinguish two classes of light (red and blue curves) that make up the bright regions. The intensity of light from both the primary and secondary rainbows entering the DB is exponentially suppressed and described by the decaying side of an Airy function. Inside the DB, there exists a critical viewing angle where the two tails have equal magnitude which is the direct analogue of the critical time in a DPT.  }
    \label{fig:rainbow_wave}
\end{figure}

To build up some intuition for the physics of Fig.~\ref{fig:caustics}, we highlight an analogy to a very different physical system: Optical rainbows. Rainbows are caustics in the angular distribution of sunlight scattered by raindrops.  In Fig.~\ref{fig:rainbow_wave} (upper panel) photographs of rainbows are displayed and we can identify three different regions in the sky from bottom to top: A bright region below the primary rainbow, a dark region between the primary and the secondary rainbow (the secondary rainbow is only faintly visible here) and finally another slightly brighter region above the secondary bow. Historically, Airy's 1838 theory of the rainbow was the first treatment of the wave theory of light near a caustic that correctly tamed the geometric ray singularity to obtain a finite result and was the origin of the eponymous Airy function \cite{Airy1838,Nussenzveig1969Jan}. The intensity of scattered light as a function of the viewing angle according to this theory is displayed schematically in Fig.~\ref{fig:rainbow_wave} (lower panel). It features two back-to-back Airy functions, one for each of the two rainbows. Each rainbow features interference fringes (supernumary arcs) on the bright side of the main lobe (the caustic) and exponential decay on the other. The dark region between the primary and secondary rainbows corresponds to Alexander's DB, and it is only populated by the two exponentially small tails. 

The field amplitude in the vicinity of the primary rainbow can be written as \cite{Nussenzveig1969Jan,van_de_Hulst_book,Olver2010}
\begin{equation}
    A_1(\theta)\approx C_1(\theta) \mathrm{Ai}(D_1 k^{2/3}(\theta-\theta_1))
    \label{eq:airy1}
\end{equation}
where $k=2\pi/\lambda$ is the wavenumber of the incident light, $D_{1}$ is a parameter with units of $\mathrm{length}^{2/3}$ determined by the physical properties of the raindrop, Ai is the Airy function and $\theta_1$ is the rainbow angle (about $42^\circ$). The prefactor $C_1(\theta)$ is non-exponential in $k$ and will not be important in the following. The $2/3$ scaling exponent on $k$ is the Berry index for the fold catastrophe and determines how the fringe spacing depends on the wavenumber of the light. The same approximation can be made for the secondary rainbow
\begin{equation}
    A_2(\theta)\approx C_2(\theta) \mathrm{Ai}(D_2 k^{2/3}(\theta_2-\theta))
\end{equation}
where $\theta_2$ is the angle of the secondary rainbow (about $50^\circ$). Within the DB we can approximate the Airy functions by exponentials, such that the full light intensity is
\begin{equation}\label{eq:rainbow_intensity}
    I(\theta)\approx \left|\tilde C_1(\theta) \exp\left(-\frac{2k}{3} D_1^{3/2} \, (\theta-\theta_1)^{3/2}\right)+\tilde C_2(\theta) \exp\left(-\frac{2k}{3} D_2^{3/2}\, (\theta_2-\theta)^{3/2}\right)\right|^2
\end{equation}
Both contributions are exponentially weak in $k$. If we take the ray-optics limit of small wavelengths $k\rightarrow\infty$ one term will exponentially dominate the other. As with the Loschmidt echo, we can define a rate function
\begin{equation}
    r(\theta)=-\frac{1}{k}\log I(\theta)\rightarrow \min\left\{\frac{2}{3} D_1^{3/2}\,(\theta-\theta_1)^{3/2},\frac{2}{3} D_2^{3/2}  \,(\theta_2-\theta)^{3/2}\right\}
\end{equation}
which yields the minimum of the two exponents, and hence a nonanalytical change when the minimum swaps from the contribution of the primary rainbow to the secondary rainbow as $\theta$ increases.  We note in passing that for a real rainbow the situation is more intricate as weak higher order angular rainbows can overlay the DB between the first and second bows, whereas this doesn't happen for our temporal DB. Nonetheless, our main point here is that the \emph{exact same mechanism} leads to a DPT in the FCTFIM. 

\begin{figure}\centering
\includegraphics{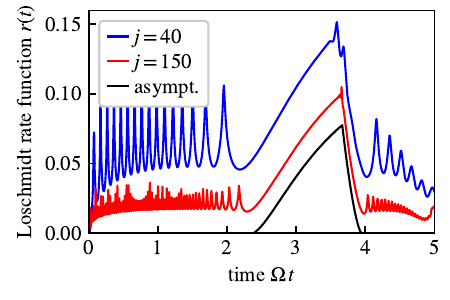}\includegraphics{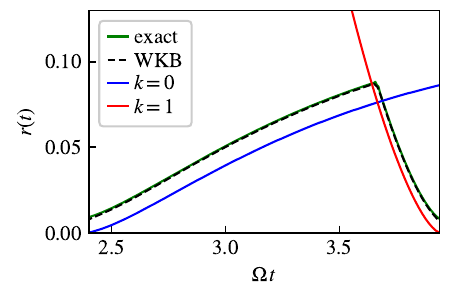}
\caption{\textbf{Left:} Loschmidt rate function (logarithm of the return probability) for the FCTFIM  and for different system sizes. A DPT occurs at $t_c\approx 3.7/\Omega$ in the dark region between the two caustics (see Fig.\ \ref{fig:caustics}). \textbf{Right:} WKB theory predictions for $j=350$ and the rate function contributions describing the overlapping exponential tails ($k=0,1$ limiting curves).}\label{fig:loschmidt}
\end{figure}

With an understanding of Airy's rainbow theory in the DB, we can now come back to our quantum Loschmidt echo from Fig.~\ref{fig:caustics}. Because of the universal properties of caustics, we expect the exact same physics to determine the Loschmidt echo. In the time interval between the two caustics, the wave function overlap is determined by two exponential tails coming from the first and second caustics 
\begin{equation}\label{eq:L_split}
    L(t)=|l_1(t)+l_2(t)|^2\, \quad , \qquad l_i(t)=\eul^{-j r_i(t)}.
\end{equation}
The system size $j$ now takes the role of the wavenumber $k$ in Eq.~\eqref{eq:rainbow_intensity}. This is because in the FCTFIM the classical (ray) limit is equivalent to the thermodynamic limit ($j$ takes the role of $1/\hbar$), in the same way that the ray optics limit corresponds to large wavenumbers $k$.  Using the same argument as above, the asymptotic rate function will behave nonanalytically as $j\rightarrow\infty$
\begin{equation}
    r(t)=-\frac{1}{j}\log L(t)\rightarrow 2\min\{r_1(t),r_2(t)\}.
\end{equation}
We can confirm this behaviour numerically as displayed in Fig.~\ref{fig:loschmidt}. In the left panel we see that the rate function acquires a finite limit in the DB region, signifying exponential suppression of the return probability. We can identify the typical sharkfin-shape that characterizes DPTs. In the bright regions the return probability has a sub-exponential scaling and hence the rate function becomes zero asymptotically.

Qualitatively this discussion already gives an intuitive explanation for the nonanalytic behaviour at a DPT in terms of the caustics forming in the Fock space wave function. For the FCTFIM we can even obtain a quantitative description using semiclassical methods. In Section \ref{sec:WKB} we employ WKB theory to relate the return probability to classical trajectories \cite{Braun1993Jan,Shchesnovich2008Aug,Nissen2010Jun,ODell2012Oct,Simon2012Nov,Simon2014May}, similar to the ray theory of geometric optics. Within semiclassics we can compute the two contributions to the Loschmidt echo in Eq.~\eqref{eq:L_split} separately. In this way we obtain the exact limiting curve (the sharkfin), as well as accurate finite size predictions, as displayed in Fig.~\ref{fig:loschmidt} (right panel). 

As presented in Sec.~\ref{sec:measurement}, the semiclassical description also gives the crucial insight for a proposed measurement scheme that can be used to signal the nonanalytic behaviour for small size systems, where overlaps can actually be measured \cite{Jurcevic2017Aug}. 
Moreover, our analysis is not restricted to the return probability, but can be applied to arbitrary Fock state overlaps. We elaborate further on this point in Sec.~\ref{sec:switching_lines}, where we introduce the more general concept of a switching line for the Fock representation of a quantum state.

\section{Rainbows: from rays to waves}
\label{sec:rainbows}

\begin{figure}[h]
    \centering
    \includegraphics{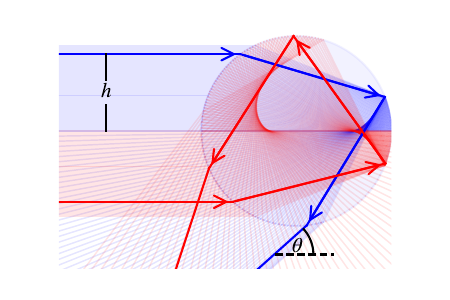}
    \includegraphics{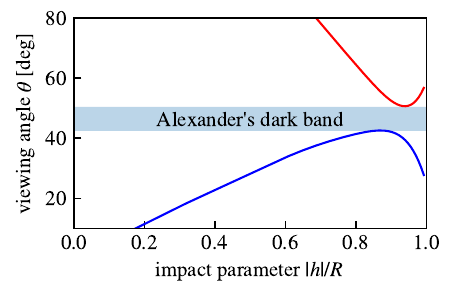}
    \caption{Geometric theory of the rainbow. \textbf{Left:} the trajectories of rays scattered by a raindrop are determined by their impact parameter $h$. After entering the droplet the rays can undergo multiple partial internal reflections before escaping. The family of rays that escape at the second reflection contribute to the primary rainbow (blue) and the family that escape at the third reflection contribute to the secondary rainbow (red). Further partial reflection events contribute to higher rainbows which are increasingly faint. The viewing angle $\theta$  measures the inclination to the horizontal of a scattered ray as seen by someone viewing the rainbow from the ground. \textbf{Right:} An input-output plot showing the viewing angle $\theta$ of outgoing rays versus the modulus of their input impact parameter $h$. Here we only include rays contributing to the primary and secondary rainbows. The lower curve (blue) has a maximum at $\theta \approx 42^{\degree}$  and 
    the upper curve (red) has a minimum at $\theta \approx 50^{\degree}$.  According to the ray theory, the intensity diverges at these two angles giving rise to angular caustics. No real rays from these two families can enter the range of angles $42^\circ <\theta<50^\circ$ 
    %$130^{\degree} < \theta_{D}<138^{\degree}$ 
    which is Alexander's DB.}
    \label{fig:raindrop}
\end{figure}

In order to gain a better appreciation of the close analogy between rainbows and the Fock space caustics that lead to the DPTs in the FCTFIM,
we now outline the mechanism of rainbow formation more fully. Fig.\ \ref{fig:raindrop} depicts the geometric ray results for a spherical raindrop. Assuming that the light source is far away in comparison to the radius $R$ of the droplet, an incoming ray can be taken as horizontal and its subsequent trajectory is entirely determined by its impact parameter $h$, which is defined as the distance of the incoming ray to the symmetry axis of the raindrop (see Fig.~\ref{fig:raindrop} left panel). After an initial refraction upon entry, the trajectory of the ray involves multiple internal reflections but at any of these the ray may escape. Rays escaping at the first reflection contribute to the forward scattered light, but rays escaping at the second and third reflections give rise to the primary (blue) and secondary (red) rainbows, respectively, as shown in the left panel of Fig.\ \ref{fig:raindrop}. The remaining light continues circulating inside the droplet and gives rise to higher rainbows that are increasingly faint. Typically only the primary and occasionally the secondary rainbows are visible to the naked eye.

The key feature of rainbows is that they are generated not by one ray but by the focusing of many rays. Parallel rays enter the raindrop at all possible impact parameters $-R \le h \le R$ and most exit at different angles $\theta$ which can be calculated using the Snell-Descartes law of refraction. However, there are certain special angles where the input/output relation between $h$ and $\theta$ is stationary as shown in the right panel of Fig.\ \ref{fig:raindrop}. In the vicinity of a stationary point a range of rays each with a slightly different impact parameter all emerge at the same angle to first order. This focusing effect gives rise to the various bows we see from the ground with the primary bow occurring at approximately $42^{\degree}$ and the second at approximately $50^{\degree}$ to the horizontal, respectively.  Within the geometric theory the angular density of rays diverges at these angles and hence rainbows are examples of caustics (focusing by natural effects). Moreover,  the stationary point of the input-output relation is independent of droplet radius and hence droplets of different sizes, as can be expected to occur naturally, reinforce each other to produce the same $42^{\degree}$ and $50^{\degree}$ intensity peaks.

Let us now turn to the wave theory. 
By imagining drawing horizontal lines across the right panel of Fig.\ \ref{fig:raindrop}, we see that as a long as we are outside of the DB exactly two different input rays contribute at each angle (except at the two rainbow angles themselves where two rays coalesce). In the wave theory this gives rise to two-wave interference, and hence interference fringes, with the highest peaks occurring just before entering the DB, see the right panel of Fig.\ \ref{fig:rainbow_wave} (wave/quantum effects give a small angular/temporal offset between the main fringe of the wave theory and the classical caustic). According to Eq.\ (\ref{eq:airy1}), the angular distribution of light due to a rainbow is proportional to the Airy function, which can be expressed in integral form as \cite{Olver2010}
\begin{equation}
    \mathrm{Ai}(x)=\frac{1}{2 \pi}\int_{-\infty}^{\infty} e^{i(s^3/3+x \, s )} \  ds .
    \label{eq:airyintegral}
\end{equation}
The Airy wave function accounts for each rainbow separately. In our case $x=D k^{2/3} \theta$ and $s=(k/D^{3})^{1/3}h$, where here the angle $\theta$ is measured from the rainbow angle, i.e.\ $\theta=0$ at the classical (ray) caustic (see Eq.~\eqref{eq:airy1}). Although the impact parameter lies in the range $-R \le h \le R$, the range of the integral is extended to $\pm \infty$, which is a good approximation in the semiclassical regime $kR \gg 1$ \cite{Nye_natural_focusing}.
For $\theta  > 0$ the Airy function is exponentially decaying corresponding to the evanescent light that leaks into the DB. For $\theta  < 0$ the Airy function displays interference fringes arising from the two wave interference. These fringes explain the supernumerary arcs that are fainter than the main rainbow but are sometimes visible and should not be confused with the geometric ray caustics. For example, the supernumerary arcs of the primary rainbow (blue fringes in the right panel of Fig.\ \ref{fig:rainbow_wave}) are quite distinct from the secondary or tertiary etc.\ classical rainbow caustics coming from a larger number of internal reflections inside the raindrop. The latter are classical effects whereas the supernumeraries are purely wave effects.

Everything we have said so far refers  to only a single colour (frequency) of light, such as would occur if the raindrops were to be illuminated by a laser. Sunlight of course consists of many frequencies and dispersion causes each frequency to be refracted by a different amount and hence to have a different deflection angle. Thus, each colour gives rise to its own independent Airy function shifted slightly from the others and the familiar sequence of colours of the rainbow corresponds only to the main fringe of the Airy function for each colour. The subsequent fringes/supernumerary arcs give rise to a repetition of the colours.

Let us now consider caustics from the point of view of catastrophe theory \cite{berry81}. This describes bifurcations of gradient maps which are essentially theories based on a minimum principle such as the principle of least action.  At a bifurcation the number of solutions changes, which in physical terms corresponds to the coalescence or birth of rays, i.e.\ a caustic. Catastrophe theory specifies a hierarchy of bifurcations of increasing dimension and complexity where the higher ones generally contain the lower ones. The simplest is the fold where two solutions coalesce as a function of a single parameter $x$. The second is the cusp where two parameters $x$ and $y$ are required to fully explore the catastrophe.  Moving across either of the two fold lines that make up the cusp the number of solutions changes by two but at the cusp point itself three solutions coalesce.  Both the fold and the cusp appear in the dynamics of the TFIM we consider in this paper. Each catastrophe is associated with a normal form or generating function, which in physical problems corresponds to the mechanical action close to the singularity.  For the fold and cusp these are
\begin{eqnarray}
\Phi_{\mathrm{fold}}(s,x) & = & s^3/3+ x \, s \\
\Phi_{\mathrm{cusp}}(s,x,y) & = & s^4/4+ y \, s^2/2+ x \, s \ .
\end{eqnarray}
Examining the Airy function in Eq.\ (\ref{eq:airyintegral}) it is clear that it is proportional to $\int_{-\infty}^{\infty} \exp[i\Phi_{\mathrm{fold}}(s,x)]ds$. From this we see that we can interpret the Airy function as a kind of elementary Feynman path integral with action $\Phi_{\mathrm{fold}}(s,x)$ where $s$ labels each ray uniquely via its impact parameter and we integrate (sum) over all rays with a uniform measure  \cite{berry81}. Because the action for the fold is a cubic function of $s$ it generally has two stationary points which correspond to two classical paths or rays, in agreement with our discussion above. These coalesce  at the classical caustic ($x=0$) and enter the complex plane, giving rise to an exponentially suppressed wave function in the DB. The Airy function is the first in a hierarchy of wave catastrophes that dress the classical catastrophes. The second is the Pearcey function $\mathrm{Pe}(x,y)= \int_{-\infty}^{\infty} \exp[i\Phi_{\mathrm{cusp}}(s,x,y)]ds$ that dresses the cusp catastrophe \cite{Olver2010}. The Airy and Pearcey functions are smooth and nonsingular, and inherit the structural stability of their underlying ray catastrophes. As they are built from an action rather than a free energy (in contrast to the partition function in thermodynamics), they are inherently dynamical and this makes them the natural candidates for being the universal wave functions near DPTs.

\section{Exact Semiclassical Description}\label{sec:WKB}

We shall now compute the Loschmidt echo for the FCTFIM within the semiclassical WKB approximation which is valid for large $j$ \cite{Braun1993Jan,Shchesnovich2008Aug,Nissen2010Jun,ODell2012Oct,Simon2012Nov,Simon2014May}. This will enable us to understand the occurrence of the nonanalyticity at the DPT and compute the exact asymptotic curve shown in Fig.~\ref{fig:loschmidt}. Moreover, the theory establishes a rigorous connection to classical mechanics, similar to the ray optics framework that we discussed for the optical rainbow in the previous section. 

We start by expressing the time dependent state vector in the energy basis  $H\ket{E_n}=E_n\ket{E_n}=j\varepsilon_n\ket{E_n}$ as
\begin{equation}
\vert \psi(t) \rangle =\sum_n c_{n} \vert E_n \rangle  \eul^{-\ii E_n t}
\end{equation} 
where $c_{n}= \braket{E_n|j,m_0}$.
For the Loschmidt echo we need to compute the overlap with the initial state
\begin{equation}
\braket{j,m_0|\psi(t)}=\sum_n \braket{j,m_0|E_n}\braket{E_n|j,m_0} \eul^{-\ii E_n t}.
\label{eq:modesum}
\end{equation} 
To evaluate this intricate sum, which converges very slowly in the semiclassical regime where there are many energy states,  we employ Poisson resummation \cite{Berry1972Feb,ODell2012Oct,ODell_2001}. This remarkable method is exact and transforms the sum over energy \textit{modes} into a sum over topologically distinct classes of \textit{paths} labelled by an integer index $k$
\begin{equation}\label{eq:poisson}
\braket{j,m_0|\psi(t)}=\sum_{k=-\infty}^{\infty}\int\diff n |\braket{j,m_0|E_n}|^2\eul^{-\ii E_n t+ 2\pi\ii n k} \ .
\end{equation}
In contrast to the energy mode sum in Eq.\ (\ref{eq:modesum}), this alternative sum converges quickly as long as we do not require long times (in comparison to the period of the classical oscillations). If the limit of large system size is taken as below, the spreading of the wave function in Fock space will become slow and a semiclassical evaluation of the sum can also be performed for long times, with only a few relevant terms that contribute significantly at each time.

\begin{figure}[t]\centering
\includegraphics{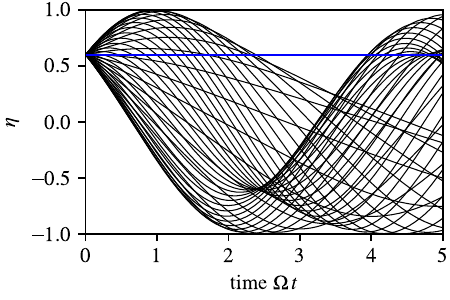}\\
\begin{tabular}{c c c c c}
    \includegraphics[height=3cm]{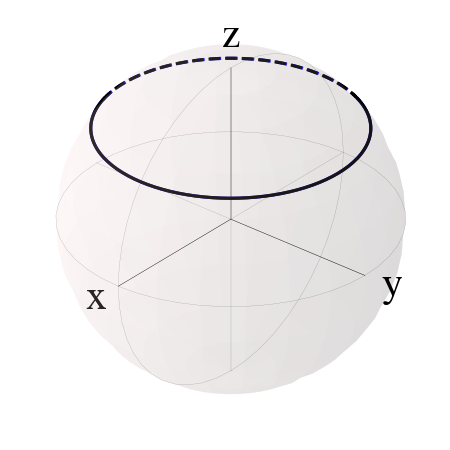} & 
\includegraphics[height=3cm]{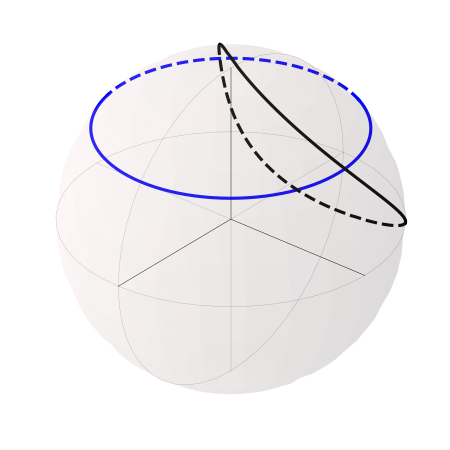} & 
\includegraphics[height=3cm]{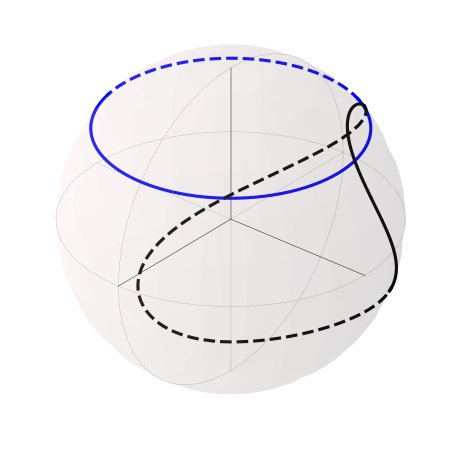} & 
\includegraphics[height=3cm]{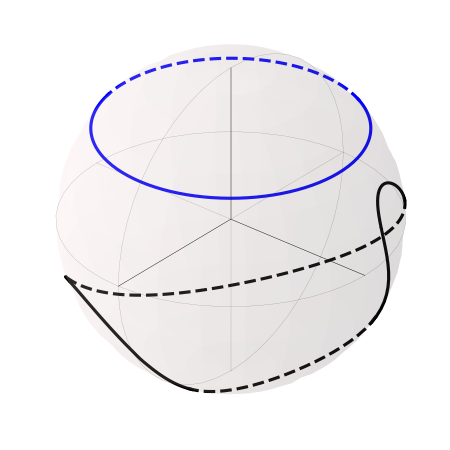} & 
\includegraphics[height=3cm]{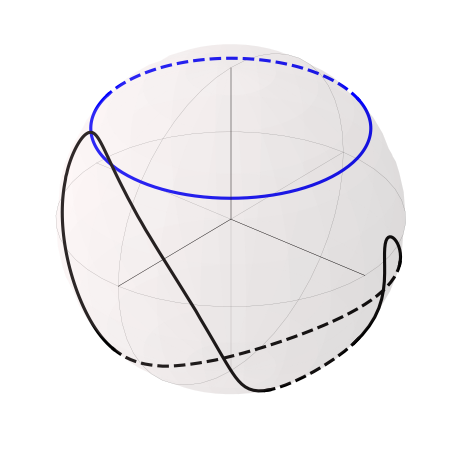}\\
    $\Omega t=0$ &     $\Omega t=1$ &    $\Omega t=2$ &     $\Omega t=3$ & $\Omega t=4$
\end{tabular}
\caption{Plots of the classical trajectories that, within the WKB approximation, contribute to the quantum wave function shown in the left panel of Fig.~\ref{fig:caustics}. \textbf{Top:} A plot of momentum $\eta$ versus time. In the FCTFIM $\eta$ is proportional to the net magentization. Each trajectory has the same initial momentum $\eta(0)=0.6$ but a different initial position which lies uniformly in the range $\phi(0)\in [0,2\pi)$ and mimics the Fock state $\vert j,m_{0}=0.6j\rangle$. \textbf{Bottom:} Snap shots of the same trajectories in the spherical phase space at different times. Each classical state is specified by a single point on the sphere with coordinates ($\phi,\theta$) which are the longitude and latitude, respectively, with the latter being measured from the equator. The initial Fock state corresponds to a ring of many classical states all at the same latitude. In all the snapshots the blue ring indicates the initial states whereas the black ring shows how the time evolved states wrap around the sphere. If we project this curve onto the $z$ axis the horizontal portions of the curve give rise to caustics in $\eta$ space.}\label{fig:clasical_trajectories}
\end{figure}

In the semiclassical regime the amplitude $\braket{j,m_0|E_n}$ can be obtained to high accuracy using the WKB approximation. We briefly recapitulate the theory in appendix \ref{app:wkb}. The WKB theory is built from solutions obtained from the Hamiltonian \begin{equation}\label{eq:Hamiltonian_classical}
    H(\eta,\phi)=\Omega\sqrt{1-\eta^2}\cos\phi-\frac{G}{2}\eta^2 ,
\end{equation}
which is the classical limit of Eq.\ (\ref{eq:FCTFI_Hamiltonian}) and mechanically corresponds to a momentum shortened pendulum \cite{Zibold10,Smerzi97}.
Here, $\eta$ is the canonical momentum which
is directly related to the quantum number $m$ that labels the Fock states via $\eta=m/j$. It
lies in the range $-1 \le \eta \le 1$ and hence can be parameterized as $\eta=\cos\theta$. $\phi$ is the corresponding canonical position or angle variable and takes values lying in the range $[0,2\pi)$. The natural phase space is therefore a sphere with the angles $(\phi,\theta)$ providing the longitude and latitude, respectively, and is the classical limit of the generalized Bloch sphere for $N$ spin-1/2 particles.  Whereas a classical state is given by a single point on the sphere, wave functions can be represented as delocalized distributions with Fock states corresponding to bands encircling the sphere at constant latitude.  The classical time evolution can be obtained by integrating Hamilton's equations
\begin{equation}
    \dot\phi=\frac{\partial H}{\partial \eta}\,,\qquad \dot\eta=-\frac{\partial H}{\partial \phi} \, ,
\end{equation}
and the resulting trajectories relevant to our problem are shown in  Fig.\ \ref{fig:clasical_trajectories}. In the top panel we plot $\eta(t)$ for multiple trajectories that all start with momentum $\eta(0)=\eta_0=0.6$ but have their initial position selected from the full range $\phi(0)\in [0,2\pi)$. This reflects the initial Fock state $\vert j,m_{0}=0.6j\rangle$ which has a precisely specified momentum and so the initial value of the conjugate variable $\phi$ is completely unknown according to the Heisenberg uncertainty relation and hence takes all possible values with equal probability. Comparison with the left panel of Fig.~\ref{fig:caustics} shows that this semiclassical picture indeed captures many of the qualitative features of the exact quantum result, including the caustics, but obviously fails to replicate the interference. In the bottom panel of Fig.\ \ref{fig:clasical_trajectories} snapshots of the trajectories at different times are plotted in the spherical phase space. Here we have included all possible trajectories for fixed $\eta(0)$, such that they form a continuous ring.

The WKB approximation relates the action of classical trajectories to the eigenstates of the quantum Hamiltonian. For the FCTFIM it gives the following general form for the overlap between energy eigenstates and Fock states 
\begin{equation}
    |\braket{j,m|E}|^2\approx |A(E,\eta)\cos( j S(E,\eta))|^2 
\end{equation}
where $A(E,\eta)$ is a prefactor that is non-exponential in $j$ and $S(E,\eta)$ is the reduced action of a trajectory with energy $E$ along the path from the classical turning point to $\eta$ (see appendix \ref{app:wkb}).   Inserting the WKB solution into Eq.\ \eqref{eq:poisson} we are left with
\begin{equation}
\begin{split}
&\braket{j,m_0|\psi(t)}\approx\sum_{k=-\infty}^{\infty}\int\diff n|A(E,\eta_0)\cos(jS(E,\eta_0))|^2\eul^{-\ii E_nt+2\pi\ii n k}.
\end{split}
\end{equation}
To complete the conversion of this expression into a bona fide semiclassical result the integral over the quantum label $n$ should be transformed into an integral over energy using the WKB (Bohr-Sommerfeld) quantization condition $jS(E_n)=2\pi(n+1/2)$, where $S(E)$ denotes the reduced action due to a full round-trip of the trajectory with energy $E$, i.e.~the phase space volume bounded by a closed torus. The resulting expression then takes the form
\begin{equation}
\begin{split}\label{eq:poisson_2}
&\braket{j,m_0|\psi(t)}\approx\sum_{k=-\infty}^{\infty}\int\diff E \frac{\diff n}{\diff E} |A(E,\eta_0)\cos(jS(E,\eta_0))|^2\eul^{\ii (jkS(E) -E t)} \ .
\end{split}
\end{equation}
Upon decomposing the cosine, for every value of $k$ one obtains four integrals over rapidly oscillating complex exponentials. For every term we identify an exponent 
\begin{equation}
2l   S(E,\eta_0)+kS(E)\,,\qquad l=\pm 1,0\qquad k\in \mathbb{Z}
\end{equation}
which can be associated with a distinct class of trajectories. The integer labels $k$ and $l$ determine how often the corresponding classical trajectories have passed through a classical turning point. For a trajectory with given energy $E$, we can define a classical return time as the derivative of the exponent with respect to energy
\begin{equation}
    T(E)=j\frac{\diff}{\diff E}\left(2l   S(E,\eta_0)+kS(E)\right).
\end{equation}
The dominant contributions of the energy integrals come from the regions around the minima of the exponents, which are determined by stationary phase conditions of the form
\begin{equation}\label{eq:stat_phase}
T(E)=t.
\end{equation}
For the relatively short times $t$ that we consider, this condition can be met only by the $k=0,1$ classes. In particular, the relevant exponents for short times come in the combinations 
\begin{eqnarray}
S_0(E) & \equiv & 2S(E,\eta_0) \qquad \qquad\qquad\, (l=+1, k=0)\\ 
S_1(E) & \equiv & -2S(E,\eta_0)+S(E)\qquad (l=-1, k=1)\, .
\end{eqnarray}
These correspond to trajectories that return to the initial condition after having passed once through a classical turning point, but on opposite sides in $\eta$ space. Note again that the time $T_{0/1}=\frac{\diff }{\diff E/j}S_{0/1}(E)$ is exactly the time that it takes for a trajectory of energy $E$ and of trajectory class $k=0$ or $k=1$ to come back to the initial condition. 

\begin{figure}\centering
\includegraphics{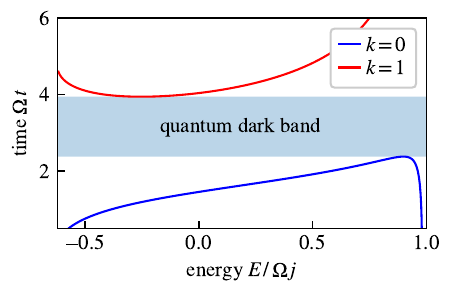}    \includegraphics{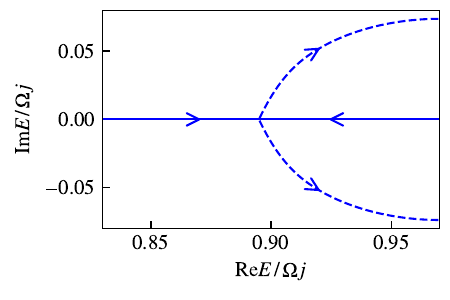}

\caption{\textbf{Left:} Energy dependent return time $T(E)$. The stationary phase condition $T(E)=t$ has no solutions in the shaded (classically forbidden) region and two solutions elsewhere. At the caustics the two solutions coalesce. This figure should be compared with its counterpart for rainbows as shown in Fig.\ \ref{fig:raindrop}. Viewing angle and impact parameter have been replaced by time and energy.  \textbf{Right:}  Solutions of the complexified steepest descent condition $\frac{\partial}{\partial z/j}\tilde S_<(z)=t$ (saddle points of $\tilde S_<(z)-zt$) for the $k=0$ trajectory class. At the caustic two real saddle points coalesce and form a conjugate pair of saddle points with nonvanishing imaginary part. Arrows indicate the progression of time $t$.}\label{fig:t_E}
\end{figure}

In our model for any time $t$ there exist different regimes where solutions $T(E)=t$ of the stationary phase condition do or do not exist. This can be seen in Fig.~\ref{fig:t_E}, where we display the $T(E)$ curves in an energy-time plot. For short times there either exist two solutions or none. The striking resemblance between this plot and the rainbow case shown in the right panel of Fig.\ \ref{fig:raindrop} is clearly evident.
At either boundary between the classically allowed and forbidden regions two stationary phase solutions coalesce signalling a breakdown of the stationary phase approximation. For the FCTFIM they are the points where fold caustics occur in Fock space. According to our earlier discussion for rainbows, the behaviour of the Loschmidt echo can then locally be described in terms of Airy functions. In recogniton of the close connection to rainbow optics, we call the region between the two caustics where no stationary phase solution exists a \textit{quantum dark band}. For times within the quantum DB the Loschmidt echo is exponentially small, as we observed previously (Fig.~\ref{fig:loschmidt}).

Continuing with the evaluation of the WKB expressions, we first focus on the first `bright' region within the $k=0$ class. Here the Loschmidt echo is given in the stationary phase approximation as the sum of the two contributions corresponding to the two classical trajectories that, at a given time, return to the initial latitude $\eta(t)=\eta_0$
\begin{equation}
    L(t)\approx|c_1\eul^{\ii (jS_0(E_1)-E_1t)}+c_2\eul^{\ii (jS_0(E_2)-E_2t)}|^2.
\end{equation}
$E_{1/2}$ are the energies of the two returning classical trajectories, i.e.~the energies for which $\frac{\diff }{\diff(E/j)}S_0=t$, and $c_{1/2}$ are nonexponential prefactors that we do not explicitly compute. Interference of the two contributions leads to the oscillatory behaviour observed in the region to the left of the caustic, as seen in Fig.~\ref{fig:loschmidt}. 

The exponential decay of the Loschmidt echo into the quantum DB region (to the right of the caustic) cannot be described in terms of standard classical mechanics. However, we can still proceed with the evaluation of the exponential integral in Eq.~\eqref{eq:poisson_2} by using the method of steepest descent rather than stationary phase. To evaluate the integral with the method of steepest descent one has to distort the contour of integration into the complex plane to pass through the saddle points of a complex extension of the exponent. Here we need a complex holomorphic extension of the energy dependent action $S_0(E)$, which we denote as $\tilde S_0(z)$, such that $\tilde S_0(E)=S_0(E)$ for real $E$ and $\partial_{z^*}\tilde S_0(z)=0$. We demonstrate in appendix \ref{app:complext} how such a complex extension can be realized in practice by considering complex classical trajectories. This approach is similar to the theory of complex rays in geometrical optics \cite{Chapman1999}. With this method it is possible to  directly compute the complex stationary points $\frac{\partial}{\partial z/j}\tilde S_0(z)=t$ (saddle points) and their corresponding action. 

The behaviour of these saddle point solutions is displayed in Fig.~\ref{fig:t_E}. Two real saddle points coalesce at the caustic and then turn into a conjugate pair of complex saddle points. Only one of these points corresponds to an exponentially decaying solution and contributes to the Loschmidt echo in the quantum DB region. From carrying out the steepest descent integration one finds that, to exponential accuracy, the Loschmidt echo is determined by $\tilde S_0$ evaluated at the saddle point
\begin{equation}\label{eq:loschmidt_1}
    L(t)\propto |\eul^{-\ii(j\tilde S_0(z_0)-tz_0)}|^2,\qquad \frac{\partial}{\partial z/j}\tilde S_0(z_0)=t.
\end{equation}
Because the exponent has a nonvanishing imaginary part, $L(t)$ is exponentially small, as expected in a regime where classical trajectories cannot enter. 

Thus far we have considered only the contribution of the $k=0$ class of trajectories that make up the first caustic. However, in the quantum DB region we have to also include a contribution from the second class (corresponding to $k=1$ in Eq.~\eqref{eq:poisson_2}). To include this contribution the procedure is identical: Find a complex extension $\tilde S_1(z)$ of the action $S_1(E)$ and compute the saddle points $\frac{\partial}{\partial z/j}\tilde S_1(z)=t$. The additional contribution to the Loschmidt echo is then
\begin{equation}\label{eq:loschmidt_2}
    L(t)\propto |\eul^{-\ii(j\tilde S_1(z_1)-tz_1)}|^2,\qquad \frac{\partial}{\partial z/j}\tilde S_1(z_1)=t.
\end{equation}
Note that this contribution increases over time as the second caustic is approached, while the first contribution \eqref{eq:loschmidt_1} decreases. Thus, there is a point in time where the dominant contribution switches. If we consider the rate function for large system size, it is determined solely by the minimum of the imaginary parts of the two exponents
\begin{equation}
r(t) \rightarrow 2\,\mathrm{ min}\{\mathrm{Im}(\tilde S_0(z_0)-z_0t),\mathrm{Im}(\tilde S_1(z_1)-z_1t)\}.\label{eq:rt_steep}
\end{equation}
This implies a nonanalytic change when the two exponentially small contributions overlap, and hence a DPT. In Fig.~\ref{fig:loschmidt} (right panel) the two exponents from the steepest descend method are shown. The dynamical phase transition arises due to an overlap of two exponential tails from the first and second caustics on opposite sides of the quantum DB, i.e.~the time interval of classically forbidden return. In the appendix \ref{app:finitesize} we show how this analysis can be generalized to account for finite size effects. Then we find excellent agreement with exact quantum calculations for large $j$, as displayed Fig.~\ref{fig:loschmidt} (right panel, dashed curve). Our rigorous geometric description makes it obvious that this phenomenon is generic in the model and robust with respect to all system parameters. In the next section, we make use of our insights from semiclassics in order to formulate a proposal for finding evidence of the DPT in experiments with finite resources.

\begin{figure}[t]
    \centering
    \includegraphics{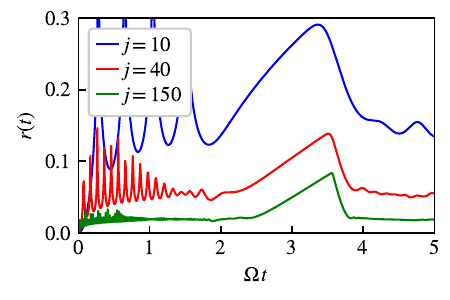}    \includegraphics{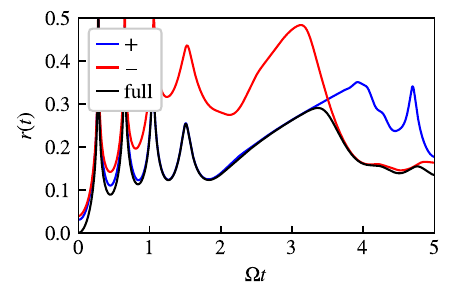}
    \caption{\textbf{Left:} Rate function of the mixed state Loschmidt echo \eqref{eq:reduced_state_Loschmidt} for different system sizes. A DPT occurs as in the original model shown in Fig.~\ref{fig:loschmidt}. \textbf{Right:} Conditioned rate functions \eqref{eq:rate_func_conditioned} corresponding to the conditioned Loschmidt echo \eqref{eq:loschmidt_conditioned} in a system with $N=20$ atoms ($j=10$). The black curve is the unconditioned rate function shown on the left.}
    \label{fig:dpt_reduced}
\end{figure}

\section{A measurement scheme}
\label{sec:measurement}

The Loschmidt echo is an exponentially small quantity that is exponentially hard to measure for a large system, see the definition given in Eq.\ (\ref{eq:loschmidt_defn}).  This makes it impossible to directly observe a dynamical phase transition, which strictly speaking only occurs at infinite system size. It is interesting to contrast this with Alexander's DB where in a single glance we are effectively observing the results of measurements on a huge number of different systems run for different `times' (angles). A light meter with a high angular resolution could perhaps pick out the angle where the switch between the dominance of the two evanescent tails occurs, although this is made difficult by the exponentially brighter regions above and below the DB.

Different strategies for measuring DPT's in small systems have been proposed and even realized in trapped ion systems \cite{Jurcevic2017Aug,Zhang17}. For the present model we propose a measurement strategy similar to that presented in Ref.~\cite{Link2021Dec}. This is based directly on our analytical understanding of the DPT as described in the previous sections. To motivate the measurement scheme, it is instructive to go back to the classical dynamics displayed in Fig.~\ref{fig:clasical_trajectories}. The origin of the caustics that occur in the Loschmidt echo are directly visible here. At short times, the time-evolved ring cuts through the initial state at two points, meaning there are two classical trajectories contributing to the Loschmidt echo. At the caustic the two trajectories coalesce which leads to an enhancement in the return probability. The DPT occurs in the time interval during which no classical trajectories reach the initial latitude (for instance $\Omega t=3$). The second caustic arises when the trajectories return to the northern hemisphere after revolving around the sphere once. The crucial observation here is that, on the sphere, the two classes of trajectories which make up the two caustics are spatially separated. The first caustic is made up of trajectories where $y>0$, whereas $y<0$ for the second caustic (see Fig.~\ref{fig:clasical_trajectories}). Thus, in this semiclassical picture, we can separately distinguish the contributions of both caustics by the $y$-coordinate, corresponding to the $J_y$ spin component. 

For the full quantum system this insight cannot be straightforwardly utilized because a measurement of the states' location in $y$ does not commute with a measurement of the Loschmidt echo. To overcome this problem we propose to split the system into two equal parts, performing a measurement of the Loschmidt echo on one part while measuring the location in phase space on the other. To formalize this, we consider a splitting $\Vec{J}=\Vec{I}+\Vec{S}$ where each subsystem contains $N$ atoms. The Hamiltonian is still given by
\begin{equation}
H=\frac{G}{2N}(I_z+S_z)^2+\Omega (I_x+S_x)
\end{equation}
For simplicity, as an initial condition we consider not the fully symmetric Dicke state of the total spin $\ket{N,2m_0}$ where $J_z\ket{N,2m_0}=2m_0\ket{N,2m_0}$, but instead the product state
\begin{equation}
    \ket{\Psi_0}=\ket{N/2,m_0}\otimes \ket{N/2,m_0}=\ket{\psi_0}^{\otimes 2} \, .
\end{equation}
We expect that this system has the same classical limit as the original FCTFIM. Now let us focus on the nonunitary dynamics of one subsystem, for instance subsystem $I$, described by the reduced state $\rho_I(t)$
\begin{equation}
    \rho_I(t)=\tr_S \ketbra{\Psi(t)}{\Psi(t)},\qquad \rho_I(0)=\ketbra{\psi_0}{\psi_0}  \, .
\end{equation}
Because the initial state $\rho_I(0)$ is chosen as a pure state, the mixed state Loschmidt echo of the subsystem $I$ is uniquely defined as \cite{Lang2018May}
\begin{equation}\label{eq:reduced_state_Loschmidt}
    L(t)=\braket{\psi_0|\rho_I(t)|\psi_0},\qquad r(t)=-\frac{1}{N}\ln L(t) \, .
\end{equation}
This function shows a DPT as in the original model, see Fig.~\ref{fig:dpt_reduced}. Crucially, one can now perform measurements on the second subsystem (subsystem $S$) in order to extract the $k=0$ and $k=1$ contributions to the Loschmidt echo which make up the two sides of the sharkfin curve in Fig.~\ref{fig:dpt_reduced}. In particular, we can think of measuring in a single experimental run the Loschmidt echo of subsystem $I$ and the $S_y$-spin orientation on subsystem $S$, because these measurements commute, given that they act on different Hilbert spaces. Then we can post-select the Loschmidt echo measurements depending on the outcomes of the $S_y$ measurements. In particular, we form two groups, a '$+$' and a '$-$' group, depending on whether the $S_y$ outcome was $S_y\geq 0$ or $S_y<0$. In a semiclassical picture, these two groups isolate the parts of the wave function on the corresponding two halves of the phase space. To formalize this, we consider a measurement of the POVM $\{E_+, E_-\}$ where $\id=E_++E_-$ and (considering $N$ even for simplicity)
\begin{equation}
    E_+=\id_I\otimes\sum_{m=0}^{N/2}\ketbra{N/2,m,\text{``y"}}{N/2,m,\text{``y"}} ,
\end{equation}
where the eigenstates of $S_y$ are
\begin{equation}
    S_y\ket{N/2,m,\text{``y"}}=m\ket{N/2,m,\text{``y"}}.
\end{equation}
The conditioned measurement outcomes for the Loschmidt echo are then obtained from the conditioned reduced state of the $I$ subsystem
\begin{equation}
    \rho_I^\pm(t)=\frac{\tr_S E_\pm \ketbra{\Psi(t)}{\Psi(t)}}{\braket{\Psi(t)|E_\pm|\Psi(t)}}.
\end{equation}
In this way the reduced state Loschmidt echo \eqref{eq:reduced_state_Loschmidt} is decomposed into two parts
\begin{equation}\label{eq:loschmidt_conditioned}
    L(t)=L_+(t)+L_-(t),\qquad L_\pm(t)=p_\pm(t)\braket{\psi_0|\rho_I^\pm(t)|\psi_0} \,,\qquad p_\pm(t)= \braket{\Psi(t)|E_\pm|\Psi(t)},
\end{equation}
where $p_\pm(t)$ is the probability of a respective measurement outcome of the $E_\pm$ measurement at time $t$.
Fig.~\ref{fig:dpt_reduced} shows the numerical results for the two contributions in a system with just $N=20$ atoms. The two curves clearly make up the two sides of the sharkfin, and cross at the DPT. Because both contributions are exponentially small it is obvious that for $N\rightarrow\infty$ the kink becomes sharp and the curve nonanalytic
\begin{equation}\label{eq:rate_func_conditioned}
    r(t)\rightarrow \mathrm{min}\{r_+(t),r_-(t)\},\quad r_\pm(t)=-\frac{1}{N}\ln L_\pm(t).
\end{equation}
In a semiclassical picture, one can think of the conditioned Loschmidt echos $L_\pm(t)$ containing, respectively, the two complex action contributions \eqref{eq:loschmidt_1} and \eqref{eq:loschmidt_2} which describe the exponential tails of the two caustics.

\section{Switching Lines in Fock Space}\label{sec:switching_lines}

\begin{figure}
    \centering
    \includegraphics{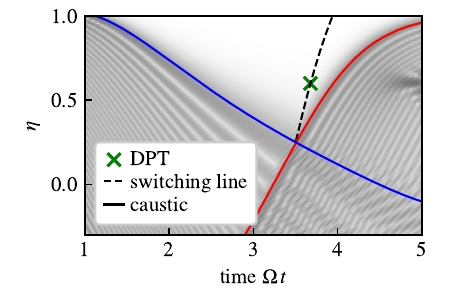}
    \caption{Fock-space overlaps of the FCTFIM wave function as in Fig.~\ref{fig:caustics}. Blue and red curves indicate the position of the fold caustics and the DPT is indicated by the green marker. Between the caustics, a switching line can be identified at which the asymptotic Loschmidt echo becomes nonanalytic. The DPT is given by the intersection of this switching line with the initial condition $\eta_0=0.6$.}
    \label{fig:switching_line}
\end{figure}

Our analysis of the Loschmidt echo can be generalized to arbitrary Fock-state overlaps. In fact, all previous arguments are independent of the particular Fock-state overlap, and a nonanalytic switching that leads to the DPT is expected to occur generically in classically forbidden regions where the wave function is exponentially suppressed. The more general concept that we identify is that of \emphquote{switching lines}, a phenomenon that was introduced in the 90s in the context of classical dissipative systems with weak fluctuations \cite{Dykman1994Nov,Kamenev2011}. 

In the FCTFIM the  DPT is merely a single point of a more general switching line where nonanalytic behaviour occurs due to overlapping exponential tails. In our example Fig.~\ref{fig:caustics} we expect a switching line to originate at the intersection of the two caustics (that separate the $k=0$ and $k=1$ trajectory classes), and passing through the DPT point. We confirm this expected behaviour numerically and the resulting switching line is shown in Fig.~\ref{fig:switching_line}.

Based on this analysis, we take the viewpoint that a DPT should properly be considered an example of large deviations \cite{Lazarescu2019Aug,Link2020Sep} within the quantum mechanical wave function. Large deviation theory \cite{Touchette2009Jul,Jack2020Apr,Causer2022Mar} aims to describe exponentially rare events in terms of a simplified physical theory, mathematically exploiting the ``rareness'' by a weak-fluctuation expansion in order to determine a rate function. In the present case the weak-fluctuation limit falls in line with semiclassics, such that the Loschmidt rate function can be described using classical actions in complexified phase space. Complex trajectories correspond to rare events that contribute to the transient return probability. When more than one trajectory exists, the exponentially dominant contribution may switch leading to a nonanalytic kink in the leading-order rate function.

Given that the switching process involves the exchange in dominance between two exponentials, it is natural to ask whether this is related to the Stokes phenomenon known from the asymptotic analysis of integrals of the form $I(\alpha)=\int_{C} \exp[\alpha f(z)]dz$, where $\alpha$ is a large positive real number which in our case is the number of spins $N$, and $C$ is a contour which in the method of steepest descents is deformed to pass through the saddles of $\Re f(z)$ in the complex plane. The integrals appearing in the exact representation of the wave function given by the  Poisson sum in Eq.\ \eqref{eq:poisson} can be evaluated in this way as explained in Sec. \ref{sec:WKB}. On an anti-Stokes set (a set of points which in the case of two dimensions would form a one dimensional curve), two saddles contributing to the integral exchange dominance (and across a Stokes set the number of contributing saddles changes by one) \cite{Wright80}. However, in our opinion a switching line does not correspond to an anti-Stokes set. This is because the Stokes phenomenon, and the anti-Stokes set, is associated with the behaviour of a single integral whereas in our case the exchange in dominance is between the exponentials from two different integrals, i.e.\ two different terms in the Poisson sum. For example, for the case  shown in the left panel of Fig.\ \ref{fig:t_E} the switching line lies between the asymptotic tails of the functions from the $k=0$ an $k=1$ terms of the Poisson sum which we take to be independent. This viewpoint is in agreement with the calculation presented in Ref.\ \cite{Berry94} which uses the Poisson sum formula in the context of quantum billiards.

\section{Discussion and Conclusions}
\label{sec:conclusion}
By explicitly calculating the Loschmidt echo in the semiclassical regime using the Poisson summation formula, we have established a novel interpretation of the origin of DPTs (or more precisely, type-II DPTs in the classification scheme given in reference \cite{Corps2023Mar}) in the FCTFIM via the switching in exponential dominance between evanescent tails originating from two neighbouring fold caustics in Fock space.
This switching occurs as a function of time and is essentially the quantum analogue of the switching that occurs as a function of angle between the primary and secondary bows on either side of Alexander's DB in ordinary rainbows seen in the sky.
We therefore have a striking analogy between the behaviour of a microscopic quantum system and a macroscopic `everyday' natural phenomenon. Motivated by this analogy, we refer to the period of time during which the Loschmidt echo is made up of only exponentially small tails as a quantum DB. 

The link between these seemingly different physical systems is the universality associated with wave singularities. According to catastrophe theory, fold caustics are structurally stable against perturbations such as variations in the initial conditions or details of the Hamiltonian, and this is the reason they occur generically in nature. Examples in classical waves include rainbows, the twinkling of starlight (due to random focusing by a turbulent atmosphere) \cite{Berry1977dec}, gravitational lensing \cite{Nye_natural_focusing}, ship's wakes \cite{Kelvin1905}, tsunamis \cite{Berry07}, tidal bores \cite{Berry2018}, and freak waves \cite{Hohmann10,metzger2014}. Caustics also occur in quantum waves such as Bose-Einstein condensates \cite{Huckans2009Oct,Rosenblum2014Mar,Mossman2021} and electron microscope beams \cite{Petersen2013Jan}. Most closely connected to the present work, caustics are predicted to occur following quenches in spin chains \cite{Kirkby2019Nov,Riddell2023sep} and in the two- and three-site Bose Hubbard models \cite{ODell2012Oct,Kirkby2022Feb}. In the latter example the caustics were also in Fock space which is the space relevant to quantum many-body states.

The interpretation of the DPT in terms of evanescent tails coming from different caustics allowed us to propose a measurement scheme that could be used to experimentally verify the existence of a DPT in this system. 
The idea is to restrict the quantum state to a local region in phase space that entails a single caustic when projected onto Fock-space. This way we can isolate the amplitude contributions due to a specific caustic. In a real system one can realize this by considering a bipartition of the system and conditioning a Fock-state measurement of the first subsystem onto a phase-space position measurement on the second subsystem. This makes it possible to measure separately the Airy-like functions that dress the caustics and cross in the quantum DB region. Such a nontrivial protocol could be realized with state-of-the-art quantum simulators, for instance trapped ion platforms \cite{Jurcevic2017Aug}. An open question in this regard is whether an observation of singular features in the many-body wave function such as DPTs could also be made visible in large collective systems, which is the natural setting for cavity-QED with cold atoms \cite{Muniz2020Apr}.

Since caustics are a common feature of nonequilibrium classical and quantum waves it would seem likely that the quantum DB mechanism proposed in this paper has wider applicability than the FCTFIM. However, the FCTFIM is an integrable model, as is the 1D TFIM which was the original setting of DPTs \cite{Heyl2013Mar}. The question then arises as to whether quantum DBs can also occur in nonintegrable models? Indeed, caustics arise from the projection down onto subspaces such as the momentum or coordinate planes of the motion around the invariant tori that foliate phase space in classical integrable systems \cite{Berry1983LesHouches}. The invariant tori themselves occur due to the complete set of conserved quantities that exist in integrable systems, and are therefore destroyed by strong chaos where these conserved quantities are absent.  However, between complete integrability and strong chaos there lies the much more typical regime of weak chaos. The celebrated Kolmogorov-Arnold-Moser (KAM) theory shows that some invariant tori survive in this intermediate regime \cite{Arnold_book}, and hence caustics do too.
In a recent paper on caustics in Heisenberg spin chains it was shown that caustics do indeed have structural stability against weak chaos which was introduced by adding non-integrable terms to the Hamiltonian in a controlled way \cite{Riddell2023sep}, and this result is in accordance with the possibility of a quantum version of the KAM theory that would preserve wave catastrophes in weakly nonintegrable regimes \cite{Brandino2015}. Moreover, DPTs have been specifically shown to also occur in nonintegrable models \cite{Karrasch2013may}. Thus, although further investigation is needed, there is substantial evidence suggesting that the quantum DB mechanism can also be used to explain DPTs in weakly chaotic systems. 

It might also be wondered whether quantum caustics can occur in mixed states and hence survive the effects of decoherence inherent in an open quantum system? The answer is yes: the fate of caustics has been investigated in a system equivalent to the FCTFIM studied here but subject to a continuous measurement of the magnetization such that the measurement-induced backaction leads to decoherence \cite{Goldberg2019Dec}. Solutions of the master equation for the density matrix reveal that caustics are still present at short times although at long times they eventually get washed away by the noise inherent to a quantum measurement. As might be expected, the rate of attrition depends on the strength of the measurement and thus the caustics last longer when the measurement is weak (the backaction in fact avoids a paradox whereby measurements could introduce singularities via the caustics \cite{Goldberg2019Dec}).

There are potentially many other systems and situations in which the quantum caustic mechanism for DPTs proposed in this paper might be tested in future work, but two seem particularly pertinent. The first type are spatially extended systems with short range interactions, such as the ordinary TFIM on a spin chain, where Hilbert space is larger than the FCTFIM and a desciption in terms of a collective macroscopic (classical) state is less applicable. For this case it is worth pointing out that caustics are known to occur in real space in the 1D TFIM and in fact they provide an explanation for the light cone-like transmission of information through the system following a quench \cite{Kirkby2019Nov}. A similar study of quenches in 1D Heisenberg spin chains revealed that caustics also occur in out-of-time-ordered correlators (OTOCs), underlining the generic nature of the phenomenon. We therefore speculate that caustics, being generic dynamical singularities in both the wave function and correlators, will continue to play a key role in the dynamics of systems with larger Hilbert spaces. Catastrophe theory guarantees the existence of higher dimensional caustics (see Ref.\ \cite{Kirkby2022Feb} for three dimensional caustics in quantum many-body dynamics), although the calculations in this case will be harder due to the higher complexity and it may be more tractable to switch to a statistical version of catastrophe theory \cite{Nye_natural_focusing}.

A second interesting future direction concerns whether quantum DBs can be seen in many-body localized states in disordered systems. These are characterized by a full set of locally conserved quantities and so have some similarties to integrable systems where constants of the motion are responsible for caustics, and yet are different because the conserved quantities in the latter are global not local \cite{Abanin2019}.

\begin{acknowledgments}
We are grateful to Professors M. V. Berry, D. M. Gangardt and C. J. Howls for discussions concerning the Stokes phenomenon, and Professor Markus Oberthaler for his question  during a seminar as to whether there might be a connection between quantum catastrophes and DPTs, which provided one of the motivations behind this work.
DO acknowledges support from the Natural Sciences and Engineering Research
Council of Canada (NSERC) [Ref. No. RGPIN-2017-06605].
\end{acknowledgments}

\bibliography{bib.bib}

%apsrev4-2.bst 2019-01-14 (MD) hand-edited version of apsrev4-1.bst
%Control: key (0)
%Control: author (8) initials jnrlst
%Control: editor formatted (1) identically to author
%Control: production of article title (0) allowed
%Control: page (0) single
%Control: year (1) truncated
%Control: production of eprint (0) enabled
\begin{thebibliography}{113}%
\makeatletter
\providecommand \@ifxundefined [1]{%
 \@ifx{#1\undefined}
}%
\providecommand \@ifnum [1]{%
 \ifnum #1\expandafter \@firstoftwo
 \else \expandafter \@secondoftwo
 \fi
}%
\providecommand \@ifx [1]{%
 \ifx #1\expandafter \@firstoftwo
 \else \expandafter \@secondoftwo
 \fi
}%
\providecommand \natexlab [1]{#1}%
\providecommand \enquote  [1]{``#1''}%
\providecommand \bibnamefont  [1]{#1}%
\providecommand \bibfnamefont [1]{#1}%
\providecommand \citenamefont [1]{#1}%
\providecommand \href@noop [0]{\@secondoftwo}%
\providecommand \href [0]{\begingroup \@sanitize@url \@href}%
\providecommand \@href[1]{\@@startlink{#1}\@@href}%
\providecommand \@@href[1]{\endgroup#1\@@endlink}%
\providecommand \@sanitize@url [0]{\catcode `\\12\catcode `\$12\catcode `\&12\catcode `\#12\catcode `\^12\catcode `\_12\catcode `\%12\relax}%
\providecommand \@@startlink[1]{}%
\providecommand \@@endlink[0]{}%
\providecommand \url  [0]{\begingroup\@sanitize@url \@url }%
\providecommand \@url [1]{\endgroup\@href {#1}{\urlprefix }}%
\providecommand \urlprefix  [0]{URL }%
\providecommand \Eprint [0]{\href }%
\providecommand \doibase [0]{https://doi.org/}%
\providecommand \selectlanguage [0]{\@gobble}%
\providecommand \bibinfo  [0]{\@secondoftwo}%
\providecommand \bibfield  [0]{\@secondoftwo}%
\providecommand \translation [1]{[#1]}%
\providecommand \BibitemOpen [0]{}%
\providecommand \bibitemStop [0]{}%
\providecommand \bibitemNoStop [0]{.\EOS\space}%
\providecommand \EOS [0]{\spacefactor3000\relax}%
\providecommand \BibitemShut  [1]{\csname bibitem#1\endcsname}%
\let\auto@bib@innerbib\@empty
%</preamble>
\bibitem [{\citenamefont {Yang}\ and\ \citenamefont {Lee}(1952)}]{yanglee1952i}%
  \BibitemOpen
  \bibfield  {author} {\bibinfo {author} {\bibfnamefont {C.~N.}\ \bibnamefont {Yang}}\ and\ \bibinfo {author} {\bibfnamefont {T.~D.}\ \bibnamefont {Lee}},\ }\bibfield  {title} {\bibinfo {title} {Statistical theory of equations of state and phase transitions. {I}. {T}heory of condensation},\ }\href {https://doi.org/10.1103/PhysRev.87.404} {\bibfield  {journal} {\bibinfo  {journal} {Phys. Rev.}\ }\textbf {\bibinfo {volume} {87}},\ \bibinfo {pages} {404} (\bibinfo {year} {1952})}\BibitemShut {NoStop}%
\bibitem [{\citenamefont {Lee}\ and\ \citenamefont {Yang}(1952)}]{yanglee1952ii}%
  \BibitemOpen
  \bibfield  {author} {\bibinfo {author} {\bibfnamefont {T.~D.}\ \bibnamefont {Lee}}\ and\ \bibinfo {author} {\bibfnamefont {C.~N.}\ \bibnamefont {Yang}},\ }\bibfield  {title} {\bibinfo {title} {Statistical theory of equations of state and phase transitions. {II}. {L}attice gas and {I}sing model},\ }\href {https://doi.org/10.1103/PhysRev.87.410} {\bibfield  {journal} {\bibinfo  {journal} {Phys. Rev.}\ }\textbf {\bibinfo {volume} {87}},\ \bibinfo {pages} {410} (\bibinfo {year} {1952})}\BibitemShut {NoStop}%
\bibitem [{\citenamefont {Sondhi}\ \emph {et~al.}(1997)\citenamefont {Sondhi}, \citenamefont {Girvin}, \citenamefont {Carini},\ and\ \citenamefont {Shahar}}]{Sondhi97}%
  \BibitemOpen
  \bibfield  {author} {\bibinfo {author} {\bibfnamefont {S.~L.}\ \bibnamefont {Sondhi}}, \bibinfo {author} {\bibfnamefont {S.~M.}\ \bibnamefont {Girvin}}, \bibinfo {author} {\bibfnamefont {J.~P.}\ \bibnamefont {Carini}},\ and\ \bibinfo {author} {\bibfnamefont {D.}~\bibnamefont {Shahar}},\ }\bibfield  {title} {\bibinfo {title} {Continuous quantum phase transitions},\ }\href {https://doi.org/10.1103/RevModPhys.69.315} {\bibfield  {journal} {\bibinfo  {journal} {Rev. Mod. Phys.}\ }\textbf {\bibinfo {volume} {69}},\ \bibinfo {pages} {315} (\bibinfo {year} {1997})}\BibitemShut {NoStop}%
\bibitem [{\citenamefont {Walls}\ \emph {et~al.}(1978)\citenamefont {Walls}, \citenamefont {Drummond}, \citenamefont {Hassan},\ and\ \citenamefont {Carmichael}}]{Walls1978}%
  \BibitemOpen
  \bibfield  {author} {\bibinfo {author} {\bibfnamefont {D.~F.}\ \bibnamefont {Walls}}, \bibinfo {author} {\bibfnamefont {P.~D.}\ \bibnamefont {Drummond}}, \bibinfo {author} {\bibfnamefont {S.~S.}\ \bibnamefont {Hassan}},\ and\ \bibinfo {author} {\bibfnamefont {H.~J.}\ \bibnamefont {Carmichael}},\ }\bibfield  {title} {\bibinfo {title} {Non-equilibrium phase transitions in cooperative atomic systems},\ }\href {https://doi.org/10.1143/PTPS.64.307} {\bibfield  {journal} {\bibinfo  {journal} {Progress of Theoretical Physics Supplement}\ }\textbf {\bibinfo {volume} {64}},\ \bibinfo {pages} {307} (\bibinfo {year} {1978})}\BibitemShut {NoStop}%
\bibitem [{\citenamefont {Torre}\ \emph {et~al.}(2013)\citenamefont {Torre}, \citenamefont {Diehl}, \citenamefont {Lukin}, \citenamefont {Sachdev},\ and\ \citenamefont {Strack}}]{Torre2013Feb}%
  \BibitemOpen
  \bibfield  {author} {\bibinfo {author} {\bibfnamefont {E.~G.~D.}\ \bibnamefont {Torre}}, \bibinfo {author} {\bibfnamefont {S.}~\bibnamefont {Diehl}}, \bibinfo {author} {\bibfnamefont {M.~D.}\ \bibnamefont {Lukin}}, \bibinfo {author} {\bibfnamefont {S.}~\bibnamefont {Sachdev}},\ and\ \bibinfo {author} {\bibfnamefont {P.}~\bibnamefont {Strack}},\ }\bibfield  {title} {\bibinfo {title} {{Keldysh approach for nonequilibrium phase transitions in quantum optics: Beyond the Dicke model in optical cavities}},\ }\href {https://doi.org/10.1103/PhysRevA.87.023831} {\bibfield  {journal} {\bibinfo  {journal} {Phys. Rev. A}\ }\textbf {\bibinfo {volume} {87}},\ \bibinfo {pages} {023831} (\bibinfo {year} {2013})}\BibitemShut {NoStop}%
\bibitem [{\citenamefont {Polkovnikov}\ \emph {et~al.}(2011)\citenamefont {Polkovnikov}, \citenamefont {Sengupta}, \citenamefont {Silva},\ and\ \citenamefont {Vengalattore}}]{Polkovnikov11}%
  \BibitemOpen
  \bibfield  {author} {\bibinfo {author} {\bibfnamefont {A.}~\bibnamefont {Polkovnikov}}, \bibinfo {author} {\bibfnamefont {K.}~\bibnamefont {Sengupta}}, \bibinfo {author} {\bibfnamefont {A.}~\bibnamefont {Silva}},\ and\ \bibinfo {author} {\bibfnamefont {M.}~\bibnamefont {Vengalattore}},\ }\bibfield  {title} {\bibinfo {title} {Colloquium: Nonequilibrium dynamics of closed interacting quantum systems},\ }\href {https://doi.org/10.1103/RevModPhys.83.863} {\bibfield  {journal} {\bibinfo  {journal} {Rev. Mod. Phys.}\ }\textbf {\bibinfo {volume} {83}},\ \bibinfo {pages} {863} (\bibinfo {year} {2011})}\BibitemShut {NoStop}%
\bibitem [{\citenamefont {Eisert}\ \emph {et~al.}(2015)\citenamefont {Eisert}, \citenamefont {Friesdorf},\ and\ \citenamefont {Gogolin}}]{Eisert15}%
  \BibitemOpen
  \bibfield  {author} {\bibinfo {author} {\bibfnamefont {J.}~\bibnamefont {Eisert}}, \bibinfo {author} {\bibfnamefont {M.}~\bibnamefont {Friesdorf}},\ and\ \bibinfo {author} {\bibfnamefont {C.}~\bibnamefont {Gogolin}},\ }\bibfield  {title} {\bibinfo {title} {Quantum many-body systems out of equilibrium},\ }\href {https://doi.org/10.1038/nphys3215} {\bibfield  {journal} {\bibinfo  {journal} {Nat. Phys.}\ }\textbf {\bibinfo {volume} {11}},\ \bibinfo {pages} {124} (\bibinfo {year} {2015})}\BibitemShut {NoStop}%
\bibitem [{\citenamefont {Pr{\ifmmode\ddot{u}\else\"{u}\fi}fer}\ \emph {et~al.}(2018)\citenamefont {Pr{\ifmmode\ddot{u}\else\"{u}\fi}fer}, \citenamefont {Kunkel}, \citenamefont {Strobel}, \citenamefont {Lannig}, \citenamefont {Linnemann}, \citenamefont {Schmied}, \citenamefont {Berges}, \citenamefont {Gasenzer},\ and\ \citenamefont {Oberthaler}}]{Prufer2018Nov}%
  \BibitemOpen
  \bibfield  {author} {\bibinfo {author} {\bibfnamefont {M.}~\bibnamefont {Pr{\ifmmode\ddot{u}\else\"{u}\fi}fer}}, \bibinfo {author} {\bibfnamefont {P.}~\bibnamefont {Kunkel}}, \bibinfo {author} {\bibfnamefont {H.}~\bibnamefont {Strobel}}, \bibinfo {author} {\bibfnamefont {S.}~\bibnamefont {Lannig}}, \bibinfo {author} {\bibfnamefont {D.}~\bibnamefont {Linnemann}}, \bibinfo {author} {\bibfnamefont {C.-M.}\ \bibnamefont {Schmied}}, \bibinfo {author} {\bibfnamefont {J.}~\bibnamefont {Berges}}, \bibinfo {author} {\bibfnamefont {T.}~\bibnamefont {Gasenzer}},\ and\ \bibinfo {author} {\bibfnamefont {M.~K.}\ \bibnamefont {Oberthaler}},\ }\bibfield  {title} {\bibinfo {title} {{Observation of universal dynamics in a spinor Bose gas far from equilibrium}},\ }\href {https://doi.org/10.1038/s41586-018-0659-0} {\bibfield  {journal} {\bibinfo  {journal} {Nature}\ }\textbf {\bibinfo {volume} {563}},\ \bibinfo {pages} {217} (\bibinfo {year} {2018})}\BibitemShut {NoStop}%
\bibitem [{\citenamefont {Eigen}\ \emph {et~al.}(2018)\citenamefont {Eigen}, \citenamefont {Glidden}, \citenamefont {Lopes}, \citenamefont {Cornell}, \citenamefont {Smith},\ and\ \citenamefont {Hadzibabic}}]{Eigen18}%
  \BibitemOpen
  \bibfield  {author} {\bibinfo {author} {\bibfnamefont {C.}~\bibnamefont {Eigen}}, \bibinfo {author} {\bibfnamefont {J.~A.~P.}\ \bibnamefont {Glidden}}, \bibinfo {author} {\bibfnamefont {R.}~\bibnamefont {Lopes}}, \bibinfo {author} {\bibfnamefont {E.~A.}\ \bibnamefont {Cornell}}, \bibinfo {author} {\bibfnamefont {R.~P.}\ \bibnamefont {Smith}},\ and\ \bibinfo {author} {\bibfnamefont {Z.}~\bibnamefont {Hadzibabic}},\ }\bibfield  {title} {\bibinfo {title} {Universal prethermal dynamics of {B}ose gases quenched to unitarity},\ }\href {https://doi.org/10.1038/s41586-018-0674-1} {\bibfield  {journal} {\bibinfo  {journal} {Nature}\ }\textbf {\bibinfo {volume} {563}},\ \bibinfo {pages} {221} (\bibinfo {year} {2018})}\BibitemShut {NoStop}%
\bibitem [{\citenamefont {Erne}\ \emph {et~al.}(2018)\citenamefont {Erne}, \citenamefont {Bücker}, \citenamefont {Gasenzer}, \citenamefont {Berges},\ and\ \citenamefont {Schmiedmayer}}]{Erne18}%
  \BibitemOpen
  \bibfield  {author} {\bibinfo {author} {\bibfnamefont {S.}~\bibnamefont {Erne}}, \bibinfo {author} {\bibfnamefont {R.}~\bibnamefont {Bücker}}, \bibinfo {author} {\bibfnamefont {T.}~\bibnamefont {Gasenzer}}, \bibinfo {author} {\bibfnamefont {J.}~\bibnamefont {Berges}},\ and\ \bibinfo {author} {\bibfnamefont {J.}~\bibnamefont {Schmiedmayer}},\ }\bibfield  {title} {\bibinfo {title} {Universal dynamics in an isolated one-dimensional {B}ose gas far from equilibrium},\ }\href {https://doi.org/10.1038/s41586-018-0667-0} {\bibfield  {journal} {\bibinfo  {journal} {Nature}\ }\textbf {\bibinfo {volume} {563}},\ \bibinfo {pages} {225} (\bibinfo {year} {2018})}\BibitemShut {NoStop}%
\bibitem [{\citenamefont {Ribeiro}\ \emph {et~al.}(2007)\citenamefont {Ribeiro}, \citenamefont {Vidal},\ and\ \citenamefont {Mosseri}}]{Ribeiro07}%
  \BibitemOpen
  \bibfield  {author} {\bibinfo {author} {\bibfnamefont {P.}~\bibnamefont {Ribeiro}}, \bibinfo {author} {\bibfnamefont {J.}~\bibnamefont {Vidal}},\ and\ \bibinfo {author} {\bibfnamefont {R.}~\bibnamefont {Mosseri}},\ }\bibfield  {title} {\bibinfo {title} {Thermodynamical limit of the {L}ipkin-{M}eshkov-{G}lick model},\ }\href {https://doi.org/10.1103/PhysRevLett.99.050402} {\bibfield  {journal} {\bibinfo  {journal} {Phys. Rev. Lett.}\ }\textbf {\bibinfo {volume} {99}},\ \bibinfo {pages} {050402} (\bibinfo {year} {2007})}\BibitemShut {NoStop}%
\bibitem [{\citenamefont {Caprio}\ \emph {et~al.}(2008)\citenamefont {Caprio}, \citenamefont {Cejnar},\ and\ \citenamefont {Iachello}}]{Caprio2008}%
  \BibitemOpen
  \bibfield  {author} {\bibinfo {author} {\bibfnamefont {M.}~\bibnamefont {Caprio}}, \bibinfo {author} {\bibfnamefont {P.}~\bibnamefont {Cejnar}},\ and\ \bibinfo {author} {\bibfnamefont {F.}~\bibnamefont {Iachello}},\ }\bibfield  {title} {\bibinfo {title} {Excited state quantum phase transitions in many-body systems},\ }\href {https://doi.org/https://doi.org/10.1016/j.aop.2007.06.011} {\bibfield  {journal} {\bibinfo  {journal} {Ann. Phys.}\ }\textbf {\bibinfo {volume} {323}},\ \bibinfo {pages} {1106} (\bibinfo {year} {2008})}\BibitemShut {NoStop}%
\bibitem [{\citenamefont {Brandes}(2013)}]{Brandes13}%
  \BibitemOpen
  \bibfield  {author} {\bibinfo {author} {\bibfnamefont {T.}~\bibnamefont {Brandes}},\ }\bibfield  {title} {\bibinfo {title} {Excited-state quantum phase transitions in {D}icke superradiance models},\ }\href {https://doi.org/10.1103/PhysRevE.88.032133} {\bibfield  {journal} {\bibinfo  {journal} {Phys. Rev. E}\ }\textbf {\bibinfo {volume} {88}},\ \bibinfo {pages} {032133} (\bibinfo {year} {2013})}\BibitemShut {NoStop}%
\bibitem [{\citenamefont {Santos}\ \emph {et~al.}(2016)\citenamefont {Santos}, \citenamefont {T\'avora},\ and\ \citenamefont {P\'erez-Bernal}}]{Santos16}%
  \BibitemOpen
  \bibfield  {author} {\bibinfo {author} {\bibfnamefont {L.~F.}\ \bibnamefont {Santos}}, \bibinfo {author} {\bibfnamefont {M.}~\bibnamefont {T\'avora}},\ and\ \bibinfo {author} {\bibfnamefont {F.}~\bibnamefont {P\'erez-Bernal}},\ }\bibfield  {title} {\bibinfo {title} {Excited-state quantum phase transitions in many-body systems with infinite-range interaction: Localization, dynamics, and bifurcation},\ }\href {https://doi.org/10.1103/PhysRevA.94.012113} {\bibfield  {journal} {\bibinfo  {journal} {Phys. Rev. A}\ }\textbf {\bibinfo {volume} {94}},\ \bibinfo {pages} {012113} (\bibinfo {year} {2016})}\BibitemShut {NoStop}%
\bibitem [{\citenamefont {Cejnar}\ \emph {et~al.}(2021)\citenamefont {Cejnar}, \citenamefont {Stránský}, \citenamefont {Macek},\ and\ \citenamefont {Kloc}}]{Cejnar2021}%
  \BibitemOpen
  \bibfield  {author} {\bibinfo {author} {\bibfnamefont {P.}~\bibnamefont {Cejnar}}, \bibinfo {author} {\bibfnamefont {P.}~\bibnamefont {Stránský}}, \bibinfo {author} {\bibfnamefont {M.}~\bibnamefont {Macek}},\ and\ \bibinfo {author} {\bibfnamefont {M.}~\bibnamefont {Kloc}},\ }\bibfield  {title} {\bibinfo {title} {Excited-state quantum phase transitions},\ }\href {https://doi.org/10.1088/1751-8121/abdfe8} {\bibfield  {journal} {\bibinfo  {journal} {J. Phys. A: Math. Theor.}\ }\textbf {\bibinfo {volume} {54}},\ \bibinfo {pages} {133001} (\bibinfo {year} {2021})}\BibitemShut {NoStop}%
\bibitem [{\citenamefont {Marino}\ \emph {et~al.}(2022)\citenamefont {Marino}, \citenamefont {Eckstein}, \citenamefont {Foster},\ and\ \citenamefont {Rey}}]{Marino_2022}%
  \BibitemOpen
  \bibfield  {author} {\bibinfo {author} {\bibfnamefont {J.}~\bibnamefont {Marino}}, \bibinfo {author} {\bibfnamefont {M.}~\bibnamefont {Eckstein}}, \bibinfo {author} {\bibfnamefont {M.~S.}\ \bibnamefont {Foster}},\ and\ \bibinfo {author} {\bibfnamefont {A.~M.}\ \bibnamefont {Rey}},\ }\bibfield  {title} {\bibinfo {title} {Dynamical phase transitions in the collisionless pre-thermal states of isolated quantum systems: theory and experiments},\ }\href {https://doi.org/10.1088/1361-6633/ac906c} {\bibfield  {journal} {\bibinfo  {journal} {Rep. Prog. Phys.}\ }\textbf {\bibinfo {volume} {85}},\ \bibinfo {pages} {116001} (\bibinfo {year} {2022})}\BibitemShut {NoStop}%
\bibitem [{\citenamefont {Smerzi}\ \emph {et~al.}(1997)\citenamefont {Smerzi}, \citenamefont {Fantoni}, \citenamefont {Giovanazzi},\ and\ \citenamefont {Shenoy}}]{Smerzi97}%
  \BibitemOpen
  \bibfield  {author} {\bibinfo {author} {\bibfnamefont {A.}~\bibnamefont {Smerzi}}, \bibinfo {author} {\bibfnamefont {S.}~\bibnamefont {Fantoni}}, \bibinfo {author} {\bibfnamefont {S.}~\bibnamefont {Giovanazzi}},\ and\ \bibinfo {author} {\bibfnamefont {S.~R.}\ \bibnamefont {Shenoy}},\ }\bibfield  {title} {\bibinfo {title} {Quantum coherent atomic tunneling between two trapped {B}ose-{E}instein condensates},\ }\href {https://doi.org/10.1103/PhysRevLett.79.4950} {\bibfield  {journal} {\bibinfo  {journal} {Phys. Rev. Lett.}\ }\textbf {\bibinfo {volume} {79}},\ \bibinfo {pages} {4950} (\bibinfo {year} {1997})}\BibitemShut {NoStop}%
\bibitem [{\citenamefont {Cataliotti}\ \emph {et~al.}(2001)\citenamefont {Cataliotti}, \citenamefont {Burger}, \citenamefont {Fort}, \citenamefont {Maddaloni}, \citenamefont {Minardi}, \citenamefont {Trombettoni}, \citenamefont {Smerzi},\ and\ \citenamefont {Inguscio}}]{Cataliotti01}%
  \BibitemOpen
  \bibfield  {author} {\bibinfo {author} {\bibfnamefont {F.}~\bibnamefont {Cataliotti}}, \bibinfo {author} {\bibfnamefont {S.}~\bibnamefont {Burger}}, \bibinfo {author} {\bibfnamefont {C.}~\bibnamefont {Fort}}, \bibinfo {author} {\bibfnamefont {P.}~\bibnamefont {Maddaloni}}, \bibinfo {author} {\bibfnamefont {F.}~\bibnamefont {Minardi}}, \bibinfo {author} {\bibfnamefont {A.}~\bibnamefont {Trombettoni}}, \bibinfo {author} {\bibfnamefont {A.}~\bibnamefont {Smerzi}},\ and\ \bibinfo {author} {\bibfnamefont {M.}~\bibnamefont {Inguscio}},\ }\bibfield  {title} {\bibinfo {title} {{J}osephson junction arrays with {B}ose-{E}instein condensates},\ }\href {https://doi.org/10.1126/science.1062612} {\bibfield  {journal} {\bibinfo  {journal} {Science}\ }\textbf {\bibinfo {volume} {293}},\ \bibinfo {pages} {843} (\bibinfo {year} {2001})}\BibitemShut {NoStop}%
\bibitem [{\citenamefont {Albiez}\ \emph {et~al.}(2005)\citenamefont {Albiez}, \citenamefont {Gati}, \citenamefont {F{\ifmmode\ddot{o}\else\"{o}\fi}lling}, \citenamefont {Hunsmann}, \citenamefont {Cristiani},\ and\ \citenamefont {Oberthaler}}]{Albiez2005Jun}%
  \BibitemOpen
  \bibfield  {author} {\bibinfo {author} {\bibfnamefont {M.}~\bibnamefont {Albiez}}, \bibinfo {author} {\bibfnamefont {R.}~\bibnamefont {Gati}}, \bibinfo {author} {\bibfnamefont {J.}~\bibnamefont {F{\ifmmode\ddot{o}\else\"{o}\fi}lling}}, \bibinfo {author} {\bibfnamefont {S.}~\bibnamefont {Hunsmann}}, \bibinfo {author} {\bibfnamefont {M.}~\bibnamefont {Cristiani}},\ and\ \bibinfo {author} {\bibfnamefont {M.~K.}\ \bibnamefont {Oberthaler}},\ }\bibfield  {title} {\bibinfo {title} {{Direct Observation of Tunneling and Nonlinear Self-Trapping in a Single Bosonic Josephson Junction}},\ }\href {https://doi.org/10.1103/PhysRevLett.95.010402} {\bibfield  {journal} {\bibinfo  {journal} {Phys. Rev. Lett.}\ }\textbf {\bibinfo {volume} {95}},\ \bibinfo {pages} {010402} (\bibinfo {year} {2005})}\BibitemShut {NoStop}%
\bibitem [{\citenamefont {Levy}\ \emph {et~al.}(2007)\citenamefont {Levy}, \citenamefont {Lahoud}, \citenamefont {Shomroni},\ and\ \citenamefont {Steinhauer}}]{Levy2007Oct}%
  \BibitemOpen
  \bibfield  {author} {\bibinfo {author} {\bibfnamefont {S.}~\bibnamefont {Levy}}, \bibinfo {author} {\bibfnamefont {E.}~\bibnamefont {Lahoud}}, \bibinfo {author} {\bibfnamefont {I.}~\bibnamefont {Shomroni}},\ and\ \bibinfo {author} {\bibfnamefont {J.}~\bibnamefont {Steinhauer}},\ }\bibfield  {title} {\bibinfo {title} {{The a.c. and d.c. Josephson effects in a Bose{\textendash}Einstein condensate}},\ }\href {https://doi.org/10.1038/nature06186} {\bibfield  {journal} {\bibinfo  {journal} {Nature}\ }\textbf {\bibinfo {volume} {449}},\ \bibinfo {pages} {579} (\bibinfo {year} {2007})}\BibitemShut {NoStop}%
\bibitem [{\citenamefont {Zibold}\ \emph {et~al.}(2010)\citenamefont {Zibold}, \citenamefont {Nicklas}, \citenamefont {Gross},\ and\ \citenamefont {Oberthaler}}]{Zibold10}%
  \BibitemOpen
  \bibfield  {author} {\bibinfo {author} {\bibfnamefont {T.}~\bibnamefont {Zibold}}, \bibinfo {author} {\bibfnamefont {E.}~\bibnamefont {Nicklas}}, \bibinfo {author} {\bibfnamefont {C.}~\bibnamefont {Gross}},\ and\ \bibinfo {author} {\bibfnamefont {M.~K.}\ \bibnamefont {Oberthaler}},\ }\bibfield  {title} {\bibinfo {title} {Classical bifurcation at the transition from {R}abi to {J}osephson dynamics},\ }\href {https://doi.org/10.1103/PhysRevLett.105.204101} {\bibfield  {journal} {\bibinfo  {journal} {Phys. Rev. Lett.}\ }\textbf {\bibinfo {volume} {105}},\ \bibinfo {pages} {204101} (\bibinfo {year} {2010})}\BibitemShut {NoStop}%
\bibitem [{\citenamefont {LeBlanc}\ \emph {et~al.}(2011)\citenamefont {LeBlanc}, \citenamefont {Bardon}, \citenamefont {McKeever}, \citenamefont {Extavour}, \citenamefont {Jervis}, \citenamefont {Thywissen}, \citenamefont {Piazza},\ and\ \citenamefont {Smerzi}}]{Leblanc2011}%
  \BibitemOpen
  \bibfield  {author} {\bibinfo {author} {\bibfnamefont {L.~J.}\ \bibnamefont {LeBlanc}}, \bibinfo {author} {\bibfnamefont {A.~B.}\ \bibnamefont {Bardon}}, \bibinfo {author} {\bibfnamefont {J.}~\bibnamefont {McKeever}}, \bibinfo {author} {\bibfnamefont {M.~H.~T.}\ \bibnamefont {Extavour}}, \bibinfo {author} {\bibfnamefont {D.}~\bibnamefont {Jervis}}, \bibinfo {author} {\bibfnamefont {J.~H.}\ \bibnamefont {Thywissen}}, \bibinfo {author} {\bibfnamefont {F.}~\bibnamefont {Piazza}},\ and\ \bibinfo {author} {\bibfnamefont {A.}~\bibnamefont {Smerzi}},\ }\bibfield  {title} {\bibinfo {title} {Dynamics of a tunable superfluid junction},\ }\href {https://doi.org/10.1103/PhysRevLett.106.025302} {\bibfield  {journal} {\bibinfo  {journal} {Phys. Rev. Lett.}\ }\textbf {\bibinfo {volume} {106}},\ \bibinfo {pages} {025302} (\bibinfo {year} {2011})}\BibitemShut {NoStop}%
\bibitem [{\citenamefont {Gerving}\ \emph {et~al.}(2012)\citenamefont {Gerving}, \citenamefont {Hoang}, \citenamefont {Land}, \citenamefont {Anquez}, \citenamefont {Hamley},\ and\ \citenamefont {Chapman}}]{Gerving12}%
  \BibitemOpen
  \bibfield  {author} {\bibinfo {author} {\bibfnamefont {C.~S.}\ \bibnamefont {Gerving}}, \bibinfo {author} {\bibfnamefont {T.~M.}\ \bibnamefont {Hoang}}, \bibinfo {author} {\bibfnamefont {B.~J.}\ \bibnamefont {Land}}, \bibinfo {author} {\bibfnamefont {M.}~\bibnamefont {Anquez}}, \bibinfo {author} {\bibfnamefont {C.~D.}\ \bibnamefont {Hamley}},\ and\ \bibinfo {author} {\bibfnamefont {M.~S.}\ \bibnamefont {Chapman}},\ }\bibfield  {title} {\bibinfo {title} {Non-equilibrium dynamics of an unstable quantum pendulum explored in a spin-1 {B}ose–{E}instein condensate},\ }\href {https://doi.org/10.1038/ncomms2179} {\bibfield  {journal} {\bibinfo  {journal} {Nature Communications}\ }\textbf {\bibinfo {volume} {3}},\ \bibinfo {pages} {1169} (\bibinfo {year} {2012})}\BibitemShut {NoStop}%
\bibitem [{\citenamefont {Trenkwalder}\ \emph {et~al.}(2016)\citenamefont {Trenkwalder}, \citenamefont {Spagnolli}, \citenamefont {Semeghini}, \citenamefont {Coop}, \citenamefont {Landini}, \citenamefont {Castilho}, \citenamefont {Pezzè}, \citenamefont {Modugno}, \citenamefont {Inguscio}, \citenamefont {Smerzi},\ and\ \citenamefont {Fattori}}]{Trenkwalder16}%
  \BibitemOpen
  \bibfield  {author} {\bibinfo {author} {\bibfnamefont {A.}~\bibnamefont {Trenkwalder}}, \bibinfo {author} {\bibfnamefont {G.}~\bibnamefont {Spagnolli}}, \bibinfo {author} {\bibfnamefont {G.}~\bibnamefont {Semeghini}}, \bibinfo {author} {\bibfnamefont {S.}~\bibnamefont {Coop}}, \bibinfo {author} {\bibfnamefont {M.}~\bibnamefont {Landini}}, \bibinfo {author} {\bibfnamefont {P.}~\bibnamefont {Castilho}}, \bibinfo {author} {\bibfnamefont {L.}~\bibnamefont {Pezzè}}, \bibinfo {author} {\bibfnamefont {G.}~\bibnamefont {Modugno}}, \bibinfo {author} {\bibfnamefont {M.}~\bibnamefont {Inguscio}}, \bibinfo {author} {\bibfnamefont {A.}~\bibnamefont {Smerzi}},\ and\ \bibinfo {author} {\bibfnamefont {M.}~\bibnamefont {Fattori}},\ }\bibfield  {title} {\bibinfo {title} {Quantum phase transitions with parity-symmetry breaking and hysteresis},\ }\href {https://doi.org/10.1038/nphys3743} {\bibfield  {journal} {\bibinfo  {journal} {Nature Physics}\ }\textbf {\bibinfo {volume} {12}},\ \bibinfo {pages} {826} (\bibinfo {year}
  {2016})}\BibitemShut {NoStop}%
\bibitem [{\citenamefont {Corps}\ \emph {et~al.}(2023)\citenamefont {Corps}, \citenamefont {Str{\ifmmode\acute{a}\else\'{a}\fi}nsk{\ifmmode\acute{y}\else\'{y}\fi}},\ and\ \citenamefont {Cejnar}}]{Corps2023Mar}%
  \BibitemOpen
  \bibfield  {author} {\bibinfo {author} {\bibfnamefont {{\ifmmode\acute{A}\else\'{A}\fi}.~L.}\ \bibnamefont {Corps}}, \bibinfo {author} {\bibfnamefont {P.}~\bibnamefont {Str{\ifmmode\acute{a}\else\'{a}\fi}nsk{\ifmmode\acute{y}\else\'{y}\fi}}},\ and\ \bibinfo {author} {\bibfnamefont {P.}~\bibnamefont {Cejnar}},\ }\bibfield  {title} {\bibinfo {title} {{Mechanism of dynamical phase transitions: The complex-time survival amplitude}},\ }\href {https://doi.org/10.1103/PhysRevB.107.094307} {\bibfield  {journal} {\bibinfo  {journal} {Phys. Rev. B}\ }\textbf {\bibinfo {volume} {107}},\ \bibinfo {pages} {094307} (\bibinfo {year} {2023})}\BibitemShut {NoStop}%
\bibitem [{\citenamefont {Heyl}\ \emph {et~al.}(2013)\citenamefont {Heyl}, \citenamefont {Polkovnikov},\ and\ \citenamefont {Kehrein}}]{Heyl2013Mar}%
  \BibitemOpen
  \bibfield  {author} {\bibinfo {author} {\bibfnamefont {M.}~\bibnamefont {Heyl}}, \bibinfo {author} {\bibfnamefont {A.}~\bibnamefont {Polkovnikov}},\ and\ \bibinfo {author} {\bibfnamefont {S.}~\bibnamefont {Kehrein}},\ }\bibfield  {title} {\bibinfo {title} {{Dynamical Quantum Phase Transitions in the Transverse-Field Ising Model}},\ }\href {https://doi.org/10.1103/PhysRevLett.110.135704} {\bibfield  {journal} {\bibinfo  {journal} {Phys. Rev. Lett.}\ }\textbf {\bibinfo {volume} {110}},\ \bibinfo {pages} {135704} (\bibinfo {year} {2013})}\BibitemShut {NoStop}%
\bibitem [{\citenamefont {Diehl}\ \emph {et~al.}(2010)\citenamefont {Diehl}, \citenamefont {Tomadin}, \citenamefont {Micheli}, \citenamefont {Fazio},\ and\ \citenamefont {Zoller}}]{Diehl2010Jul}%
  \BibitemOpen
  \bibfield  {author} {\bibinfo {author} {\bibfnamefont {S.}~\bibnamefont {Diehl}}, \bibinfo {author} {\bibfnamefont {A.}~\bibnamefont {Tomadin}}, \bibinfo {author} {\bibfnamefont {A.}~\bibnamefont {Micheli}}, \bibinfo {author} {\bibfnamefont {R.}~\bibnamefont {Fazio}},\ and\ \bibinfo {author} {\bibfnamefont {P.}~\bibnamefont {Zoller}},\ }\bibfield  {title} {\bibinfo {title} {{Dynamical Phase Transitions and Instabilities in Open Atomic Many-Body Systems}},\ }\href {https://doi.org/10.1103/PhysRevLett.105.015702} {\bibfield  {journal} {\bibinfo  {journal} {Phys. Rev. Lett.}\ }\textbf {\bibinfo {volume} {105}},\ \bibinfo {pages} {015702} (\bibinfo {year} {2010})}\BibitemShut {NoStop}%
\bibitem [{\citenamefont {Heyl}(2019)}]{Heyl2019Feb}%
  \BibitemOpen
  \bibfield  {author} {\bibinfo {author} {\bibfnamefont {M.}~\bibnamefont {Heyl}},\ }\bibfield  {title} {\bibinfo {title} {{Dynamical quantum phase transitions: A brief survey}},\ }\href {https://doi.org/10.1209/0295-5075/125/26001} {\bibfield  {journal} {\bibinfo  {journal} {EPL}\ }\textbf {\bibinfo {volume} {125}},\ \bibinfo {pages} {26001} (\bibinfo {year} {2019})}\BibitemShut {NoStop}%
\bibitem [{\citenamefont {Hunyadi}\ \emph {et~al.}(2004)\citenamefont {Hunyadi}, \citenamefont {R\'acz},\ and\ \citenamefont {Sasv\'ari}}]{Hunyadi04}%
  \BibitemOpen
  \bibfield  {author} {\bibinfo {author} {\bibfnamefont {V.}~\bibnamefont {Hunyadi}}, \bibinfo {author} {\bibfnamefont {Z.}~\bibnamefont {R\'acz}},\ and\ \bibinfo {author} {\bibfnamefont {L.}~\bibnamefont {Sasv\'ari}},\ }\bibfield  {title} {\bibinfo {title} {Dynamic scaling of fronts in the quantum {$XX$} chain},\ }\href {https://doi.org/10.1103/PhysRevE.69.066103} {\bibfield  {journal} {\bibinfo  {journal} {Phys. Rev. E}\ }\textbf {\bibinfo {volume} {69}},\ \bibinfo {pages} {066103} (\bibinfo {year} {2004})}\BibitemShut {NoStop}%
\bibitem [{\citenamefont {Eisler}\ and\ \citenamefont {Maislinger}(2018)}]{Eisler18}%
  \BibitemOpen
  \bibfield  {author} {\bibinfo {author} {\bibfnamefont {V.}~\bibnamefont {Eisler}}\ and\ \bibinfo {author} {\bibfnamefont {F.}~\bibnamefont {Maislinger}},\ }\bibfield  {title} {\bibinfo {title} {Hydrodynamical phase transition for domain-wall melting in the {XY} chain},\ }\href {https://doi.org/10.1103/PhysRevB.98.161117} {\bibfield  {journal} {\bibinfo  {journal} {Phys. Rev. B}\ }\textbf {\bibinfo {volume} {98}},\ \bibinfo {pages} {161117} (\bibinfo {year} {2018})}\BibitemShut {NoStop}%
\bibitem [{\citenamefont {Trapin}\ and\ \citenamefont {Heyl}(2018)}]{Trapin2018May}%
  \BibitemOpen
  \bibfield  {author} {\bibinfo {author} {\bibfnamefont {D.}~\bibnamefont {Trapin}}\ and\ \bibinfo {author} {\bibfnamefont {M.}~\bibnamefont {Heyl}},\ }\bibfield  {title} {\bibinfo {title} {{Constructing effective free energies for dynamical quantum phase transitions in the transverse-field Ising chain}},\ }\href {https://doi.org/10.1103/PhysRevB.97.174303} {\bibfield  {journal} {\bibinfo  {journal} {Phys. Rev. B}\ }\textbf {\bibinfo {volume} {97}},\ \bibinfo {pages} {174303} (\bibinfo {year} {2018})}\BibitemShut {NoStop}%
\bibitem [{\citenamefont {Lacki}\ and\ \citenamefont {Heyl}(2019)}]{Lacki2019Mar}%
  \BibitemOpen
  \bibfield  {author} {\bibinfo {author} {\bibfnamefont {M.}~\bibnamefont {Lacki}}\ and\ \bibinfo {author} {\bibfnamefont {M.}~\bibnamefont {Heyl}},\ }\bibfield  {title} {\bibinfo {title} {{Dynamical quantum phase transitions in collapse and revival oscillations of a quenched superfluid}},\ }\href {https://doi.org/10.1103/PhysRevB.99.121107} {\bibfield  {journal} {\bibinfo  {journal} {Phys. Rev. B}\ }\textbf {\bibinfo {volume} {99}},\ \bibinfo {pages} {121107} (\bibinfo {year} {2019})}\BibitemShut {NoStop}%
\bibitem [{\citenamefont {Khasseh}\ \emph {et~al.}(2020)\citenamefont {Khasseh}, \citenamefont {Russomanno}, \citenamefont {Schmitt}, \citenamefont {Heyl},\ and\ \citenamefont {Fazio}}]{Khasseh2020Jul}%
  \BibitemOpen
  \bibfield  {author} {\bibinfo {author} {\bibfnamefont {R.}~\bibnamefont {Khasseh}}, \bibinfo {author} {\bibfnamefont {A.}~\bibnamefont {Russomanno}}, \bibinfo {author} {\bibfnamefont {M.}~\bibnamefont {Schmitt}}, \bibinfo {author} {\bibfnamefont {M.}~\bibnamefont {Heyl}},\ and\ \bibinfo {author} {\bibfnamefont {R.}~\bibnamefont {Fazio}},\ }\bibfield  {title} {\bibinfo {title} {{Discrete truncated Wigner approach to dynamical phase transitions in Ising models after a quantum quench}},\ }\href {https://doi.org/10.1103/PhysRevB.102.014303} {\bibfield  {journal} {\bibinfo  {journal} {Phys. Rev. B}\ }\textbf {\bibinfo {volume} {102}},\ \bibinfo {pages} {014303} (\bibinfo {year} {2020})}\BibitemShut {NoStop}%
\bibitem [{\citenamefont {Halimeh}\ \emph {et~al.}(2021)\citenamefont {Halimeh}, \citenamefont {Van~Damme}, \citenamefont {Guo}, \citenamefont {Lang},\ and\ \citenamefont {Hauke}}]{Halimeh2021Sep}%
  \BibitemOpen
  \bibfield  {author} {\bibinfo {author} {\bibfnamefont {J.~C.}\ \bibnamefont {Halimeh}}, \bibinfo {author} {\bibfnamefont {M.}~\bibnamefont {Van~Damme}}, \bibinfo {author} {\bibfnamefont {L.}~\bibnamefont {Guo}}, \bibinfo {author} {\bibfnamefont {J.}~\bibnamefont {Lang}},\ and\ \bibinfo {author} {\bibfnamefont {P.}~\bibnamefont {Hauke}},\ }\bibfield  {title} {\bibinfo {title} {{Dynamical phase transitions in quantum spin models with antiferromagnetic long-range interactions}},\ }\href {https://doi.org/10.1103/PhysRevB.104.115133} {\bibfield  {journal} {\bibinfo  {journal} {Phys. Rev. B}\ }\textbf {\bibinfo {volume} {104}},\ \bibinfo {pages} {115133} (\bibinfo {year} {2021})}\BibitemShut {NoStop}%
\bibitem [{\citenamefont {Pallister}\ \emph {et~al.}(2022)\citenamefont {Pallister}, \citenamefont {Gangardt},\ and\ \citenamefont {Abanov}}]{Pallister22}%
  \BibitemOpen
  \bibfield  {author} {\bibinfo {author} {\bibfnamefont {J.~S.}\ \bibnamefont {Pallister}}, \bibinfo {author} {\bibfnamefont {D.~M.}\ \bibnamefont {Gangardt}},\ and\ \bibinfo {author} {\bibfnamefont {A.~G.}\ \bibnamefont {Abanov}},\ }\bibfield  {title} {\bibinfo {title} {Limit shape phase transitions: a merger of arctic circles},\ }\href {https://doi.org/10.1088/1751-8121/ac79ad} {\bibfield  {journal} {\bibinfo  {journal} {J. Phys. A: Math. Theor.}\ }\textbf {\bibinfo {volume} {55}},\ \bibinfo {pages} {304001} (\bibinfo {year} {2022})}\BibitemShut {NoStop}%
\bibitem [{\citenamefont {Van~Damme}\ \emph {et~al.}(2023)\citenamefont {Van~Damme}, \citenamefont {Desaules}, \citenamefont {Papi{\ifmmode\acute{c}\else\'{c}\fi}},\ and\ \citenamefont {Halimeh}}]{VanDamme2023Aug}%
  \BibitemOpen
  \bibfield  {author} {\bibinfo {author} {\bibfnamefont {M.}~\bibnamefont {Van~Damme}}, \bibinfo {author} {\bibfnamefont {J.-Y.}\ \bibnamefont {Desaules}}, \bibinfo {author} {\bibfnamefont {Z.}~\bibnamefont {Papi{\ifmmode\acute{c}\else\'{c}\fi}}},\ and\ \bibinfo {author} {\bibfnamefont {J.~C.}\ \bibnamefont {Halimeh}},\ }\bibfield  {title} {\bibinfo {title} {{Anatomy of dynamical quantum phase transitions}},\ }\href {https://doi.org/10.1103/PhysRevResearch.5.033090} {\bibfield  {journal} {\bibinfo  {journal} {Phys. Rev. Res.}\ }\textbf {\bibinfo {volume} {5}},\ \bibinfo {pages} {033090} (\bibinfo {year} {2023})}\BibitemShut {NoStop}%
\bibitem [{\citenamefont {Cheraghi}\ and\ \citenamefont {Sedlmayr}(2023)}]{Cheraghi2023Oct}%
  \BibitemOpen
  \bibfield  {author} {\bibinfo {author} {\bibfnamefont {H.}~\bibnamefont {Cheraghi}}\ and\ \bibinfo {author} {\bibfnamefont {N.}~\bibnamefont {Sedlmayr}},\ }\bibfield  {title} {\bibinfo {title} {Dynamical quantum phase transitions following double quenches: Persistence of the initial state vs dynamical phases},\ }\href {https://doi.org/10.1088/1367-2630/ad016e} {\bibfield  {journal} {\bibinfo  {journal} {New J. Phys.}\ }\textbf {\bibinfo {volume} {25}},\ \bibinfo {pages} {103035} (\bibinfo {year} {2023})}\BibitemShut {NoStop}%
\bibitem [{\citenamefont {Jurcevic}\ \emph {et~al.}(2017)\citenamefont {Jurcevic}, \citenamefont {Shen}, \citenamefont {Hauke}, \citenamefont {Maier}, \citenamefont {Brydges}, \citenamefont {Hempel}, \citenamefont {Lanyon}, \citenamefont {Heyl}, \citenamefont {Blatt},\ and\ \citenamefont {Roos}}]{Jurcevic2017Aug}%
  \BibitemOpen
  \bibfield  {author} {\bibinfo {author} {\bibfnamefont {P.}~\bibnamefont {Jurcevic}}, \bibinfo {author} {\bibfnamefont {H.}~\bibnamefont {Shen}}, \bibinfo {author} {\bibfnamefont {P.}~\bibnamefont {Hauke}}, \bibinfo {author} {\bibfnamefont {C.}~\bibnamefont {Maier}}, \bibinfo {author} {\bibfnamefont {T.}~\bibnamefont {Brydges}}, \bibinfo {author} {\bibfnamefont {C.}~\bibnamefont {Hempel}}, \bibinfo {author} {\bibfnamefont {B.~P.}\ \bibnamefont {Lanyon}}, \bibinfo {author} {\bibfnamefont {M.}~\bibnamefont {Heyl}}, \bibinfo {author} {\bibfnamefont {R.}~\bibnamefont {Blatt}},\ and\ \bibinfo {author} {\bibfnamefont {C.~F.}\ \bibnamefont {Roos}},\ }\bibfield  {title} {\bibinfo {title} {{Direct Observation of Dynamical Quantum Phase Transitions in an Interacting Many-Body System}},\ }\href {https://doi.org/10.1103/PhysRevLett.119.080501} {\bibfield  {journal} {\bibinfo  {journal} {Phys. Rev. Lett.}\ }\textbf {\bibinfo {volume} {119}},\ \bibinfo {pages} {080501} (\bibinfo {year} {2017})}\BibitemShut {NoStop}%
\bibitem [{\citenamefont {Zhang}\ \emph {et~al.}(2017)\citenamefont {Zhang}, \citenamefont {Pagano}, \citenamefont {Hess}, \citenamefont {Kyprianidis}, \citenamefont {Becker}, \citenamefont {Kaplan}, \citenamefont {Gorshkov}, \citenamefont {Gong},\ and\ \citenamefont {Monroe}}]{Zhang17}%
  \BibitemOpen
  \bibfield  {author} {\bibinfo {author} {\bibfnamefont {J.}~\bibnamefont {Zhang}}, \bibinfo {author} {\bibfnamefont {G.}~\bibnamefont {Pagano}}, \bibinfo {author} {\bibfnamefont {P.~W.}\ \bibnamefont {Hess}}, \bibinfo {author} {\bibfnamefont {A.}~\bibnamefont {Kyprianidis}}, \bibinfo {author} {\bibfnamefont {P.}~\bibnamefont {Becker}}, \bibinfo {author} {\bibfnamefont {H.}~\bibnamefont {Kaplan}}, \bibinfo {author} {\bibfnamefont {A.~V.}\ \bibnamefont {Gorshkov}}, \bibinfo {author} {\bibfnamefont {Z.~X.}\ \bibnamefont {Gong}},\ and\ \bibinfo {author} {\bibfnamefont {C.}~\bibnamefont {Monroe}},\ }\bibfield  {title} {\bibinfo {title} {Observation of a many-body dynamical phase transition with a 53-qubit quantum simulator},\ }\href {https://doi.org/10.1038/nature24654} {\bibfield  {journal} {\bibinfo  {journal} {Nature}\ }\textbf {\bibinfo {volume} {551}},\ \bibinfo {pages} {601} (\bibinfo {year} {2017})}\BibitemShut {NoStop}%
\bibitem [{\citenamefont {Fl{\ifmmode\ddot{a}\else\"{a}\fi}schner}\ \emph {et~al.}(2018)\citenamefont {Fl{\ifmmode\ddot{a}\else\"{a}\fi}schner}, \citenamefont {Vogel}, \citenamefont {Tarnowski}, \citenamefont {Rem}, \citenamefont {L{\ifmmode\ddot{u}\else\"{u}\fi}hmann}, \citenamefont {Heyl}, \citenamefont {Budich}, \citenamefont {Mathey}, \citenamefont {Sengstock},\ and\ \citenamefont {Weitenberg}}]{Flaschner2018Mar}%
  \BibitemOpen
  \bibfield  {author} {\bibinfo {author} {\bibfnamefont {N.}~\bibnamefont {Fl{\ifmmode\ddot{a}\else\"{a}\fi}schner}}, \bibinfo {author} {\bibfnamefont {D.}~\bibnamefont {Vogel}}, \bibinfo {author} {\bibfnamefont {M.}~\bibnamefont {Tarnowski}}, \bibinfo {author} {\bibfnamefont {B.~S.}\ \bibnamefont {Rem}}, \bibinfo {author} {\bibfnamefont {D.-S.}\ \bibnamefont {L{\ifmmode\ddot{u}\else\"{u}\fi}hmann}}, \bibinfo {author} {\bibfnamefont {M.}~\bibnamefont {Heyl}}, \bibinfo {author} {\bibfnamefont {J.~C.}\ \bibnamefont {Budich}}, \bibinfo {author} {\bibfnamefont {L.}~\bibnamefont {Mathey}}, \bibinfo {author} {\bibfnamefont {K.}~\bibnamefont {Sengstock}},\ and\ \bibinfo {author} {\bibfnamefont {C.}~\bibnamefont {Weitenberg}},\ }\bibfield  {title} {\bibinfo {title} {{Observation of dynamical vortices after quenches in a system with topology}},\ }\href {https://doi.org/10.1038/s41567-017-0013-8} {\bibfield  {journal} {\bibinfo  {journal} {Nat. Phys.}\ }\textbf {\bibinfo {volume} {14}},\ \bibinfo {pages} {265}
  (\bibinfo {year} {2018})}\BibitemShut {NoStop}%
\bibitem [{\citenamefont {Janas}\ \emph {et~al.}(2016)\citenamefont {Janas}, \citenamefont {Kamenev},\ and\ \citenamefont {Meerson}}]{Janas2016Sep}%
  \BibitemOpen
  \bibfield  {author} {\bibinfo {author} {\bibfnamefont {M.}~\bibnamefont {Janas}}, \bibinfo {author} {\bibfnamefont {A.}~\bibnamefont {Kamenev}},\ and\ \bibinfo {author} {\bibfnamefont {B.}~\bibnamefont {Meerson}},\ }\bibfield  {title} {\bibinfo {title} {{Dynamical phase transition in large-deviation statistics of the Kardar-Parisi-Zhang equation}},\ }\href {https://doi.org/10.1103/PhysRevE.94.032133} {\bibfield  {journal} {\bibinfo  {journal} {Phys. Rev. E}\ }\textbf {\bibinfo {volume} {94}},\ \bibinfo {pages} {032133} (\bibinfo {year} {2016})}\BibitemShut {NoStop}%
\bibitem [{\citenamefont {Smith}\ \emph {et~al.}(2018)\citenamefont {Smith}, \citenamefont {Kamenev},\ and\ \citenamefont {Meerson}}]{Smith2018Apr}%
  \BibitemOpen
  \bibfield  {author} {\bibinfo {author} {\bibfnamefont {N.~R.}\ \bibnamefont {Smith}}, \bibinfo {author} {\bibfnamefont {A.}~\bibnamefont {Kamenev}},\ and\ \bibinfo {author} {\bibfnamefont {B.}~\bibnamefont {Meerson}},\ }\bibfield  {title} {\bibinfo {title} {{Landau theory of the short-time dynamical phase transitions of the Kardar-Parisi-Zhang interface}},\ }\href {https://doi.org/10.1103/PhysRevE.97.042130} {\bibfield  {journal} {\bibinfo  {journal} {Phys. Rev. E}\ }\textbf {\bibinfo {volume} {97}},\ \bibinfo {pages} {042130} (\bibinfo {year} {2018})}\BibitemShut {NoStop}%
\bibitem [{\citenamefont {Baek}\ \emph {et~al.}(2019)\citenamefont {Baek}, \citenamefont {Kafri},\ and\ \citenamefont {Lecomte}}]{Baek2019Oct}%
  \BibitemOpen
  \bibfield  {author} {\bibinfo {author} {\bibfnamefont {Y.}~\bibnamefont {Baek}}, \bibinfo {author} {\bibfnamefont {Y.}~\bibnamefont {Kafri}},\ and\ \bibinfo {author} {\bibfnamefont {V.}~\bibnamefont {Lecomte}},\ }\bibfield  {title} {\bibinfo {title} {{Finite-size and finite-time effects in large deviation functions near dynamical symmetry breaking transitions}},\ }\href {https://doi.org/10.1088/1742-5468/ab43d5} {\bibfield  {journal} {\bibinfo  {journal} {J. Stat. Mech.: Theory Exp.}\ }\textbf {\bibinfo {volume} {2019}}\bibinfo  {number} { (10)},\ \bibinfo {pages} {103202}}\BibitemShut {NoStop}%
\bibitem [{\citenamefont {Lang}\ \emph {et~al.}(2018{\natexlab{a}})\citenamefont {Lang}, \citenamefont {Frank},\ and\ \citenamefont {Halimeh}}]{Lang2018Sep}%
  \BibitemOpen
\bibfield  {number} {  }\bibfield  {author} {\bibinfo {author} {\bibfnamefont {J.}~\bibnamefont {Lang}}, \bibinfo {author} {\bibfnamefont {B.}~\bibnamefont {Frank}},\ and\ \bibinfo {author} {\bibfnamefont {J.~C.}\ \bibnamefont {Halimeh}},\ }\bibfield  {title} {\bibinfo {title} {{Dynamical Quantum Phase Transitions: A Geometric Picture}},\ }\href {https://doi.org/10.1103/PhysRevLett.121.130603} {\bibfield  {journal} {\bibinfo  {journal} {Phys. Rev. Lett.}\ }\textbf {\bibinfo {volume} {121}},\ \bibinfo {pages} {130603} (\bibinfo {year} {2018}{\natexlab{a}})}\BibitemShut {NoStop}%
\bibitem [{\citenamefont {Nussenzveig}(1977)}]{Nussenzveig1977}%
  \BibitemOpen
  \bibfield  {author} {\bibinfo {author} {\bibfnamefont {H.~M.}\ \bibnamefont {Nussenzveig}},\ }\bibfield  {title} {\bibinfo {title} {The theory of the rainbow},\ }\href {https://doi.org/10.1038/scientificamerican0477-116} {\bibfield  {journal} {\bibinfo  {journal} {Physics Today}\ }\textbf {\bibinfo {volume} {236}},\ \bibinfo {pages} {116} (\bibinfo {year} {1977})}\BibitemShut {NoStop}%
\bibitem [{\citenamefont {Nye}(1999)}]{Nye_natural_focusing}%
  \BibitemOpen
  \bibfield  {author} {\bibinfo {author} {\bibfnamefont {J.~F.}\ \bibnamefont {Nye}},\ }\href {https://aapt.scitation.org/doi/abs/10.1119/1.19543} {\emph {\bibinfo {title} {Natural Focusing and Fine Structure of Light: Caustics and Wave Dislocations}}}\ (\bibinfo  {publisher} {Institute of Physics Publishing: Bristol and Philadelphia},\ \bibinfo {year} {1999})\BibitemShut {NoStop}%
\bibitem [{\citenamefont {Berry}(1981)}]{berry81}%
  \BibitemOpen
  \bibfield  {author} {\bibinfo {author} {\bibfnamefont {M.~V.}\ \bibnamefont {Berry}},\ }\bibfield  {title} {\bibinfo {title} {Singularities in waves and rays},\ }in\ \href@noop {} {\emph {\bibinfo {booktitle} {Physics of Defects (1980)}}},\ Vol.\ \bibinfo {volume} {XXXV},\ \bibinfo {editor} {edited by\ \bibinfo {editor} {\bibfnamefont {R.}~\bibnamefont {Balian}}\ and\ \bibinfo {editor} {\bibnamefont {et~al.}}}\ (\bibinfo  {publisher} {North-Holland Publishing, Amsterdam},\ \bibinfo {year} {1981})\BibitemShut {NoStop}%
\bibitem [{\citenamefont {Thom}(1975)}]{Thom1975}%
  \BibitemOpen
  \bibfield  {author} {\bibinfo {author} {\bibfnamefont {R.}~\bibnamefont {Thom}},\ }\href@noop {} {\emph {\bibinfo {title} {Structural Stability and Morphogenesis}}}\ (\bibinfo  {publisher} {Benjamin},\ \bibinfo {address} {Reading, MA},\ \bibinfo {year} {1975})\BibitemShut {NoStop}%
\bibitem [{\citenamefont {Arnol’d}(1975)}]{Arnold1975}%
  \BibitemOpen
  \bibfield  {author} {\bibinfo {author} {\bibfnamefont {V.~I.}\ \bibnamefont {Arnol’d}},\ }\bibfield  {title} {\bibinfo {title} {Critical points of smooth functions and their normal forms},\ }\href {https://doi.org/10.1070/rm1975v030n05abeh001521} {\bibfield  {journal} {\bibinfo  {journal} {Russ. Math. Surv.}\ }\textbf {\bibinfo {volume} {30}},\ \bibinfo {pages} {1} (\bibinfo {year} {1975})}\BibitemShut {NoStop}%
\bibitem [{\citenamefont {Backhaus}\ \emph {et~al.}(1998)\citenamefont {Backhaus}, \citenamefont {Pereverzev}, \citenamefont {Simmonds}, \citenamefont {Loshak}, \citenamefont {Davis},\ and\ \citenamefont {Packard}}]{Backhaus98}%
  \BibitemOpen
  \bibfield  {author} {\bibinfo {author} {\bibfnamefont {S.}~\bibnamefont {Backhaus}}, \bibinfo {author} {\bibfnamefont {S.}~\bibnamefont {Pereverzev}}, \bibinfo {author} {\bibfnamefont {R.~W.}\ \bibnamefont {Simmonds}}, \bibinfo {author} {\bibfnamefont {A.}~\bibnamefont {Loshak}}, \bibinfo {author} {\bibfnamefont {J.~C.}\ \bibnamefont {Davis}},\ and\ \bibinfo {author} {\bibfnamefont {R.~E.}\ \bibnamefont {Packard}},\ }\bibfield  {title} {\bibinfo {title} {Discovery of a metastable $\pi$-state in a superfluid $^3${H}e weak link},\ }\href {https://doi.org/10.1038/33629} {\bibfield  {journal} {\bibinfo  {journal} {Nature}\ }\textbf {\bibinfo {volume} {392}},\ \bibinfo {pages} {687} (\bibinfo {year} {1998})}\BibitemShut {NoStop}%
\bibitem [{\citenamefont {Das}\ \emph {et~al.}(2006)\citenamefont {Das}, \citenamefont {Sengupta}, \citenamefont {Sen},\ and\ \citenamefont {Chakrabarti}}]{Das2006}%
  \BibitemOpen
  \bibfield  {author} {\bibinfo {author} {\bibfnamefont {A.}~\bibnamefont {Das}}, \bibinfo {author} {\bibfnamefont {K.}~\bibnamefont {Sengupta}}, \bibinfo {author} {\bibfnamefont {D.}~\bibnamefont {Sen}},\ and\ \bibinfo {author} {\bibfnamefont {B.~K.}\ \bibnamefont {Chakrabarti}},\ }\bibfield  {title} {\bibinfo {title} {Infinite-range ising ferromagnet in a time-dependent transverse magnetic field: Quench and ac dynamics near the quantum critical point},\ }\href {https://doi.org/10.1103/PhysRevB.74.144423} {\bibfield  {journal} {\bibinfo  {journal} {Phys. Rev. B}\ }\textbf {\bibinfo {volume} {74}},\ \bibinfo {pages} {144423} (\bibinfo {year} {2006})}\BibitemShut {NoStop}%
\bibitem [{\citenamefont {Lagoudakis}\ \emph {et~al.}(2010)\citenamefont {Lagoudakis}, \citenamefont {Pietka}, \citenamefont {Wouters}, \citenamefont {Andr\'e},\ and\ \citenamefont {Deveaud-Pl\'edran}}]{Lagoudakis10}%
  \BibitemOpen
  \bibfield  {author} {\bibinfo {author} {\bibfnamefont {K.~G.}\ \bibnamefont {Lagoudakis}}, \bibinfo {author} {\bibfnamefont {B.}~\bibnamefont {Pietka}}, \bibinfo {author} {\bibfnamefont {M.}~\bibnamefont {Wouters}}, \bibinfo {author} {\bibfnamefont {R.}~\bibnamefont {Andr\'e}},\ and\ \bibinfo {author} {\bibfnamefont {B.}~\bibnamefont {Deveaud-Pl\'edran}},\ }\bibfield  {title} {\bibinfo {title} {Coherent oscillations in an exciton-polariton {J}osephson junction},\ }\href {https://doi.org/10.1103/PhysRevLett.105.120403} {\bibfield  {journal} {\bibinfo  {journal} {Phys. Rev. Lett.}\ }\textbf {\bibinfo {volume} {105}},\ \bibinfo {pages} {120403} (\bibinfo {year} {2010})}\BibitemShut {NoStop}%
\bibitem [{\citenamefont {Abbarchi}\ \emph {et~al.}(2013)\citenamefont {Abbarchi}, \citenamefont {Amo}, \citenamefont {Sala}, \citenamefont {Solnyshkov}, \citenamefont {Flayac}, \citenamefont {Ferrier}, \citenamefont {Sagnes}, \citenamefont {Galopin}, \citenamefont {Lemaître}, \citenamefont {Malpuech},\ and\ \citenamefont {Bloch}}]{Abbarchi13}%
  \BibitemOpen
  \bibfield  {author} {\bibinfo {author} {\bibfnamefont {M.}~\bibnamefont {Abbarchi}}, \bibinfo {author} {\bibfnamefont {A.}~\bibnamefont {Amo}}, \bibinfo {author} {\bibfnamefont {V.~G.}\ \bibnamefont {Sala}}, \bibinfo {author} {\bibfnamefont {D.~D.}\ \bibnamefont {Solnyshkov}}, \bibinfo {author} {\bibfnamefont {H.}~\bibnamefont {Flayac}}, \bibinfo {author} {\bibfnamefont {L.}~\bibnamefont {Ferrier}}, \bibinfo {author} {\bibfnamefont {I.}~\bibnamefont {Sagnes}}, \bibinfo {author} {\bibfnamefont {E.}~\bibnamefont {Galopin}}, \bibinfo {author} {\bibfnamefont {A.}~\bibnamefont {Lemaître}}, \bibinfo {author} {\bibfnamefont {G.}~\bibnamefont {Malpuech}},\ and\ \bibinfo {author} {\bibfnamefont {J.}~\bibnamefont {Bloch}},\ }\bibfield  {title} {\bibinfo {title} {Macroscopic quantum self-trapping and {J}osephson oscillations of exciton polaritons},\ }\href {https://doi.org/10.1038/nphys2609} {\bibfield  {journal} {\bibinfo  {journal} {Nature Physics}\ }\textbf {\bibinfo {volume} {9}},\ \bibinfo {pages} {275}
  (\bibinfo {year} {2013})}\BibitemShut {NoStop}%
\bibitem [{\citenamefont {Guti\'errez-Cuevas}\ \emph {et~al.}(2023)\citenamefont {Guti\'errez-Cuevas}, \citenamefont {O'Dell}, \citenamefont {Dennis},\ and\ \citenamefont {Alonso}}]{Gutierrez23}%
  \BibitemOpen
  \bibfield  {author} {\bibinfo {author} {\bibfnamefont {R.}~\bibnamefont {Guti\'errez-Cuevas}}, \bibinfo {author} {\bibfnamefont {D.~H.~J.}\ \bibnamefont {O'Dell}}, \bibinfo {author} {\bibfnamefont {M.~R.}\ \bibnamefont {Dennis}},\ and\ \bibinfo {author} {\bibfnamefont {M.~A.}\ \bibnamefont {Alonso}},\ }\bibfield  {title} {\bibinfo {title} {Exactly solvable model behind {B}ose-{H}ubbard dimers, {I}nce-{G}auss beams, and aberrated optical cavities},\ }\href {https://doi.org/10.1103/PhysRevA.107.L031502} {\bibfield  {journal} {\bibinfo  {journal} {Phys. Rev. A}\ }\textbf {\bibinfo {volume} {107}},\ \bibinfo {pages} {L031502} (\bibinfo {year} {2023})}\BibitemShut {NoStop}%
\bibitem [{\citenamefont {Britton}\ \emph {et~al.}(2012)\citenamefont {Britton}, \citenamefont {Sawyer}, \citenamefont {Keith}, \citenamefont {Wang}, \citenamefont {Freericks}, \citenamefont {Uys}, \citenamefont {Biercuk},\ and\ \citenamefont {Bollinger}}]{Britton2012}%
  \BibitemOpen
  \bibfield  {author} {\bibinfo {author} {\bibfnamefont {J.~W.}\ \bibnamefont {Britton}}, \bibinfo {author} {\bibfnamefont {B.~C.}\ \bibnamefont {Sawyer}}, \bibinfo {author} {\bibfnamefont {A.~C.}\ \bibnamefont {Keith}}, \bibinfo {author} {\bibfnamefont {C.-C.~J.}\ \bibnamefont {Wang}}, \bibinfo {author} {\bibfnamefont {J.~K.}\ \bibnamefont {Freericks}}, \bibinfo {author} {\bibfnamefont {H.}~\bibnamefont {Uys}}, \bibinfo {author} {\bibfnamefont {M.~J.}\ \bibnamefont {Biercuk}},\ and\ \bibinfo {author} {\bibfnamefont {J.~J.}\ \bibnamefont {Bollinger}},\ }\bibfield  {title} {\bibinfo {title} {Engineered two-dimensional {I}sing interactions in a trapped-ion quantum simulator with hundreds of spins},\ }\href {https://doi.org/10.1038/nature10981} {\bibfield  {journal} {\bibinfo  {journal} {Nature (London)}\ }\textbf {\bibinfo {volume} {484}},\ \bibinfo {pages} {489} (\bibinfo {year} {2012})}\BibitemShut {NoStop}%
\bibitem [{\citenamefont {Jurcevic}\ \emph {et~al.}(2014)\citenamefont {Jurcevic}, \citenamefont {Lanyon}, \citenamefont {Hauke}, \citenamefont {Hempel}, \citenamefont {Zoller}, \citenamefont {Blatt},\ and\ \citenamefont {Roos}}]{Jurcevic14}%
  \BibitemOpen
  \bibfield  {author} {\bibinfo {author} {\bibfnamefont {P.}~\bibnamefont {Jurcevic}}, \bibinfo {author} {\bibfnamefont {B.~P.}\ \bibnamefont {Lanyon}}, \bibinfo {author} {\bibfnamefont {P.}~\bibnamefont {Hauke}}, \bibinfo {author} {\bibfnamefont {C.}~\bibnamefont {Hempel}}, \bibinfo {author} {\bibfnamefont {P.}~\bibnamefont {Zoller}}, \bibinfo {author} {\bibfnamefont {R.}~\bibnamefont {Blatt}},\ and\ \bibinfo {author} {\bibfnamefont {C.~F.}\ \bibnamefont {Roos}},\ }\bibfield  {title} {\bibinfo {title} {Quasiparticle engineering and entanglement propagation in a quantum many-body system},\ }\href {https://doi.org/10.1038/nature13461} {\bibfield  {journal} {\bibinfo  {journal} {Nature}\ }\textbf {\bibinfo {volume} {511}},\ \bibinfo {pages} {202} (\bibinfo {year} {2014})}\BibitemShut {NoStop}%
\bibitem [{\citenamefont {Bohnet}\ \emph {et~al.}(2016)\citenamefont {Bohnet}, \citenamefont {Sawyer}, \citenamefont {Britton}, \citenamefont {Wall}, \citenamefont {Rey}, \citenamefont {Foss-Feig},\ and\ \citenamefont {Bollinger}}]{Bohnet2016}%
  \BibitemOpen
  \bibfield  {author} {\bibinfo {author} {\bibfnamefont {J.~G.}\ \bibnamefont {Bohnet}}, \bibinfo {author} {\bibfnamefont {B.~C.}\ \bibnamefont {Sawyer}}, \bibinfo {author} {\bibfnamefont {J.~W.}\ \bibnamefont {Britton}}, \bibinfo {author} {\bibfnamefont {M.~L.}\ \bibnamefont {Wall}}, \bibinfo {author} {\bibfnamefont {A.~M.}\ \bibnamefont {Rey}}, \bibinfo {author} {\bibfnamefont {M.}~\bibnamefont {Foss-Feig}},\ and\ \bibinfo {author} {\bibfnamefont {J.}~\bibnamefont {Bollinger}},\ }\bibfield  {title} {\bibinfo {title} {Quantum spin dynamics and entanglement generation with hundreds of trapped ions},\ }\href {https://doi.org/10.1126/science.aad9958} {\bibfield  {journal} {\bibinfo  {journal} {Science}\ }\textbf {\bibinfo {volume} {352}},\ \bibinfo {pages} {1297} (\bibinfo {year} {2016})}\BibitemShut {NoStop}%
\bibitem [{\citenamefont {Baumann}\ \emph {et~al.}(2010)\citenamefont {Baumann}, \citenamefont {Guerlin}, \citenamefont {Brennecke},\ and\ \citenamefont {Esslinger}}]{baumann2010}%
  \BibitemOpen
  \bibfield  {author} {\bibinfo {author} {\bibfnamefont {K.}~\bibnamefont {Baumann}}, \bibinfo {author} {\bibfnamefont {C.}~\bibnamefont {Guerlin}}, \bibinfo {author} {\bibfnamefont {F.}~\bibnamefont {Brennecke}},\ and\ \bibinfo {author} {\bibfnamefont {T.}~\bibnamefont {Esslinger}},\ }\bibfield  {title} {\bibinfo {title} {Dicke quantum phase transition with a superfluid gas in an optical cavity},\ }\href {https://doi.org/10.1038/nature09009} {\bibfield  {journal} {\bibinfo  {journal} {Nature}\ }\textbf {\bibinfo {volume} {464}},\ \bibinfo {pages} {1301} (\bibinfo {year} {2010})}\BibitemShut {NoStop}%
\bibitem [{\citenamefont {Klinder}\ \emph {et~al.}(2015)\citenamefont {Klinder}, \citenamefont {Keßler}, \citenamefont {Wolke},\ and\ \citenamefont {Hemmerich}}]{Klinder15}%
  \BibitemOpen
  \bibfield  {author} {\bibinfo {author} {\bibfnamefont {J.}~\bibnamefont {Klinder}}, \bibinfo {author} {\bibfnamefont {H.}~\bibnamefont {Keßler}}, \bibinfo {author} {\bibfnamefont {M.}~\bibnamefont {Wolke}},\ and\ \bibinfo {author} {\bibfnamefont {A.}~\bibnamefont {Hemmerich}},\ }\bibfield  {title} {\bibinfo {title} {Dynamical phase transition in the open {D}icke model},\ }\href {https://doi.org/10.1073/pnas.141713211} {\bibfield  {journal} {\bibinfo  {journal} {Proc. Natl. Acad. Sci. U.S.A.}\ }\textbf {\bibinfo {volume} {112}},\ \bibinfo {pages} {3290} (\bibinfo {year} {2015})}\BibitemShut {NoStop}%
\bibitem [{\citenamefont {Defenu}\ \emph {et~al.}(2018)\citenamefont {Defenu}, \citenamefont {Enss}, \citenamefont {Kastner},\ and\ \citenamefont {Morigi}}]{Defenu18}%
  \BibitemOpen
  \bibfield  {author} {\bibinfo {author} {\bibfnamefont {N.}~\bibnamefont {Defenu}}, \bibinfo {author} {\bibfnamefont {T.}~\bibnamefont {Enss}}, \bibinfo {author} {\bibfnamefont {M.}~\bibnamefont {Kastner}},\ and\ \bibinfo {author} {\bibfnamefont {G.}~\bibnamefont {Morigi}},\ }\bibfield  {title} {\bibinfo {title} {Dynamical critical scaling of long-range interacting quantum magnets},\ }\href {https://doi.org/10.1103/PhysRevLett.121.240403} {\bibfield  {journal} {\bibinfo  {journal} {Phys. Rev. Lett.}\ }\textbf {\bibinfo {volume} {121}},\ \bibinfo {pages} {240403} (\bibinfo {year} {2018})}\BibitemShut {NoStop}%
\bibitem [{\citenamefont {Muniz}\ \emph {et~al.}(2020)\citenamefont {Muniz}, \citenamefont {Barberena}, \citenamefont {Lewis-Swan}, \citenamefont {Young}, \citenamefont {Cline}, \citenamefont {Rey},\ and\ \citenamefont {Thompson}}]{Muniz2020Apr}%
  \BibitemOpen
  \bibfield  {author} {\bibinfo {author} {\bibfnamefont {J.~A.}\ \bibnamefont {Muniz}}, \bibinfo {author} {\bibfnamefont {D.}~\bibnamefont {Barberena}}, \bibinfo {author} {\bibfnamefont {R.~J.}\ \bibnamefont {Lewis-Swan}}, \bibinfo {author} {\bibfnamefont {D.~J.}\ \bibnamefont {Young}}, \bibinfo {author} {\bibfnamefont {J.~R.~K.}\ \bibnamefont {Cline}}, \bibinfo {author} {\bibfnamefont {A.~M.}\ \bibnamefont {Rey}},\ and\ \bibinfo {author} {\bibfnamefont {J.~K.}\ \bibnamefont {Thompson}},\ }\bibfield  {title} {\bibinfo {title} {{Exploring dynamical phase transitions with cold atoms in an optical cavity}},\ }\href {https://doi.org/10.1038/s41586-020-2224-x} {\bibfield  {journal} {\bibinfo  {journal} {Nature}\ }\textbf {\bibinfo {volume} {580}},\ \bibinfo {pages} {602} (\bibinfo {year} {2020})}\BibitemShut {NoStop}%
\bibitem [{\citenamefont {Lang}\ \emph {et~al.}(2018{\natexlab{b}})\citenamefont {Lang}, \citenamefont {Frank},\ and\ \citenamefont {Halimeh}}]{Lang2018May}%
  \BibitemOpen
  \bibfield  {author} {\bibinfo {author} {\bibfnamefont {J.}~\bibnamefont {Lang}}, \bibinfo {author} {\bibfnamefont {B.}~\bibnamefont {Frank}},\ and\ \bibinfo {author} {\bibfnamefont {J.~C.}\ \bibnamefont {Halimeh}},\ }\bibfield  {title} {\bibinfo {title} {{Concurrence of dynamical phase transitions at finite temperature in the fully connected transverse-field Ising model}},\ }\href {https://doi.org/10.1103/PhysRevB.97.174401} {\bibfield  {journal} {\bibinfo  {journal} {Phys. Rev. B}\ }\textbf {\bibinfo {volume} {97}},\ \bibinfo {pages} {174401} (\bibinfo {year} {2018}{\natexlab{b}})}\BibitemShut {NoStop}%
\bibitem [{\citenamefont {Sehrawat}\ \emph {et~al.}(2021)\citenamefont {Sehrawat}, \citenamefont {Srivastava},\ and\ \citenamefont {Sen}}]{Sehrawat2021Aug}%
  \BibitemOpen
  \bibfield  {author} {\bibinfo {author} {\bibfnamefont {A.}~\bibnamefont {Sehrawat}}, \bibinfo {author} {\bibfnamefont {C.}~\bibnamefont {Srivastava}},\ and\ \bibinfo {author} {\bibfnamefont {U.}~\bibnamefont {Sen}},\ }\bibfield  {title} {\bibinfo {title} {{Dynamical phase transitions in the fully connected quantum Ising model: Time period and critical time}},\ }\href {https://doi.org/10.1103/PhysRevB.104.085105} {\bibfield  {journal} {\bibinfo  {journal} {Phys. Rev. B}\ }\textbf {\bibinfo {volume} {104}},\ \bibinfo {pages} {085105} (\bibinfo {year} {2021})}\BibitemShut {NoStop}%
\bibitem [{\citenamefont {Corps}\ and\ \citenamefont {Rela{\ifmmode\tilde{n}\else\~{n}\fi}o}(2022)}]{Corps2022Jul}%
  \BibitemOpen
  \bibfield  {author} {\bibinfo {author} {\bibfnamefont {{\ifmmode\acute{A}\else\'{A}\fi}.~L.}\ \bibnamefont {Corps}}\ and\ \bibinfo {author} {\bibfnamefont {A.}~\bibnamefont {Rela{\ifmmode\tilde{n}\else\~{n}\fi}o}},\ }\bibfield  {title} {\bibinfo {title} {{Dynamical and excited-state quantum phase transitions in collective systems}},\ }\href {https://doi.org/10.1103/PhysRevB.106.024311} {\bibfield  {journal} {\bibinfo  {journal} {Phys. Rev. B}\ }\textbf {\bibinfo {volume} {106}},\ \bibinfo {pages} {024311} (\bibinfo {year} {2022})}\BibitemShut {NoStop}%
\bibitem [{\citenamefont {Mumford}\ \emph {et~al.}(2019)\citenamefont {Mumford}, \citenamefont {Turner}, \citenamefont {Sprung},\ and\ \citenamefont {O{'}Dell}}]{Mumford2019May}%
  \BibitemOpen
  \bibfield  {author} {\bibinfo {author} {\bibfnamefont {J.}~\bibnamefont {Mumford}}, \bibinfo {author} {\bibfnamefont {E.}~\bibnamefont {Turner}}, \bibinfo {author} {\bibfnamefont {D.~W.~L.}\ \bibnamefont {Sprung}},\ and\ \bibinfo {author} {\bibfnamefont {D.~H.~J.}\ \bibnamefont {O{'}Dell}},\ }\bibfield  {title} {\bibinfo {title} {{Quantum Spin Dynamics in Fock Space Following Quenches: Caustics and Vortices}},\ }\href {https://doi.org/10.1103/PhysRevLett.122.170402} {\bibfield  {journal} {\bibinfo  {journal} {Phys. Rev. Lett.}\ }\textbf {\bibinfo {volume} {122}},\ \bibinfo {pages} {170402} (\bibinfo {year} {2019})}\BibitemShut {NoStop}%
\bibitem [{\citenamefont {Kirkby}\ \emph {et~al.}(2019)\citenamefont {Kirkby}, \citenamefont {Mumford},\ and\ \citenamefont {O'Dell}}]{Kirkby2019Nov}%
  \BibitemOpen
  \bibfield  {author} {\bibinfo {author} {\bibfnamefont {W.}~\bibnamefont {Kirkby}}, \bibinfo {author} {\bibfnamefont {J.}~\bibnamefont {Mumford}},\ and\ \bibinfo {author} {\bibfnamefont {D.~H.~J.}\ \bibnamefont {O'Dell}},\ }\bibfield  {title} {\bibinfo {title} {Quantum caustics and the hierarchy of light cones in quenched spin chains},\ }\href {https://doi.org/10.1103/PhysRevResearch.1.033135} {\bibfield  {journal} {\bibinfo  {journal} {Phys. Rev. Res.}\ }\textbf {\bibinfo {volume} {1}},\ \bibinfo {pages} {033135} (\bibinfo {year} {2019})}\BibitemShut {NoStop}%
\bibitem [{\citenamefont {Kirkby}\ \emph {et~al.}(2022)\citenamefont {Kirkby}, \citenamefont {Yee}, \citenamefont {Shi},\ and\ \citenamefont {O'Dell}}]{Kirkby2022Feb}%
  \BibitemOpen
  \bibfield  {author} {\bibinfo {author} {\bibfnamefont {W.}~\bibnamefont {Kirkby}}, \bibinfo {author} {\bibfnamefont {Y.}~\bibnamefont {Yee}}, \bibinfo {author} {\bibfnamefont {K.}~\bibnamefont {Shi}},\ and\ \bibinfo {author} {\bibfnamefont {D.~H.~J.}\ \bibnamefont {O'Dell}},\ }\bibfield  {title} {\bibinfo {title} {Caustics in quantum many-body dynamics},\ }\href {https://doi.org/10.1103/PhysRevResearch.4.013105} {\bibfield  {journal} {\bibinfo  {journal} {Phys. Rev. Res.}\ }\textbf {\bibinfo {volume} {4}},\ \bibinfo {pages} {013105} (\bibinfo {year} {2022})}\BibitemShut {NoStop}%
\bibitem [{\citenamefont {H\"ohmann}\ \emph {et~al.}(2010)\citenamefont {H\"ohmann}, \citenamefont {Kuhl}, \citenamefont {St\"ockmann}, \citenamefont {Kaplan},\ and\ \citenamefont {Heller}}]{Hohmann10}%
  \BibitemOpen
  \bibfield  {author} {\bibinfo {author} {\bibfnamefont {R.}~\bibnamefont {H\"ohmann}}, \bibinfo {author} {\bibfnamefont {U.}~\bibnamefont {Kuhl}}, \bibinfo {author} {\bibfnamefont {H.-J.}\ \bibnamefont {St\"ockmann}}, \bibinfo {author} {\bibfnamefont {L.}~\bibnamefont {Kaplan}},\ and\ \bibinfo {author} {\bibfnamefont {E.~J.}\ \bibnamefont {Heller}},\ }\bibfield  {title} {\bibinfo {title} {Freak waves in the linear regime: A microwave study},\ }\href {https://doi.org/10.1103/PhysRevLett.104.093901} {\bibfield  {journal} {\bibinfo  {journal} {Phys. Rev. Lett.}\ }\textbf {\bibinfo {volume} {104}},\ \bibinfo {pages} {093901} (\bibinfo {year} {2010})}\BibitemShut {NoStop}%
\bibitem [{\citenamefont {Metzger}\ \emph {et~al.}(2014)\citenamefont {Metzger}, \citenamefont {Fleischmann},\ and\ \citenamefont {Geisel}}]{metzger2014}%
  \BibitemOpen
  \bibfield  {author} {\bibinfo {author} {\bibfnamefont {J.~J.}\ \bibnamefont {Metzger}}, \bibinfo {author} {\bibfnamefont {R.}~\bibnamefont {Fleischmann}},\ and\ \bibinfo {author} {\bibfnamefont {T.}~\bibnamefont {Geisel}},\ }\bibfield  {title} {\bibinfo {title} {Statistics of extreme waves in random media},\ }\href {https://doi.org/10.1103/PhysRevLett.112.203903} {\bibfield  {journal} {\bibinfo  {journal} {Phys. Rev. Lett.}\ }\textbf {\bibinfo {volume} {112}},\ \bibinfo {pages} {203903} (\bibinfo {year} {2014})}\BibitemShut {NoStop}%
\bibitem [{\citenamefont {Berry}(2018)}]{Berry2018}%
  \BibitemOpen
  \bibfield  {author} {\bibinfo {author} {\bibfnamefont {M.~V.}\ \bibnamefont {Berry}},\ }\bibfield  {title} {\bibinfo {title} {Minimal analytical model for undular tidal bore profile; quantum and {H}awking effect analogies},\ }\href {https://doi.org/10.1088/1367-2630/aac285} {\bibfield  {journal} {\bibinfo  {journal} {New J. Phys.}\ }\textbf {\bibinfo {volume} {20}},\ \bibinfo {pages} {053066} (\bibinfo {year} {2018})}\BibitemShut {NoStop}%
\bibitem [{\citenamefont {Berry}(2007)}]{Berry07}%
  \BibitemOpen
  \bibfield  {author} {\bibinfo {author} {\bibfnamefont {M.~V.}\ \bibnamefont {Berry}},\ }\bibfield  {title} {\bibinfo {title} {Focused tsunami waves},\ }\href {https://doi.org/10.1098rspa.2007.0051} {\bibfield  {journal} {\bibinfo  {journal} {Proc. R. Soc. A}\ }\textbf {\bibinfo {volume} {463}},\ \bibinfo {pages} {3055} (\bibinfo {year} {2007})}\BibitemShut {NoStop}%
\bibitem [{\citenamefont {Degueldre}\ \emph {et~al.}(2016)\citenamefont {Degueldre}, \citenamefont {Metzger}, \citenamefont {Geisel},\ and\ \citenamefont {Fleischmann}}]{Degueldre16}%
  \BibitemOpen
  \bibfield  {author} {\bibinfo {author} {\bibfnamefont {H.}~\bibnamefont {Degueldre}}, \bibinfo {author} {\bibfnamefont {J.~J.}\ \bibnamefont {Metzger}}, \bibinfo {author} {\bibfnamefont {T.}~\bibnamefont {Geisel}},\ and\ \bibinfo {author} {\bibfnamefont {R.}~\bibnamefont {Fleischmann}},\ }\bibfield  {title} {\bibinfo {title} {Random focusing of tsunami waves},\ }\href {https://doi.org/10.1038/nphys3557} {\bibfield  {journal} {\bibinfo  {journal} {Nat. Phys.}\ }\textbf {\bibinfo {volume} {12}},\ \bibinfo {pages} {259} (\bibinfo {year} {2016})}\BibitemShut {NoStop}%
\bibitem [{\citenamefont {O{'}Dell}(2012)}]{ODell2012Oct}%
  \BibitemOpen
  \bibfield  {author} {\bibinfo {author} {\bibfnamefont {D.~H.~J.}\ \bibnamefont {O{'}Dell}},\ }\bibfield  {title} {\bibinfo {title} {{Quantum Catastrophes and Ergodicity in the Dynamics of Bosonic Josephson Junctions}},\ }\href {https://doi.org/10.1103/PhysRevLett.109.150406} {\bibfield  {journal} {\bibinfo  {journal} {Phys. Rev. Lett.}\ }\textbf {\bibinfo {volume} {109}},\ \bibinfo {pages} {150406} (\bibinfo {year} {2012})}\BibitemShut {NoStop}%
\bibitem [{\citenamefont {Mumford}\ \emph {et~al.}(2017)\citenamefont {Mumford}, \citenamefont {Kirkby},\ and\ \citenamefont {O'Dell}}]{Mumford2017}%
  \BibitemOpen
  \bibfield  {author} {\bibinfo {author} {\bibfnamefont {J.}~\bibnamefont {Mumford}}, \bibinfo {author} {\bibfnamefont {W.}~\bibnamefont {Kirkby}},\ and\ \bibinfo {author} {\bibfnamefont {D.~H.~J.}\ \bibnamefont {O'Dell}},\ }\bibfield  {title} {\bibinfo {title} {Catastrophes in non-equilibrium many-particle wave functions: universality and critical scaling},\ }\href {https://doi.org/10.1088/1361-6455/aa56af} {\bibfield  {journal} {\bibinfo  {journal} {J. Phys. B: At. Mol. Opt. Phys.}\ }\textbf {\bibinfo {volume} {50}},\ \bibinfo {pages} {044005} (\bibinfo {year} {2017})}\BibitemShut {NoStop}%
\bibitem [{\citenamefont {Goldberg}\ \emph {et~al.}(2019)\citenamefont {Goldberg}, \citenamefont {Al-Qasimi}, \citenamefont {Mumford},\ and\ \citenamefont {O'Dell}}]{Goldberg2019Dec}%
  \BibitemOpen
  \bibfield  {author} {\bibinfo {author} {\bibfnamefont {A.~Z.}\ \bibnamefont {Goldberg}}, \bibinfo {author} {\bibfnamefont {A.}~\bibnamefont {Al-Qasimi}}, \bibinfo {author} {\bibfnamefont {J.}~\bibnamefont {Mumford}},\ and\ \bibinfo {author} {\bibfnamefont {D.~H.~J.}\ \bibnamefont {O'Dell}},\ }\bibfield  {title} {\bibinfo {title} {{Emergence of singularities from decoherence: Quantum catastrophes}},\ }\href {https://doi.org/10.1103/PhysRevA.100.063628} {\bibfield  {journal} {\bibinfo  {journal} {Phys. Rev. A}\ }\textbf {\bibinfo {volume} {100}},\ \bibinfo {pages} {063628} (\bibinfo {year} {2019})}\BibitemShut {NoStop}%
\bibitem [{\citenamefont {Link}\ and\ \citenamefont {Strunz}(2020)}]{Link2020Sep}%
  \BibitemOpen
  \bibfield  {author} {\bibinfo {author} {\bibfnamefont {V.}~\bibnamefont {Link}}\ and\ \bibinfo {author} {\bibfnamefont {W.~T.}\ \bibnamefont {Strunz}},\ }\bibfield  {title} {\bibinfo {title} {{Dynamical Phase Transitions in Dissipative Quantum Dynamics with Quantum Optical Realization}},\ }\href {https://doi.org/10.1103/PhysRevLett.125.143602} {\bibfield  {journal} {\bibinfo  {journal} {Phys. Rev. Lett.}\ }\textbf {\bibinfo {volume} {125}},\ \bibinfo {pages} {143602} (\bibinfo {year} {2020})}\BibitemShut {NoStop}%
\bibitem [{\citenamefont {Berry}\ and\ \citenamefont {Upstill}(1980)}]{Berry&Upstill1980}%
  \BibitemOpen
  \bibfield  {author} {\bibinfo {author} {\bibfnamefont {M.~V.}\ \bibnamefont {Berry}}\ and\ \bibinfo {author} {\bibfnamefont {C.}~\bibnamefont {Upstill}},\ }\bibfield  {title} {\bibinfo {title} {Catastrophe optics: morphologies of caustics and their diffraction patterns},\ }\href {https://doi.org/10.1016/S0079-6638(08)70215-4} {\bibfield  {journal} {\bibinfo  {journal} {Prog. Opt.}\ }\textbf {\bibinfo {volume} {18}},\ \bibinfo {pages} {257} (\bibinfo {year} {1980})}\BibitemShut {NoStop}%
\bibitem [{\citenamefont {Olver}\ \emph {et~al.}(2010)\citenamefont {Olver}, \citenamefont {Lozier}, \citenamefont {Boisvert},\ and\ \citenamefont {Clark}}]{Olver2010}%
  \BibitemOpen
  \bibinfo {editor} {\bibfnamefont {F.~W.~J.}\ \bibnamefont {Olver}}, \bibinfo {editor} {\bibfnamefont {D.~W.}\ \bibnamefont {Lozier}}, \bibinfo {editor} {\bibfnamefont {R.~F.}\ \bibnamefont {Boisvert}},\ and\ \bibinfo {editor} {\bibfnamefont {C.~W.}\ \bibnamefont {Clark}},\ eds.,\ \href {dlmf.nist.gov} {\emph {\bibinfo {title} {NIST Handbook of Mathematical Functions}}}\ (\bibinfo  {publisher} {Cambridge University Press},\ \bibinfo {address} {New York, NY},\ \bibinfo {year} {2010})\BibitemShut {NoStop}%
\bibitem [{\citenamefont {Pearcey}(1946)}]{Pearcey46}%
  \BibitemOpen
  \bibfield  {author} {\bibinfo {author} {\bibfnamefont {T.}~\bibnamefont {Pearcey}},\ }\bibfield  {title} {\bibinfo {title} {The structure of an electromagnetic field in the neighbourhood of a cusp of a caustic},\ }\href {https://doi.org/10.1080/14786444608561335} {\bibfield  {journal} {\bibinfo  {journal} {Phil. Mag.}\ }\textbf {\bibinfo {volume} {37}},\ \bibinfo {pages} {311} (\bibinfo {year} {1946})}\BibitemShut {NoStop}%
\bibitem [{\citenamefont {Airy}(1838)}]{Airy1838}%
  \BibitemOpen
  \bibfield  {author} {\bibinfo {author} {\bibfnamefont {G.~B.}\ \bibnamefont {Airy}},\ }\bibfield  {title} {\bibinfo {title} {On the intensity of light in the neighbourhood of a caustic},\ }\href@noop {} {\bibfield  {journal} {\bibinfo  {journal} {Trans. Camb. Phil. Soc.}\ }\textbf {\bibinfo {volume} {6}},\ \bibinfo {pages} {379} (\bibinfo {year} {1838})}\BibitemShut {NoStop}%
\bibitem [{\citenamefont {Nussenzveig}(1969)}]{Nussenzveig1969Jan}%
  \BibitemOpen
  \bibfield  {author} {\bibinfo {author} {\bibfnamefont {H.~M.}\ \bibnamefont {Nussenzveig}},\ }\bibfield  {title} {\bibinfo {title} {{High{-}Frequency Scattering by a Transparent Sphere. II. Theory of the Rainbow and the Glory}},\ }\href {https://doi.org/10.1063/1.1664747} {\bibfield  {journal} {\bibinfo  {journal} {J. Math. Phys.}\ }\textbf {\bibinfo {volume} {10}},\ \bibinfo {pages} {125} (\bibinfo {year} {1969})}\BibitemShut {NoStop}%
\bibitem [{\citenamefont {van~de Hulst}(1981)}]{van_de_Hulst_book}%
  \BibitemOpen
  \bibfield  {author} {\bibinfo {author} {\bibfnamefont {H.~C.}\ \bibnamefont {van~de Hulst}},\ }\href@noop {} {\emph {\bibinfo {title} {Light scattering by small particles}}}\ (\bibinfo  {publisher} {Dover},\ \bibinfo {address} {New York, USA},\ \bibinfo {year} {1981})\BibitemShut {NoStop}%
\bibitem [{\citenamefont {Braun}(1993)}]{Braun1993Jan}%
  \BibitemOpen
  \bibfield  {author} {\bibinfo {author} {\bibfnamefont {P.~A.}\ \bibnamefont {Braun}},\ }\bibfield  {title} {\bibinfo {title} {{Discrete semiclassical methods in the theory of Rydberg atoms in external fields}},\ }\href {https://doi.org/10.1103/RevModPhys.65.115} {\bibfield  {journal} {\bibinfo  {journal} {Rev. Mod. Phys.}\ }\textbf {\bibinfo {volume} {65}},\ \bibinfo {pages} {115} (\bibinfo {year} {1993})}\BibitemShut {NoStop}%
\bibitem [{\citenamefont {Shchesnovich}\ and\ \citenamefont {Trippenbach}(2008)}]{Shchesnovich2008Aug}%
  \BibitemOpen
  \bibfield  {author} {\bibinfo {author} {\bibfnamefont {V.~S.}\ \bibnamefont {Shchesnovich}}\ and\ \bibinfo {author} {\bibfnamefont {M.}~\bibnamefont {Trippenbach}},\ }\bibfield  {title} {\bibinfo {title} {Fock-space {WKB} method for the boson {J}osephson model describing a {B}ose-{E}instein condensate trapped in a double-well potential},\ }\href {https://doi.org/10.1103/PhysRevA.78.023611} {\bibfield  {journal} {\bibinfo  {journal} {Phys. Rev. A}\ }\textbf {\bibinfo {volume} {78}},\ \bibinfo {pages} {023611} (\bibinfo {year} {2008})}\BibitemShut {NoStop}%
\bibitem [{\citenamefont {Nissen}\ and\ \citenamefont {Keeling}(2010)}]{Nissen2010Jun}%
  \BibitemOpen
  \bibfield  {author} {\bibinfo {author} {\bibfnamefont {F.}~\bibnamefont {Nissen}}\ and\ \bibinfo {author} {\bibfnamefont {J.}~\bibnamefont {Keeling}},\ }\bibfield  {title} {\bibinfo {title} {{Wentzel-Kramers-Brillouin approach and quantum corrections to classical dynamics in the Josephson problem}},\ }\href {https://doi.org/10.1103/PhysRevA.81.063628} {\bibfield  {journal} {\bibinfo  {journal} {Phys. Rev. A}\ }\textbf {\bibinfo {volume} {81}},\ \bibinfo {pages} {063628} (\bibinfo {year} {2010})}\BibitemShut {NoStop}%
\bibitem [{\citenamefont {Simon}\ and\ \citenamefont {Strunz}(2012)}]{Simon2012Nov}%
  \BibitemOpen
  \bibfield  {author} {\bibinfo {author} {\bibfnamefont {L.}~\bibnamefont {Simon}}\ and\ \bibinfo {author} {\bibfnamefont {W.~T.}\ \bibnamefont {Strunz}},\ }\bibfield  {title} {\bibinfo {title} {{Analytical results for Josephson dynamics of ultracold bosons}},\ }\href {https://doi.org/10.1103/PhysRevA.86.053625} {\bibfield  {journal} {\bibinfo  {journal} {Phys. Rev. A}\ }\textbf {\bibinfo {volume} {86}},\ \bibinfo {pages} {053625} (\bibinfo {year} {2012})}\BibitemShut {NoStop}%
\bibitem [{\citenamefont {Simon}\ and\ \citenamefont {Strunz}(2014)}]{Simon2014May}%
  \BibitemOpen
  \bibfield  {author} {\bibinfo {author} {\bibfnamefont {L.}~\bibnamefont {Simon}}\ and\ \bibinfo {author} {\bibfnamefont {W.~T.}\ \bibnamefont {Strunz}},\ }\bibfield  {title} {\bibinfo {title} {{Time-dependent semiclassics for ultracold bosons}},\ }\href {https://doi.org/10.1103/PhysRevA.89.052112} {\bibfield  {journal} {\bibinfo  {journal} {Phys. Rev. A}\ }\textbf {\bibinfo {volume} {89}},\ \bibinfo {pages} {052112} (\bibinfo {year} {2014})}\BibitemShut {NoStop}%
\bibitem [{\citenamefont {Berry}\ and\ \citenamefont {Mount}(1972)}]{Berry1972Feb}%
  \BibitemOpen
  \bibfield  {author} {\bibinfo {author} {\bibfnamefont {M.~V.}\ \bibnamefont {Berry}}\ and\ \bibinfo {author} {\bibfnamefont {K.~E.}\ \bibnamefont {Mount}},\ }\href {https://doi.org/10.1088/0034-4885/35/1/306} {\bibfield  {journal} {\bibinfo  {journal} {Rep. Prog. Phys.}\ }\textbf {\bibinfo {volume} {35}},\ \bibinfo {pages} {315} (\bibinfo {year} {1972})}\BibitemShut {NoStop}%
\bibitem [{\citenamefont {O'Dell}(2001)}]{ODell_2001}%
  \BibitemOpen
  \bibfield  {author} {\bibinfo {author} {\bibfnamefont {D.~H.~J.}\ \bibnamefont {O'Dell}},\ }\bibfield  {title} {\bibinfo {title} {Dynamical diffraction in sinusoidal potentials: uniform approximations for {M}athieu functions},\ }\href {https://doi.org/10.1088/0305-4470/34/18/316} {\bibfield  {journal} {\bibinfo  {journal} {J. Phys. A: Math. Gen.}\ }\textbf {\bibinfo {volume} {34}},\ \bibinfo {pages} {3897} (\bibinfo {year} {2001})}\BibitemShut {NoStop}%
\bibitem [{\citenamefont {Chapman}\ \emph {et~al.}(1999)\citenamefont {Chapman}, \citenamefont {Lawry}, \citenamefont {Ockendon},\ and\ \citenamefont {Tew}}]{Chapman1999}%
  \BibitemOpen
  \bibfield  {author} {\bibinfo {author} {\bibfnamefont {S.~J.}\ \bibnamefont {Chapman}}, \bibinfo {author} {\bibfnamefont {J.~M.~H.}\ \bibnamefont {Lawry}}, \bibinfo {author} {\bibfnamefont {J.~R.}\ \bibnamefont {Ockendon}},\ and\ \bibinfo {author} {\bibfnamefont {R.~H.}\ \bibnamefont {Tew}},\ }\bibfield  {title} {\bibinfo {title} {On the theory of complex rays},\ }\href {http://www.jstor.org/stable/2653264} {\bibfield  {journal} {\bibinfo  {journal} {SIAM Review}\ }\textbf {\bibinfo {volume} {41}},\ \bibinfo {pages} {417} (\bibinfo {year} {1999})}\BibitemShut {NoStop}%
\bibitem [{\citenamefont {Link}\ \emph {et~al.}(2022)\citenamefont {Link}, \citenamefont {M{\ifmmode\ddot{u}\else\"{u}\fi}ller}, \citenamefont {Lena}, \citenamefont {Luoma}, \citenamefont {Damanet}, \citenamefont {Strunz},\ and\ \citenamefont {Daley}}]{Link2021Dec}%
  \BibitemOpen
  \bibfield  {author} {\bibinfo {author} {\bibfnamefont {V.}~\bibnamefont {Link}}, \bibinfo {author} {\bibfnamefont {K.}~\bibnamefont {M{\ifmmode\ddot{u}\else\"{u}\fi}ller}}, \bibinfo {author} {\bibfnamefont {R.~G.}\ \bibnamefont {Lena}}, \bibinfo {author} {\bibfnamefont {K.}~\bibnamefont {Luoma}}, \bibinfo {author} {\bibfnamefont {F.}~\bibnamefont {Damanet}}, \bibinfo {author} {\bibfnamefont {W.~T.}\ \bibnamefont {Strunz}},\ and\ \bibinfo {author} {\bibfnamefont {A.~J.}\ \bibnamefont {Daley}},\ }\bibfield  {title} {\bibinfo {title} {{Non-Markovian Quantum Dynamics in Strongly Coupled Multimode Cavities Conditioned on Continuous Measurement}},\ }\href {https://doi.org/10.1103/PRXQuantum.3.020348} {\bibfield  {journal} {\bibinfo  {journal} {PRX Quantum}\ }\textbf {\bibinfo {volume} {3}},\ \bibinfo {pages} {020348} (\bibinfo {year} {2022})}\BibitemShut {NoStop}%
\bibitem [{\citenamefont {Dykman}\ \emph {et~al.}(1994)\citenamefont {Dykman}, \citenamefont {Millonas},\ and\ \citenamefont {Smelyanskiy}}]{Dykman1994Nov}%
  \BibitemOpen
  \bibfield  {author} {\bibinfo {author} {\bibfnamefont {M.~I.}\ \bibnamefont {Dykman}}, \bibinfo {author} {\bibfnamefont {M.~M.}\ \bibnamefont {Millonas}},\ and\ \bibinfo {author} {\bibfnamefont {V.~N.}\ \bibnamefont {Smelyanskiy}},\ }\bibfield  {title} {\bibinfo {title} {{Observable and hidden singular features of large fluctuations in nonequilibrium systems}},\ }\href {https://doi.org/10.1016/0375-9601(94)90426-X} {\bibfield  {journal} {\bibinfo  {journal} {Phys. Lett. A}\ }\textbf {\bibinfo {volume} {195}},\ \bibinfo {pages} {53} (\bibinfo {year} {1994})}\BibitemShut {NoStop}%
\bibitem [{\citenamefont {Kamenev}(2011)}]{Kamenev2011}%
  \BibitemOpen
  \bibfield  {author} {\bibinfo {author} {\bibfnamefont {A.}~\bibnamefont {Kamenev}},\ }\href@noop {} {\emph {\bibinfo {title} {Field Theory of Non-Equilibrium Systems}}}\ (\bibinfo  {publisher} {Cambridge University Press},\ \bibinfo {year} {2011})\BibitemShut {NoStop}%
\bibitem [{\citenamefont {Lazarescu}\ \emph {et~al.}(2019)\citenamefont {Lazarescu}, \citenamefont {Cossetto}, \citenamefont {Falasco},\ and\ \citenamefont {Esposito}}]{Lazarescu2019Aug}%
  \BibitemOpen
  \bibfield  {author} {\bibinfo {author} {\bibfnamefont {A.}~\bibnamefont {Lazarescu}}, \bibinfo {author} {\bibfnamefont {T.}~\bibnamefont {Cossetto}}, \bibinfo {author} {\bibfnamefont {G.}~\bibnamefont {Falasco}},\ and\ \bibinfo {author} {\bibfnamefont {M.}~\bibnamefont {Esposito}},\ }\bibfield  {title} {\bibinfo {title} {{Large deviations and dynamical phase transitions in stochastic chemical networks}},\ }\href {https://doi.org/10.1063/1.5111110} {\bibfield  {journal} {\bibinfo  {journal} {J. Chem. Phys.}\ }\textbf {\bibinfo {volume} {151}},\ \bibinfo {pages} {064117} (\bibinfo {year} {2019})}\BibitemShut {NoStop}%
\bibitem [{\citenamefont {Touchette}(2009)}]{Touchette2009Jul}%
  \BibitemOpen
  \bibfield  {author} {\bibinfo {author} {\bibfnamefont {H.}~\bibnamefont {Touchette}},\ }\bibfield  {title} {\bibinfo {title} {{The large deviation approach to statistical mechanics}},\ }\href {https://doi.org/10.1016/j.physrep.2009.05.002} {\bibfield  {journal} {\bibinfo  {journal} {Phys. Rep.}\ }\textbf {\bibinfo {volume} {478}},\ \bibinfo {pages} {1} (\bibinfo {year} {2009})}\BibitemShut {NoStop}%
\bibitem [{\citenamefont {Jack}(2020)}]{Jack2020Apr}%
  \BibitemOpen
  \bibfield  {author} {\bibinfo {author} {\bibfnamefont {R.~L.}\ \bibnamefont {Jack}},\ }\bibfield  {title} {\bibinfo {title} {{Ergodicity and large deviations in physical systems with stochastic dynamics}},\ }\href {https://doi.org/10.1140/epjb/e2020-100605-3} {\bibfield  {journal} {\bibinfo  {journal} {Eur. Phys. J. B}\ }\textbf {\bibinfo {volume} {93}},\ \bibinfo {pages} {1} (\bibinfo {year} {2020})}\BibitemShut {NoStop}%
\bibitem [{\citenamefont {Causer}\ \emph {et~al.}(2022)\citenamefont {Causer}, \citenamefont {Ba{\ifmmode\tilde{n}\else\~{n}\fi}uls},\ and\ \citenamefont {Garrahan}}]{Causer2022Mar}%
  \BibitemOpen
  \bibfield  {author} {\bibinfo {author} {\bibfnamefont {L.}~\bibnamefont {Causer}}, \bibinfo {author} {\bibfnamefont {M.~C.}\ \bibnamefont {Ba{\ifmmode\tilde{n}\else\~{n}\fi}uls}},\ and\ \bibinfo {author} {\bibfnamefont {J.~P.}\ \bibnamefont {Garrahan}},\ }\bibfield  {title} {\bibinfo {title} {{Finite Time Large Deviations via Matrix Product States}},\ }\href {https://doi.org/10.1103/PhysRevLett.128.090605} {\bibfield  {journal} {\bibinfo  {journal} {Phys. Rev. Lett.}\ }\textbf {\bibinfo {volume} {128}},\ \bibinfo {pages} {090605} (\bibinfo {year} {2022})}\BibitemShut {NoStop}%
\bibitem [{\citenamefont {Wright}(1980)}]{Wright80}%
  \BibitemOpen
  \bibfield  {author} {\bibinfo {author} {\bibfnamefont {F.~J.}\ \bibnamefont {Wright}},\ }\bibfield  {title} {\bibinfo {title} {The {S}tokes set of the cusp diffraction catastrophe},\ }\href {https://doi.org/10.1088/0305-4470/13/9/018} {\bibfield  {journal} {\bibinfo  {journal} {J. Phys. A: Math. Gen.}\ }\textbf {\bibinfo {volume} {13}},\ \bibinfo {pages} {2913} (\bibinfo {year} {1980})}\BibitemShut {NoStop}%
\bibitem [{\citenamefont {Berry}\ and\ \citenamefont {Howls}(1994)}]{Berry94}%
  \BibitemOpen
  \bibfield  {author} {\bibinfo {author} {\bibfnamefont {M.~V.}\ \bibnamefont {Berry}}\ and\ \bibinfo {author} {\bibfnamefont {C.~J.}\ \bibnamefont {Howls}},\ }\bibfield  {title} {\bibinfo {title} {High orders of the weyl expansion for quantum billiards: resurgence of periodic orbits, and the {S}tokes phenomenon},\ }\href {https://doi.org/10.1098/rspa.1994.0154} {\bibfield  {journal} {\bibinfo  {journal} {Proc. R. Soc. Lond. A}\ }\textbf {\bibinfo {volume} {447}},\ \bibinfo {pages} {527} (\bibinfo {year} {1994})}\BibitemShut {NoStop}%
\bibitem [{\citenamefont {Berry}(1977)}]{Berry1977dec}%
  \BibitemOpen
  \bibfield  {author} {\bibinfo {author} {\bibfnamefont {M.~V.}\ \bibnamefont {Berry}},\ }\bibfield  {title} {\bibinfo {title} {Focusing and twinkling: critical exponents from catastrophes in non-{G}aussian random short waves},\ }\href {https://doi.org/10.1088/0305-4470/10/12/015} {\bibfield  {journal} {\bibinfo  {journal} {J. Phys. A: Math Gen.}\ }\textbf {\bibinfo {volume} {10}},\ \bibinfo {pages} {2061} (\bibinfo {year} {1977})}\BibitemShut {NoStop}%
\bibitem [{\citenamefont {Kelvin}(1905)}]{Kelvin1905}%
  \BibitemOpen
  \bibfield  {author} {\bibinfo {author} {\bibfnamefont {L.}~\bibnamefont {Kelvin}},\ }\bibfield  {title} {\bibinfo {title} {Deep water ship-waves},\ }\href {https://doi.org/10.1080/14786440509463327} {\bibfield  {journal} {\bibinfo  {journal} {Philos. Mag.}\ }\textbf {\bibinfo {volume} {9}},\ \bibinfo {pages} {733} (\bibinfo {year} {1905})}\BibitemShut {NoStop}%
\bibitem [{\citenamefont {Huckans}\ \emph {et~al.}(2009)\citenamefont {Huckans}, \citenamefont {Spielman}, \citenamefont {Tolra}, \citenamefont {Phillips},\ and\ \citenamefont {Porto}}]{Huckans2009Oct}%
  \BibitemOpen
  \bibfield  {author} {\bibinfo {author} {\bibfnamefont {J.~H.}\ \bibnamefont {Huckans}}, \bibinfo {author} {\bibfnamefont {I.~B.}\ \bibnamefont {Spielman}}, \bibinfo {author} {\bibfnamefont {B.~L.}\ \bibnamefont {Tolra}}, \bibinfo {author} {\bibfnamefont {W.~D.}\ \bibnamefont {Phillips}},\ and\ \bibinfo {author} {\bibfnamefont {J.~V.}\ \bibnamefont {Porto}},\ }\bibfield  {title} {\bibinfo {title} {Quantum and classical dynamics of a {B}ose-{E}instein condensate in a large-period optical lattice},\ }\href {https://doi.org/10.1103/PhysRevA.80.043609} {\bibfield  {journal} {\bibinfo  {journal} {Phys. Rev. A}\ }\textbf {\bibinfo {volume} {80}},\ \bibinfo {pages} {043609} (\bibinfo {year} {2009})}\BibitemShut {NoStop}%
\bibitem [{\citenamefont {Rosenblum}\ \emph {et~al.}(2014)\citenamefont {Rosenblum}, \citenamefont {Bechler}, \citenamefont {Shomroni}, \citenamefont {Kaner}, \citenamefont {Arusi-Parpar}, \citenamefont {Raz},\ and\ \citenamefont {Dayan}}]{Rosenblum2014Mar}%
  \BibitemOpen
  \bibfield  {author} {\bibinfo {author} {\bibfnamefont {S.}~\bibnamefont {Rosenblum}}, \bibinfo {author} {\bibfnamefont {O.}~\bibnamefont {Bechler}}, \bibinfo {author} {\bibfnamefont {I.}~\bibnamefont {Shomroni}}, \bibinfo {author} {\bibfnamefont {R.}~\bibnamefont {Kaner}}, \bibinfo {author} {\bibfnamefont {T.}~\bibnamefont {Arusi-Parpar}}, \bibinfo {author} {\bibfnamefont {O.}~\bibnamefont {Raz}},\ and\ \bibinfo {author} {\bibfnamefont {B.}~\bibnamefont {Dayan}},\ }\bibfield  {title} {\bibinfo {title} {Demonstration of fold and cusp catastrophes in an atomic cloud reflected from an optical barrier in the presence of gravity},\ }\href {https://doi.org/10.1103/PhysRevLett.112.120403} {\bibfield  {journal} {\bibinfo  {journal} {Phys. Rev. Lett.}\ }\textbf {\bibinfo {volume} {112}},\ \bibinfo {pages} {120403} (\bibinfo {year} {2014})}\BibitemShut {NoStop}%
\bibitem [{\citenamefont {Mossman}\ \emph {et~al.}(2021)\citenamefont {Mossman}, \citenamefont {Bersano}, \citenamefont {Forbes},\ and\ \citenamefont {Engels}}]{Mossman2021}%
  \BibitemOpen
  \bibfield  {author} {\bibinfo {author} {\bibfnamefont {M.~E.}\ \bibnamefont {Mossman}}, \bibinfo {author} {\bibfnamefont {T.~M.}\ \bibnamefont {Bersano}}, \bibinfo {author} {\bibfnamefont {M.~M.}\ \bibnamefont {Forbes}},\ and\ \bibinfo {author} {\bibfnamefont {P.}~\bibnamefont {Engels}},\ }\bibfield  {title} {\bibinfo {title} {Gravitational caustics in an atom laser},\ }\href {https://doi.org/10.1038/s41467-021-27555-3} {\bibfield  {journal} {\bibinfo  {journal} {Nat. Commun.}\ }\textbf {\bibinfo {volume} {12}},\ \bibinfo {pages} {7226} (\bibinfo {year} {2021})}\BibitemShut {NoStop}%
\bibitem [{\citenamefont {Petersen}\ \emph {et~al.}(2013)\citenamefont {Petersen}, \citenamefont {Weyland}, \citenamefont {Paganin}, \citenamefont {Simula}, \citenamefont {Eastwood},\ and\ \citenamefont {Morgan}}]{Petersen2013Jan}%
  \BibitemOpen
  \bibfield  {author} {\bibinfo {author} {\bibfnamefont {T.~C.}\ \bibnamefont {Petersen}}, \bibinfo {author} {\bibfnamefont {M.}~\bibnamefont {Weyland}}, \bibinfo {author} {\bibfnamefont {D.~M.}\ \bibnamefont {Paganin}}, \bibinfo {author} {\bibfnamefont {T.~P.}\ \bibnamefont {Simula}}, \bibinfo {author} {\bibfnamefont {S.~A.}\ \bibnamefont {Eastwood}},\ and\ \bibinfo {author} {\bibfnamefont {M.~J.}\ \bibnamefont {Morgan}},\ }\bibfield  {title} {\bibinfo {title} {Electron vortex production and control using aberration induced diffraction catastrophes},\ }\href {https://doi.org/10.1103/PhysRevLett.110.033901} {\bibfield  {journal} {\bibinfo  {journal} {Phys. Rev. Lett.}\ }\textbf {\bibinfo {volume} {110}},\ \bibinfo {pages} {033901} (\bibinfo {year} {2013})}\BibitemShut {NoStop}%
\bibitem [{\citenamefont {Riddell}\ \emph {et~al.}(2023)\citenamefont {Riddell}, \citenamefont {Kirkby}, \citenamefont {O'Dell},\ and\ \citenamefont {S\o{}rensen}}]{Riddell2023sep}%
  \BibitemOpen
  \bibfield  {author} {\bibinfo {author} {\bibfnamefont {J.}~\bibnamefont {Riddell}}, \bibinfo {author} {\bibfnamefont {W.}~\bibnamefont {Kirkby}}, \bibinfo {author} {\bibfnamefont {D.~H.~J.}\ \bibnamefont {O'Dell}},\ and\ \bibinfo {author} {\bibfnamefont {E.~S.}\ \bibnamefont {S\o{}rensen}},\ }\bibfield  {title} {\bibinfo {title} {Scaling at the out-of-time-ordered correlator wavefront: Free versus chaotic models},\ }\href {https://doi.org/10.1103/PhysRevB.108.L121108} {\bibfield  {journal} {\bibinfo  {journal} {Phys. Rev. B}\ }\textbf {\bibinfo {volume} {108}},\ \bibinfo {pages} {L121108} (\bibinfo {year} {2023})}\BibitemShut {NoStop}%
\bibitem [{\citenamefont {Berry}(1983)}]{Berry1983LesHouches}%
  \BibitemOpen
  \bibfield  {author} {\bibinfo {author} {\bibfnamefont {M.~V.}\ \bibnamefont {Berry}},\ }\bibfield  {title} {\bibinfo {title} {Semiclassical mechanics of regular and irregular motion},\ }in\ \href@noop {} {\emph {\bibinfo {booktitle} {Chaotic Behaviour of Deterministic Systems}}},\ \bibinfo {series} {Proceedings of the Les Houches Summer School of Theoretical Physics}, Vol.~\bibinfo {volume} {35},\ \bibinfo {editor} {edited by\ \bibinfo {editor} {\bibfnamefont {G.}~\bibnamefont {Iooss}}, \bibinfo {editor} {\bibfnamefont {R.~H.~G.}\ \bibnamefont {Helleman}},\ and\ \bibinfo {editor} {\bibfnamefont {R.}~\bibnamefont {Stora}}}\ (\bibinfo  {publisher} {North-Holland},\ \bibinfo {address} {Amsterdam},\ \bibinfo {year} {1983})\ p.\ \bibinfo {pages} {171–271}\BibitemShut {NoStop}%
\bibitem [{\citenamefont {Arnold}(1989)}]{Arnold_book}%
  \BibitemOpen
  \bibfield  {author} {\bibinfo {author} {\bibfnamefont {V.~I.}\ \bibnamefont {Arnold}},\ }\href@noop {} {\emph {\bibinfo {title} {Mathematical Methods of Classical Mechanics}}}\ (\bibinfo  {publisher} {Springer},\ \bibinfo {address} {New York},\ \bibinfo {year} {1989})\BibitemShut {NoStop}%
\bibitem [{\citenamefont {Brandino}\ \emph {et~al.}(2015)\citenamefont {Brandino}, \citenamefont {Caux},\ and\ \citenamefont {Konik}}]{Brandino2015}%
  \BibitemOpen
  \bibfield  {author} {\bibinfo {author} {\bibfnamefont {G.~P.}\ \bibnamefont {Brandino}}, \bibinfo {author} {\bibfnamefont {J.-S.}\ \bibnamefont {Caux}},\ and\ \bibinfo {author} {\bibfnamefont {R.~M.}\ \bibnamefont {Konik}},\ }\bibfield  {title} {\bibinfo {title} {Glimmers of a quantum {KAM} theorem: Insights from quantum quenches in one-dimensional {B}ose gases},\ }\href {https://doi.org/10.1103/PhysRevX.5.041043} {\bibfield  {journal} {\bibinfo  {journal} {Phys. Rev. X}\ }\textbf {\bibinfo {volume} {5}},\ \bibinfo {pages} {041043} (\bibinfo {year} {2015})}\BibitemShut {NoStop}%
\bibitem [{\citenamefont {Karrasch}\ and\ \citenamefont {Schuricht}(2013)}]{Karrasch2013may}%
  \BibitemOpen
  \bibfield  {author} {\bibinfo {author} {\bibfnamefont {C.}~\bibnamefont {Karrasch}}\ and\ \bibinfo {author} {\bibfnamefont {D.}~\bibnamefont {Schuricht}},\ }\bibfield  {title} {\bibinfo {title} {Dynamical phase transitions after quenches in nonintegrable models},\ }\href {https://doi.org/10.1103/PhysRevB.87.195104} {\bibfield  {journal} {\bibinfo  {journal} {Phys. Rev. B}\ }\textbf {\bibinfo {volume} {87}},\ \bibinfo {pages} {195104} (\bibinfo {year} {2013})}\BibitemShut {NoStop}%
\bibitem [{\citenamefont {Abanin}\ \emph {et~al.}(2019)\citenamefont {Abanin}, \citenamefont {Altman}, \citenamefont {Bloch},\ and\ \citenamefont {Serbyn}}]{Abanin2019}%
  \BibitemOpen
  \bibfield  {author} {\bibinfo {author} {\bibfnamefont {D.~A.}\ \bibnamefont {Abanin}}, \bibinfo {author} {\bibfnamefont {E.}~\bibnamefont {Altman}}, \bibinfo {author} {\bibfnamefont {I.}~\bibnamefont {Bloch}},\ and\ \bibinfo {author} {\bibfnamefont {M.}~\bibnamefont {Serbyn}},\ }\bibfield  {title} {\bibinfo {title} {Colloquium: Many-body localization, thermalization, and entanglement},\ }\href {https://doi.org/10.1103/RevModPhys.91.021001} {\bibfield  {journal} {\bibinfo  {journal} {Rev. Mod. Phys.}\ }\textbf {\bibinfo {volume} {91}},\ \bibinfo {pages} {021001} (\bibinfo {year} {2019})}\BibitemShut {NoStop}%
\bibitem [{\citenamefont {Van~Vleck}(1928)}]{VanVleck1928Feb}%
  \BibitemOpen
  \bibfield  {author} {\bibinfo {author} {\bibfnamefont {J.~H.}\ \bibnamefont {Van~Vleck}},\ }\bibfield  {title} {\bibinfo {title} {{The Correspondence Principle in the Statistical Interpretation of Quantum Mechanics}},\ }\href {https://doi.org/10.1073/pnas.14.2.178} {\bibfield  {journal} {\bibinfo  {journal} {Proc. Natl. Acad. Sci. U.S.A.}\ }\textbf {\bibinfo {volume} {14}},\ \bibinfo {pages} {178} (\bibinfo {year} {1928})}\BibitemShut {NoStop}%
\bibitem [{\citenamefont {Cao}\ and\ \citenamefont {Voth}(1996)}]{Cao1996Jan}%
  \BibitemOpen
  \bibfield  {author} {\bibinfo {author} {\bibfnamefont {J.}~\bibnamefont {Cao}}\ and\ \bibinfo {author} {\bibfnamefont {G.~A.}\ \bibnamefont {Voth}},\ }\bibfield  {title} {\bibinfo {title} {{Semiclassical approximations to quantum dynamical time correlation functions}},\ }\href {https://doi.org/10.1063/1.470898} {\bibfield  {journal} {\bibinfo  {journal} {J. Chem. Phys.}\ }\textbf {\bibinfo {volume} {104}},\ \bibinfo {pages} {273} (\bibinfo {year} {1996})}\BibitemShut {NoStop}%
\end{thebibliography}%
\newpage
\appendix

\section{WKB theory}\label{app:wkb}

In this appendix we give the details of the application of WKB theory to the FCTFIM \cite{Braun1993Jan,Shchesnovich2008Aug,Nissen2010Jun,Simon2012Nov}. The full Hamiltonian for the model reads
\begin{equation}
H=\frac{G}{2j}J_z^2+\Omega J_x.
\end{equation}
In WKB theory the goal is to determine approximations for the wave function of the energy eigenstates $\braket{j,m|E_n}$ where $H\ket{E_n}=E_n\ket{E_n}$ and $J_z\ket{j,m}=m\ket{j,m}$. For a convenient notation we set $\eta=m/j$ and $\psi_n(\eta)=\braket{j,m|E_n}$. To find a differential equation for $\psi_n$ consider
\begin{equation}
\begin{split}
\bra{j,m}H\ket{E_n}&=j\frac{G}{2}\eta^2\psi_n(\eta)+\frac{\Omega}{2}\bra{j,m}(J_++J_-)\ket{E_n}
\\&=j\frac{G}{2}\eta^2\psi_n(\eta)+j\frac{\Omega}{2}\sqrt{1-\eta^2+\frac{1}{j}(1+\eta)}\psi_n(\eta-\frac{1}{j})+j\frac{\Omega}{2}\sqrt{1-\eta^2+\frac{1}{j}(1-\eta)}\psi_n(\eta+\frac{1}{j}).
\end{split}
\end{equation}
We insert the Taylor series $\psi_n(\eta\pm\frac{1}{j})=\eul^{\pm\frac{1}{j} \partial_\eta}\psi_n(\eta)$ which yields
\begin{equation}
\begin{split}
E_n\psi_n(\eta)= j\frac{G}{2}\eta^2\psi_n(\eta)+j\frac{\Omega}{2}\sqrt{1-\eta^2+\frac{1}{j}(1+\eta)}\eul^{-\frac{1}{j} \partial_\eta}\psi_n(\eta)+j\frac{\Omega}{2}\sqrt{1-\eta^2+\frac{1}{j}(1-\eta)}\eul^{+\frac{1}{j} \partial_\eta}\psi_n(\eta).
\end{split}
\end{equation}
\quad \\
For large $j$ we can use the WKB ansatz $\psi_n(\eta)=A(E_n,\eta)\eul^{\ii j S(E_n,\eta)}$ and consider all terms of order $j^0$. For this we write $F'=\partial_\eta F$ and note that
\begin{equation}
\begin{split}
\left(\eul^{ \frac{1}{j}\partial_\eta}+\eul^{- \frac{1}{j}\partial_\eta}\right)A\eul^{\ii j S}&=
2\cos(S')A\eul^{\ii j S}+\mathcal{O}(j^{-1}).
\end{split}
\end{equation}
We also expand the square root term
\begin{equation}
\sqrt{1+\frac{1}{j}-\eta^2\pm\frac{1}{j}\eta}=\sqrt{1-\eta^2}+\mathcal{O}(j^{-1}),
\end{equation}
which gives in total
\begin{equation}
\begin{split}
    E_n= &j\frac{G}{2}\eta^2+j \Omega\sqrt{1-\eta^2}\cos(S')+\frac{\Omega}{2}\frac{1}{\sqrt{1-\eta^2}}\cos(S')\\&+\Omega\sqrt{1-\eta^2}\left(\ii \frac{A'}{A}\sin(S')+\ii \frac{1}{2} S''\cos(S')\right)-\ii\frac{\Omega}{2}\sin(S')\frac{\eta}{\sqrt{1-\eta^2}}+\mathcal{O}(j^{-1}).
\end{split}
\end{equation}\label{eq:WKB}
Real and imaginary parts of this give equations for $A$ and $S$ respectively. Note that for our purposes we do not have to determine the amplitude $A$ explicitly. For the action we obtain
\begin{equation}
\begin{split}
E_n/j =H(\partial_\eta S(E_n,\eta),\eta)+\mathcal{O}(j^{-1})\,,\qquad H(\phi,\eta)=\frac{G}{2}\eta^2+\Omega \sqrt{1-\eta^2}\cos\phi.
\end{split}
\end{equation}
Neglecting subleading terms, the solution of the resulting Hamilton-Jacobi equation is given by the classical action
\begin{equation}
\begin{split}
S(E_n,\eta)=\int\limits_{\eta^*}^\eta \phi(E_n,\eta')\diff \eta'\,,\qquad \phi(E,\eta)=\pm\arccos\left(\frac{E/j-\frac{G}{2}\eta^2}{\Omega\sqrt{1-\eta^2}}\right)
\end{split}
\end{equation}
where $\phi(E,\eta)$ is determined by $H(\phi(E,\eta),\eta)=E/j$. To ensure that boundary contributions vanish one has to choose $\eta^*$ as a classical turning point for which $\phi=0$. We choose the positive turning point $\eta^*>0$ here. Note that we have both positive and negative solutions for $S$. Therefore, we can write the wave function as
\begin{equation}
    \psi_n(\eta)\approx A(E_n,\eta)\cos(jS(E_n,\eta)).
\end{equation}
To obtain discrete energy values $E_n$ we can use the semiclassical (Bohr-Sommerfeld) quantization condition 
\begin{eqnarray}
    j S(E_n)=2\pi (n+1/2), \qquad n=0,1,2,...
\end{eqnarray}
where $S(E_n)$ denotes the classical action along a closed torus. In the present case one has to additionally account for a compact phase space. For $E<\frac{G}{2}$ one has to take
\begin{equation}
S(E)=\oint\phi(E,\eta)\diff \eta,
\end{equation}
whereas for $E>\frac{G}{2}$
\begin{equation}
S(E)=4\pi+\oint(\phi(E,\eta)-\pi)\diff \eta.
\end{equation}

\begin{figure}[b]\centering
\includegraphics{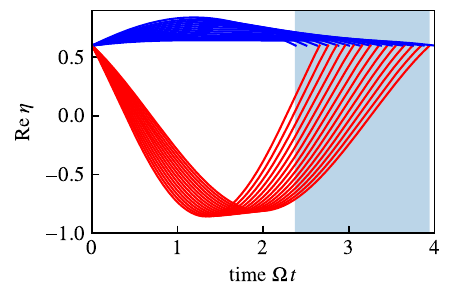}
\includegraphics{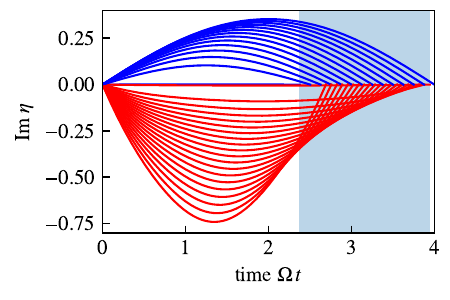}
\caption{Complex trajectories $\eta_{cpx}(t)$ that connect $\eta_{cpx}(0)=\eta_0$ and $\eta_{cpx}(t)=\eta_0$ in the DB region of classically forbidden times (shaded area). These trajectories connect smoothly to the real trajectory at the corresponding classical caustic. Blue trajectories correspond to the tail of the first caustic and the red trajectories correspond to the tail of the second caustic. The imaginary parts of the corresponding action along the trajectory yield the Loschmidt rate function in the DB region. }\label{fig:etacpx}
\end{figure} 

\section{Analytical continuation via complex mechanics}\label{app:complext}

In this appendix we show how an analytical continuation of the classical action can be obtained using complex classical trajectories.
For this consider the Hamilton-Jacobi equation for the time-dependent action $S(t,\eta)$
\begin{equation}
\partial_t S(t,\eta)=-H(\partial_\eta S(t,\eta),\eta)\,,\qquad  H(\phi,\eta)=\frac{G}{2}\eta^2+\Omega \sqrt{1-\eta^2}\cos\phi.
\end{equation}
This equation is valid for all $\eta$ and $t$ even in classically forbidden regions where a solution in terms of a real-valued classical action does not exist. In classically allowed regions the solution of the Hamilton-Jacobi equation can be obtained by finding the trajectory $\phi_{cl}(t),\,\eta_{cl}(t)$ from $\eta_{cl}(0)=\eta_0$ to $\eta_{cl}(t)=\eta$ obeying Hamiltons equations. This is a characteristic curve of the Hamilton-Jacobi equation and the action $S(t,\eta)$ reads
\begin{equation}
S(t,\eta)=\int_0^t\dot\eta_{cl}(s) \phi_{cl}(s)\diff s -H(\eta,\phi_{cl}(t))t\,.\label{eq:action_t}
\end{equation}
Note that by fixing the initial and final conditions one recovers the WKB action $S(E,\eta)$ from appendix~\ref{sec:WKB}, where the energy $E$ has to be chosen in correspondence to the time $t$. In the classically forbidden region no classical trajectories exist that connect the given initial condition with the given final condition at time $t$, and we expect the wave function to be exponentially decaying. Consequently, $S(t,\eta)$ will no longer be real-valued. In order to still solve the Hamilton-Jacobi equation one can consider complex initial conditions $\phi_{cpx}(0)\in \mathbb{C}$ and search for complex solutions of Hamilton's equations satisfying the boundary condition $\eta_{cpx}(0)=\eta_0\in \mathbb{R}$ and $\eta_{cpx}(t)=\eta\in \mathbb{R}$. This leads to both complex energies and a complex action that can be obtained from equation (\ref{eq:action_t}), upon replacing $\phi_{cl},\,\eta_{cl}$ with $\phi_{cpx},\,\eta_{cpx}$. We indeed find that at any time in the forbidden region two complex trajectories exist with a positive imaginary part of the action corresponding to an exponentially small wave function, see Fig.~\ref{fig:etacpx}. These two families of trajectories originate from the first and second caustic respectively, i.e.~they are continuously connected to the classical trajectories on the bright side of the corresponding caustic. Since we work directly in a time-dependent framework, the action \eqref{eq:action_t} obeys the steepest descent condition \eqref{eq:stat_phase} by construction. Thus, it is not necessary to compute energy-dependent actions for arbitrary complex energies $E$. The imaginary parts of the time-dependent complex actions directly yield the exact asymptotic rate function as displayed in Fig.~\ref{fig:loschmidt} (right panel, blue and red curves).

\section{Finite size}\label{app:finitesize}
Obtaining finite size corrections for semiclassics is nontrivial here, and the following arguments should not be seen as a rigorous derivation. Despite this word of caution, we do find good agreement of the result with exact finite-size numerics, as displayed in Fig.~\ref{fig:loschmidt} (right panel, dashed curve). For the finite size correction we consider the Fourier representation of the time-dependent wave-function 
\begin{eqnarray}
    \psi(\eta,t)=\frac{\sqrt{2\pi}}{\sqrt{2j+1}}\int \diff\phi \tilde\psi(\phi,t)\eul^{\ii j\phi\eta}.
\end{eqnarray}
Assigning $\eta$ as momentum and $\phi$ as the configuration-space coordinate, we assume the Van-Vleck form for the $\phi$-space wave function \cite{VanVleck1928Feb,Cao1996Jan}
\begin{equation}
\tilde\psi(\phi,t)=A(\phi,t)\eul^{\ii j S(t,\phi)}\,,\qquad A(\phi(t),t)=A(\phi(0),0)\sqrt{\frac{\partial \phi(0)}{\partial \phi(t)}}.
\end{equation}
Choosing a Fock state $\braket{\phi|\eta_0}=\frac{1}{\sqrt{2j+1}}\eul^{-\ii j \eta_0 \phi}$ as the initial state fixes the initial value $A(\phi(0),0)$. Evaluating the Fourier-integral using steepest descent we obtain for $\psi(\eta,t)$ 
\begin{equation}
\psi(\eta,t)=\frac{\sqrt{2\pi}}{\sqrt{2j+1}}\int \diff\phi \eul^{\ii j (\eta \phi+S(\phi,t))}A(\phi(t),t)\approx \frac{1}{\sqrt{2j+1}}\sqrt{\frac{j}{2j+1}}\frac{\sqrt{\frac{\partial \phi(0)}{\partial \phi(t)}}}{\sqrt{\frac{\partial \eta(t)}{\partial \phi(t)}}}\eul^{-\ii j \eta_0 \phi(0)} \eul^{\ii j (\eta(t) \phi(t)+S(\phi(t),t))},
\end{equation}
where we use $\frac{\partial^2S}{\partial \phi^2}=\frac{\partial \eta}{\partial \phi}$. Further replacing the exponent by the corresponding $\eta$-dependent action our result becomes
\begin{equation}
\psi(\eta,t)=\frac{1}{\sqrt{2j+1}}\sqrt{\frac{j}{2j+1}}\sqrt{\frac{\partial \phi(0)}{\partial \eta(t)}}\eul^{\ii j S(t,\eta(t))}.
\end{equation}
The prefactor then yields the most relevant finite-size corrections.
\end{document}